\documentclass[fleqn,usenatbib]{mnras}

\usepackage{newtxtext,newtxmath}

\usepackage[T1]{fontenc}

\usepackage{color}

\usepackage{graphicx}	
\usepackage{amsmath}	

\def\msun{\rm{\,M_{\odot}}}

\def\mtwo{M_{\rm 200}}
\def\rtwo{R_{\rm 200}}

\def\dtog{\rm{M_{dust}/M_{gas}}}

\definecolor{orange}{rgb}{1,0.5,0}

\definecolor{amethyst}{rgb}{0.6, 0.4, 0.8}
\definecolor{ao(english)}{rgb}{0.0, 0.5, 0.0}

\newcommand{\edi}[1]{{#1}}
\newcommand{\rev}[1]{{#1}}
\newcommand{\revrev}[1]{{#1}}
\definecolor{red}{rgb}{1.0, 0.0, 0.0}

\definecolor{turco}{rgb}{0.0, 0.5, 1.0}

\definecolor{applegreen}{rgb}{0.0, 0.5, 0.0}

\definecolor{viola}{rgb}{1.0, 0.0, 1.0}

\title[Dust evolution in cosmological simulations]{Dust evolution in zoom-in cosmological simulations of galaxy formation}

\author[G.L.\ Granato et al.]{
Gian Luigi Granato$^{1,2,3}$\thanks{E-mail: gian.granato@inaf.it},
Cinthia Ragone-Figueroa$^{2,1}$,
Antonela Taverna$^{2,1}$,
\\~\\
{\rm {\LARGE
Laura Silva$^{1,3}$,
Milena Valentini$^{4,5,1}$,
Stefano Borgani$^{6,1,7,3}$,
Pierluigi Monaco$^{6,1,3}$,}}
\\~\\
{\rm {\LARGE
Giuseppe Murante$^{1,3}$ and
Luca Tornatore$^{1,3}$}}
\\
$^{1}$ INAF, Osservatorio Astronomico di Trieste, via Tiepolo 11, I-34131, Trieste, Italy \\
$^{2}$ Instituto de Astronom\'ia Te\'orica y Experimental (IATE), Consejo Nacional de Investigaciones Cient\'ificas y T\'ecnicas de la\\ Rep\'ublica Argentina (CONICET), Universidad Nacional de C\'ordoba, Laprida 854, X5000BGR, C\'ordoba, Argentina\\
$^{3}$ IFPU - Institute for Fundamental Physics of the Universe, Via Beirut 2, 34014 Trieste, Italy\\
$^{4}$ Universit{\"a}ts-Sternwarte M{\"u}nchen, Fakult{\"a}t f{\"u}r Physik, LMU Munich, Scheinerstr. 1, 81679 M{\"u}nchen, Germany\\
$^{5}$ Excellence Cluster ORIGINS, Boltzmannstr. 2, D-85748 Garching, Germany\\
$^6$ Dipartimento di Fisica dell' Universit\`a di Trieste, Sezione di Astronomia, via Tiepolo 11, I-34131 Trieste, Italy
\\
$^7$ INFN - National Institute for Nuclear Physics, Via Valerio 2, I-34127 Trieste, Italy\\
}

\date{Accepted XXX. Received YYY; in original form ZZZ}

\pubyear{2021}

\begin{document}
\label{firstpage}
\pagerange{\pageref{firstpage}--\pageref{lastpage}}
\maketitle

\begin{abstract}
We present cosmological zoom-in hydro-dynamical simulations for the formation of disc galaxies, implementing dust evolution and dust promoted cooling of hot gas.
We couple an improved version of our previous treatment of dust evolution, which adopts the two-size approximation to estimate the grain size distribution, with the MUPPI star formation and feedback sub-resolution model. Our dust evolution model follows carbon and silicate dust separately. To distinguish differences induced by the chaotic behaviour of simulations from those genuinely due to different simulation set-up, we run each model six times, after introducing tiny perturbations in the initial conditions. With this method, we discuss the role of various dust-related physical processes and the effect of a few possible approximations adopted in the literature.

Metal depletion and dust cooling affect the evolution of the system, causing substantial variations in its stellar, gas and dust content.
We discuss possible effects on the Spectral Energy Distribution of the significant variations of the size distribution and chemical composition of grains, as predicted by our simulations during the evolution of the galaxy. We compare dust surface density, dust-to-gas ratio and small-to-large grain mass ratio as a function of galaxy radius and gas metallicity predicted by our fiducial run with recent observational estimates for three disc galaxies of different masses. The general agreement is good, in particular taking into account that we have not adjusted our model for this purpose.
\end{abstract}




\section{Introduction}
The interstellar medium (ISM) contains small solid particles, whose size may range from a few Angstrom to a few microns, and incorporate ({\it deplete}) a substantial fraction of the ISM metals.
We refer to them as (astrophysical) dust.
It is widely believed that dust grains consist of two main chemical classes: one carbon-based, and another dubbed ``astronomical silicates", whose composition is dominated by four elements, O, Si, Mg and Fe.  However, dust is a complex component of the ISM. Unsurprisingly, even its most basic physical properties, such as the chemical composition and the size distribution, depend on the environment, and evolve with time. Constraints on these properties derive mostly from dust reprocessing of the electromagnetic radiation produced by stars and active galactic nuclei, and from depletion studies \citep[e.g.][and references therein]{draine03,jenkins09}. However, in general dust reprocessing complicates the interpretation of observations and theoretical predictions.

Dust is an active component of the ISM and its presence substantially affects galaxy evolution. For instance, the chemical species that dust depletes from gas are important ISM coolants while in gaseous form. On the other hand, collision of ions with grains in the hot plasma contribute to its cooling \citep[e.g.][]{burke74,dwek81,montier04,vogelsberger19}.
Dust grain surfaces catalyze the formation of  H${_2}$ \citep[for a recent review see][]{wakelam17}, a key coolant and the primary constituent of molecular clouds (MCs) where new  stars form. Dust enhances by orders of magnitude the capability of interstellar matter (ISM) to receive radiation pressure, thus contributing to its dynamics in general, and in particular to the onset of galactic winds \cite[e.g.][]{Murray2005}. Dust can also ease the loss of angular momentum by gas in forming spheroidal galaxies, as required to fuel their central super-massive black holes \citep[e.g.][]{granato04}.

Despite its importance, dust has received for a long while relatively little attention in the context of galaxy formation models, partly due to the very complex and uncertain processes that regulate its life-cycle in the ISM. Until a few years ago, most of the efforts have been in the context of one zone, non-cosmological, chemical evolution models \citep[e,g,][]{dwek98,calura08,asano13,Feldmann2015,hirashita15}. These works paved the way to the recent inclusion of dust-related processes in some semi-analytic work \cite[e.g.][]{valiante11,popping16,vijayan19,triani20} and hydrodynamical simulations
\citep[e.g.][]{bekki13,Valentini2015,mckinnon16,mckinnon17,aoyama17,hou17,gjergo18,aoyama18,mckinnon18,li19,vogelsberger19,aoyama19,hou19,aoyama20,graziani20,osman2020}. Another possibility is to investigate the problem by means of post-processing computations based on results of simulations not including dust evolution
\citep[e.g.][]{zhukovska16,hirashita19}. On the one hand, the latter approach allows the fast exploration of sophisticated dust evolution model. However, it cannot capture the effects of dust on the simulation evolution.
Concurrent progress
and availability of computing power and simulation codes are making
this young field a rapidly evolving one, albeit plagued by strong uncertainties and important, almost unavoidable, simplifications.

The most studied effect of dust in astrophysics is by far radiation reprocessing, which can profoundly modify observational properties of stellar systems. As a general rule, this effect is an increasing function of the star formation activity and, as a consequence, of redshift. For a self-consistent comparison of galaxy formation models with observation, it would be ideal to have from the former ones a prediction of the relevant dust properties. These include not only the dust mass and its spatial distribution, but also the grain size distribution and the chemical composition. Indeed, the optical properties of grains strongly depend on their size and material. Nevertheless, the majority of the published simulations so far do not include a prediction either of the grain size or of its composition, or both.

The first model including a fairly complete treatment of the grain size distribution and its evolution, separately for carbonaceous and silicate dust, was \cite{asano13}. However, this work was in the framework of one-zone non cosmological models. It  predicted that the size distribution of grains should change drastically while galaxies evolve, because the dominant processes regulating it also change over time.
The general trend, confirmed by later works, is for an evolution from a distribution initially dominated by large grains ($\gtrsim 0.1 \mu$m) to a broad size distribution dominated in number, and to a lesser extent in mass, by the small radius end.
\cite{hirashita15} pointed out that the main features of the treatment by \cite{asano13} can be reproduced with a simple and numerically light approximation. He replaced the continuum size distribution with just two representative grain radii, henceforth referred to as small and large grains, whose limiting radius is at $a=0.03 \mu$m. The work was still in the context of one zone models, but this {\it two-size approximation} was explicitly thought for more numerically demanding galaxy evolution models, in particular for numerical simulations where memory and CPU time consumption are an issue\footnote{For other possible numerically effective approaches, so far much less exploited in galaxy formation simulations, see \cite{mattsson16} and \citet{Sumpter2020}.}. Indeed, soon later \cite{aoyama17} applied the two-size approximation to an SPH simulation of an idealized isolated galaxy, not including the distinction between the two main chemical species of dust grains. They found encouraging results such as a broad agreement with observed profiles of dust-to-gas and dust-to-metals ratio. In \cite{gjergo18} our group implemented the two-size approximation in simulations of galaxy cluster formation, separately for carbon and silicate dust.
We pointed out that high-z proto-clusters are expected to have a dust population significantly different from those inferred for the {\it average} Milky Way dust, and commonly adopted in dust reprocessing computations. In these initial phases dust is predicted to have much smaller grains and a lower fraction of silicate grains.

The two-size approximation has now been applied in quite a few simulations works. \cite{hou17}
improved the simulation by \cite{aoyama17} by separating dust species into carbonaceous and silicate, with the aim of predicting the extinction curve evolution. The approximation has also been employed in simulations of cosmological boxes by \cite{aoyama18} and \cite{hou19}, finding a broad agreement with the local dust mass functions of galaxies, and predicting that galaxies tend to have the highest dust content at $z=1-2$.  Interestingly, \cite{aoyama20} presented recently a simulation of an isolated disc implementing a full grain size distribution sampled with 32 grid point. They found results in very good agreement with those previously obtained by the same group with the two-size approximation \citep[e.g.][]{aoyama17,hou17,aoyama19}, further validating the method.

In this work, we implement, with a few improvements, the two-size approximation in a version of our simulation code which has been already shown to produce galaxies with a realistic disc morphology and chemical properties. The code relies on the star formation and feedback sub resolution model MUPPI \citep[MUlti Phase Particle Integrator,][]{murante10,murante15,Valentini2017,valentini19}. Moreover, we added the treatment of hot gas cooling via collisions of energetic electrons with dust grains, following the computations by \cite{dwek81}. We devote this paper mostly on assessing the effects of different approximations and assumptions, and to some preliminary comparison with observations, avoiding deliberately any adjustment to the previously selected values of the model parameters.

The plan of the paper is the following. Section \ref{sec:simu} is devoted to everything that concerns the numerical setup. In particular, we present in Subsection \ref{sec:dust} a thorough description of our dust creation and evolution model, and in Subsection \ref{sec:chaos} the method we use to cope with the chaotic behaviour of numerical simulation runs. Our results are described in Sections \ref{sec:results} and \ref{sec:obs}. In the former we study the impact of changing some relevant parameters, assumptions and approximations of different dust properties; while in the latter we present comparisons with recent quantities derived from observations. Finally, Section \ref{sec:summary} is dedicated to a brief summary of our work and its prospects.

We adopt a $\Lambda$CDM cosmology, with $\Omega_{\rm m}=0.25$, $\Omega_{\rm \Lambda}=0.75$,
$\Omega_{\rm baryon}=0.04$, $\sigma_8 = 0.9$, $n_s=1$,
and $H_{\rm 0}=100 \,h$ km s$^{-1}$ Mpc$^{-1}=73$ km s$^{-1}$ Mpc$^{-1}$.

\section{Numerical simulations}
\label{sec:simu}
The simulations have been performed with the GADGET3 code,
a non-public evolution of the GADGET2 \citep{springel05}. Our version of the code adopts the improved SPH (smoothed particle hydrodynamics) formulation by \cite{beck16}. The unresolved processes of star formation and stellar feedback are treated through the sub-resolution model MUPPI \citep[MUlti Phase Particle Integrator,][]{murante10,murante15,Valentini2017}, briefly described in the next section.
Moreover, in this work, we have implemented in MUPPI, with some relevant modifications (see Section \ref{sec:dust}), a treatment of dust formation and evolution similar to that by \cite{gjergo18}. Thus, we take into account the chemical composition and size distribution of dust grain. For the latter, we use the two-size approximation devised by \cite{hirashita15}. We have also included the process of hot gas cooling promoted by collisions of ions with dust particles, following \cite{dwek81}.

\subsection{Initial conditions and gravitational softenings}
\label{ICs}

We perform cosmological hydrodynamical simulations from zoom-in initial conditions (ICs) which produce at $z=0$ an isolated dark matter (DM) halo of  mass $ \simeq 2 \times 10^{12}$~M$_{\odot}$.
The details on these ICs, identified as AqC, can be found in \cite{Springel2008,scannapieco12}. In particular, baryons and high resolution DM particles define a Lagrangian volume that, by z=0, includes a sphere of $\sim 3$~Mpc radius centred on the main galaxy.
\edi{The general expectation is that the simulated halo forms a disc galaxy at low redshift, since it does not experience major mergers at late time. }

In this paper, we present simulations performed at two different resolutions. Low-resolution (LR, usually dubbed in this work C6) simulations are analyzed to explore the effect of varying parameters and physical assumptions in the treatment of dust evolution. High-resolution (HR or C5) simulations are considered when comparing with observations our fiducial model. In the language of \cite{scannapieco12} ({\it the Aquila comparison project}), LR-C6 and HR-C5 simulations correspond to level-6 and level-5, respectively.
At HR-C5 (LR-C6) the mass resolution for the DM is $2.2 \times 10^6$~M$_{\odot}$ ($1.8 \times 10^7 $~M$_{\odot}$). The masses of gas particles change during the simulation. They can decrease due to star formation, or increase because of gas return by neighbour star particles. Initially, they are $4.1 \times 10^5$~M$_{\odot}$ at HR and $6.6 \times 10^6$~M$_{\odot}$ at LR.
As for the computation of the gravitational force, we use a Plummer-equivalent softening length of $445$~pc ($890$~pc) for HR (LR) resolution simulations, constant in comoving units at $z>6$, and constant in physical units at $z \le 6$.

\subsection{Star formation, Stellar feedback and Chemical enrichment}
\label{sec:muppi}
To describe the processes occurring on unresolved scales, in particular star formation and stellar feedback, we adopt the sub-resolution model MUPPI, in the specific version described by \cite{valentini19}. In this section, we recall its main features, while we refer the reader to the latter paper, and references therein, for a full account.

MUPPI describes a multi-phase (MP) ISM.
\edi{An SPH particle is treated as MP if it increases its density above a threshold ($n_{\rm thres}=0.01$ cm$^{-3}$) and its temperature drops below another threshold ($T_{\rm thres}=10 ^5$ K).} A MP particle consists of a hot and a cold gas phases in pressure equilibrium, plus a "virtual" stellar component. \edi{Ordinary differential equations describe mass and energy exchanges between these components.}
We indicate with $f_{\rm hot}$ and $f_{\rm cold}$ the mass fraction of the gas particles in the two phases.
Hot gas condenses into a cold phase (whose assumed temperature is $T_{c}=300$ K) due to radiative cooling,
while some cold gas evaporates due to the destruction of molecular clouds. \rev{The densities of the two phases, used in the dust modelling of this paper, are computed from their filling factor, as described in section 3.1 of \cite{murante15}.}
A fraction $f_{\rm mol}$ of the cold gas mass is in the molecular phase, from which a virtual stellar component forms, according to a given efficiency.
We rely on the phenomenological prescription by \citet{blitz06} to estimate $f_{\rm mol}$.
This is an (almost) linear relation between the ratio of molecular and neutral hydrogen surface densities and the midplane pressure of galaxy discs, and is implemented in our code to compute the molecular fraction from hydrodynamical pressure; it can be seen as a convenient bypass to the complex process of molecular cloud formation, that produces a good order-of-magnitude estimate of the molecular fraction, at least at solar metallicities. Over time, MP gas particles generate new stellar particles, using the virtual stellar mass accumulated previously as a consequence of its star formation rate (SFR). The latter process uses a stochastic algorithm \citep{springel03}.

Besides radiative cooling, including here also that promoted by dust (Section \ref{sec:cool}), the energy budget of the hot phase is affected by a pure hydrodynamical term, accounting for shocks and heating or cooling due to gravitational compression or expansion of the gas, and by thermal stellar feedback. The latter includes both the contribution of supernovae explosions within the stellar component of the particle itself, as well as that coming from neighbouring particles.

\edi{A gas particle stays MP for a maximum allowed time given by the dynamical time of its cold phase, provided that the density keeps above the threshold.
When exiting from the MP state, the particle has a probability (a parameter set to 3\%) to be kicked and to becomes a wind particle for a given time interval. In this case, it is decoupled from the hydrodynamic forces due to the surrounding medium, but it can receive kinetic feedback energy from other particles.} This is used to increase their velocity along their own least resistance path, defined by the opposite direction of the gas density gradient. This sub-resolution model leads to the formation of disc galaxies with
morphological, kinematic and chemical properties in reasonable agreement with observations \citep[][and references therein]{valentini19}.

Star formation and evolution also produce a chemical feedback. Chemical evolution and enrichment processes follow \citet{tornatore07}.  Star particles act as simple stellar populations (SSPs).
The production of heavy elements include  the contributions from Asymptotic Giant Branch (AGB) stellar winds, Type Ia Supernovae (SNIa) and Type II Supernovae (SNII). We follow the production and evolution of 15 chemical elements: H, He, C, N, O, Ne, Na, Mg, Al, Si, S, Ar, Ca, Fe and Ni \citep[see][for the adopted stellar ejecta] {valentini19}, and radiative gas cooling takes into account their contribution. As we will see in the next section, the starting point of our treatment of dust evolution consists in assuming that a certain fraction of some of these elements, produced by the three stellar channels, are given back to medium in form of dust grains rather than in gaseous form.

\edi{The elements produced and ejected  by stars, both in gaseous and dust form (see next section for the latter) are distributed to the surrounding gas particles by using the SPH kernel. This spreading is not meant to constitute a specific physical treatment of their diffusion, but to avoid a noisy estimation of metal-dependent cooling rates.  Therefore heavy elements and dust can be spatially distributed after that only via dynamical processes involving the enriched gas particles. }

\subsection{Dust formation and evolution}
\label{sec:dust}
The dust evolution model used here is similar to that we presented in \cite{gjergo18}, with a few improvements detailed in the rest of this Section. In that paper the scope was to study the evolution of dust in galaxy clusters, and the star formation and stellar feedback model was a simpler one, namely an updated version of that proposed by \cite{springel03}. Our dust treatment builds on the proposal by \cite{hirashita15}, who demonstrated that it is possible to follow reasonably well the evolution of the size distribution of grains and to estimate the effectiveness of size-dependent processes just considering two representative sizes, which we refer in the following as large and small grains.
The boundary between them was set by \cite{hirashita15} at $a\simeq 0.03 \, \mu$m.
This two-size approximation is clearly advantageous in terms of computational cost, compared to more detailed treatments \citep[e.g.][]{mckinnon18,aoyama20}. Interestingly, the latter paper, based on a full treatment of the size distribution and its evolution, confirmed the results obtained previously by means of the two-size approximation.
In the present application, the overhead of computing time arising from the inclusion of dust evolution is limited to less than 15 per cent.

We exploit our advanced treatment of chemical evolution (Section \ref{sec:muppi}), which follow individually several elements, to trace two chemically distinct dust grain populations, namely carbonaceous and "astronomical silicate" grains, the latter composed by O, Si, Mg and Fe.
\edi{Therefore, we adopt the standard view \citep[e.g.][and references therein]{draine03,jenkins09} that dust is dominated by these two chemical species.
More specifically, we represent astronomical silicates with the element partition of Olivine MgFeSiO$_4$ \citep{draine03}.\footnote{More properly, the term {\it olivine} in mineralogy indicates a composition in which Fe and Mg atoms are mutually interchangeable, which can be expressed as Mg$_X$Fe$_{2-X}$SiO$_4$. The two extreme cases are {\it forsterite} Mg$_2$SiO$_4$ and {\it fayalite} Fe$_2$SiO$_4$. Here we follow the common practise in dust astrophysics of considering as representative the intermediate case in which $X=1$.}
Actually, most radiative transfer calculations in astrophysical dusty environment adopt optical properties of dust grains calculated by assuming this chemical composition. However, our treatment can be trivially adapted to different dust composition models, provided that the relevant elements are followed by the chemical evolution. In conclusion, we predict the evolution of four classes of dust grains: large and small C grains, and large and small silicate grains.}

For the sake of clearness, we briefly anticipate here, before digging into the details, the improvements
that we have implemented for the present work compared to \cite{gjergo18}: (i) we replaced the step-like transition between shattering and coagulation, occurring when density increases, with a more gentle and realistic one (Section \ref{sec:shattering}). Shattering or coagulation are two possible opposite results of grain-grain collision, and we found that a smoother transition is required to get small over large grain ratios compatible with observations (Section \ref{sec:sharp_sha}); (ii) we modified the treatment of accretion of gas onto seed silicate dust grains with respect to the same desired abundance ratios of the various elements in the grains (Section \ref{sec:acc_cou}) assumed at stellar production of dust (Section \ref{dsynth}). Indeed, while in the cluster simulations presented by \cite{gjergo18} accretion was not important enough to strongly modify these ratios, because cluster galaxies tend to lose their gas relatively soon
their gas, the same is not true for disc galaxies (Section \ref{sec:fre}), and a more careful treatment is required; (iii) to estimate the fraction of multiphase SPH gas particle in molecular clouds, where the processes of accretion and coagulation can occur efficiently (Section \ref{sec:acc_cou}), we adopted for consistency the same prescription used by the star formation and feedback model MUPPI (Section \ref{sec:muppi});
(iv) we introduced the process of gas cooling promoted by ions-grains collision (Section \ref{sec:cool});

\begin{figure*}
    \includegraphics[width=14cm]{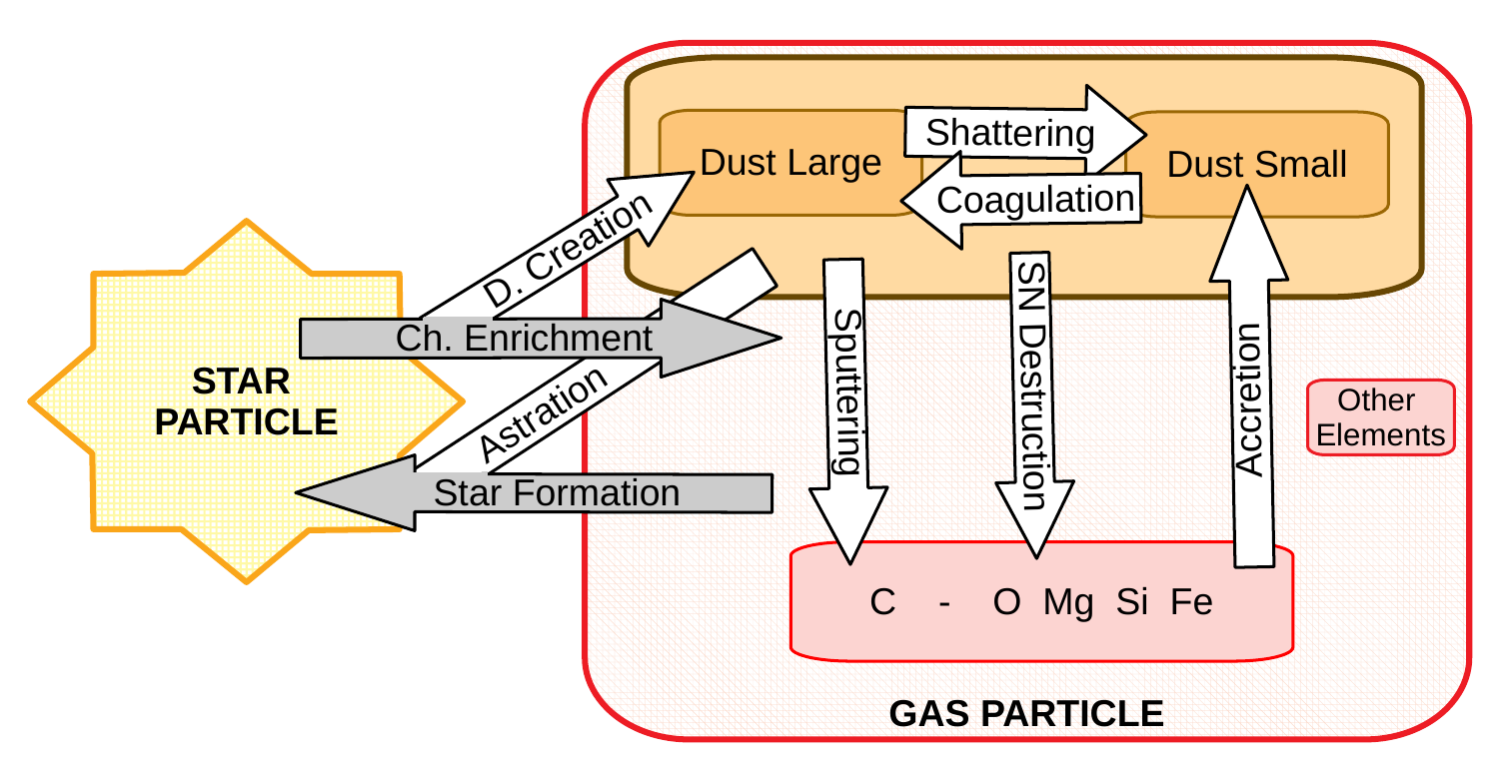}
    \caption{Schematic view of the mass flows produced by the different dust processes (white arrows). Each gas particle has a dust mass budget which is divided in large and small grains. The grain cycle begins with ${\it \boldsymbol{Dust~Creation}}$ of large grains in the ejecta of stars (AGB, SNII and SNIa). Once injected into the ISM, dust grains are affected by various evolutionary processes.
    Depending on the collision velocity, grain-grain collisions can either result in grain Shattering or Coagulation. ${\it \boldsymbol{Shattering}}$ dominates in the diffuse ISM, leading to the fragmentation of large grains in small grains. It gives rise to the seeds of small grains. ${\it \boldsymbol{Coagulation}}$ instead, dominates in the dense ISM where velocities are low, moving the grain size distribution towards larger grain sizes. Small grains can gain mass due to metal ${\it \boldsymbol{Accretion}}$ from the surrounding gas.
    On the contrary there is also a mass flow from dust to gas due to grain collisions with energetic ions. The dust atom erosion can be caused by ${\it \boldsymbol{SN~ destruction}}$, where grains are swept by SNae shocks, and by ${\it \boldsymbol{Sputtering}}$, where grains are subject to violent ion collisions in the ISM. Finally, when a gas particle spawns a new stellar particle a corresponding amount of dust is returned to the stellar component, this phenomenon is known as ${\it \boldsymbol{Astration}}$.}
    \label{fig:cartoon}
\end{figure*}

\edi{A treatment of dust evolution including its size distribution, rather than simply the total dust mass, turns out to be more reliable, because some processes are size dependent. Moreover, since the optical properties of grains strongly depend on their dimension, it is also beneficial for post-processing computations involving radiative transfer}, such as those presented in \cite{hou17}, \cite{gjergo18} and in Section \ref{sec:SED}.

The main processes affecting the dust content of galaxies can be summarized as follows. The life cycle of grains begins with {\it dust production} in the ejecta of stars, including AGB star winds as well as SNII and possibly SNIa explosions, although recent observational and theoretical work cast doubts on a significant contribution from SNIa (see Sections \ref{dsynth} and \ref{sec:IIagbIa}). In any case, it can be assumed that dust production affects directly only large grains \citep[e.g.][]{nozawa07,bianchi07}.

Once in the ISM, grains
undergo several physical processes influencing their size and chemical composition.
Gas metal atoms can stick on their surface. \citep{dwek98,hirashita99}. Being this {\it accretion} a surface process, it is more important for small grains, because of their larger total surface per unit mass. In the two size approximation framework, its {\it direct} effect is to increase the mass of small grains, while both the mass increase of large grains as well as the evolution of small grains to large ones can be safely neglected \citep{hirashita15}. However, an increased abundance of small grains causes a more effective production of big ones. Indeed, grain-grain collisions can either result in grain {\it coagulation} or {\it shattering}, depending on the collision velocity. The former process dominates in dense ISM where velocities are low, and it acts as a sink mechanism for small grains and a source mechanism for large ones.
Accretion on small grains in combination with their coagulation into larger ones increases the mass budget of the latter ones.
In this sense, accretion affects {\it indirectly} large grains mass as well.
Shattering dominates in the diffuse ISM, acting as a source of small grains and a sink of large ones.

{\it Sputtering} is the grain erosion due to collisions with energetic ions. It is another surface process, more effective on small grains. The eroded atoms are given back to the gas phase. It occurs both when grains are swept by SNae shocks, and when they are subject to violent ion collisions in the hot diffuse gas. On top of SNae dust destruction, caused both by kinetic and thermal relative motions, we include the latter thermal sputtering, which is important only when grains are surrounded by hot gas $T\gtrsim 5 \times 10^5$ K. \cite{hirashita15} and \cite{aoyama17} did not consider this form of sputtering, because they were interested only in galactic ISM. However, our results show that it can be relevant also when modelling the evolution of dust in disc galaxies (Section \ref{sec:results}). We refer to this process simply as {\it sputtering}, while we call the former one  {\it SNae destruction}.

When gas is turned into stars, its dust content is subtracted from the ISM, a process usually dubbed {\it astration}.
We found that it is not negligible compared to other dust destruction mechanisms (Section \ref{sec:no_ast}), contrariwise to recent claims \citep{mckinnon18}.
A visual summary of how the various processes act on gas and stellar particles in the simulation is presented in Fig. \ref{fig:cartoon}.

\edi{This processes are implemented in our simulations by mean of sub-resolution prescriptions.
A fraction of some of the metals ejected by star particles into the ISM is assumed to be given in form of grains rather than gas. This fraction depends on the specific metal, is computed as described in Section \ref{dsynth} and is greater than zero only for metals entering into the grain composition. To track these metals locked in grains,} the gas particle numerical structure includes two extra vectors,  which store the amount of each element that is in large and in small dust grains individually, rather than in gas. \edi{The total fraction of each "gas" particle in grains never surpasses a few per cent, and we still refer to them simply as SPH particles.} Therefore in our code it is assumed a perfect coupling between gas and dust. At the relatively coarse resolution of galaxy formation simulations, and for the typical conditions of galactic ISM this assumption is well justified \citep[e.g.][]{draine79,Murray2005,mckinnon18}. In particular, \cite{mckinnon18} presented a careful numerical framework including a state-of-the-art treatment of dust evolution and its size distribution. So far, this model has been tested only on a very idealized isolated disc galaxy simulation, with cooling and star formation but without any feedback. At variance with most numerical works on dust evolution in galaxies (including our own), dust is not assumed by construction to be perfectly dynamically coupled to gas. However, their galaxy simulations show that the drag force between the two components turns out to be strong enough to grant their effective coupling.
Taking into account all the aforementioned processes, the rate of change  of the dust mass, for each SPH particle in the two-size populations of small grains $M_{d,S}$ and large grains $M_{d,L}$, is:

\begin{equation}
\begin{split}
    \frac{d {M_{d,L}}}{dt} & =     \frac{d {M_{p*}}}{dt}   - \frac{M_{d,L}}{\tau_{sh}} + \frac{M_{d,S}}{\tau_{co}}  - \frac{M_{d,L}}{\tau_{SN,L}} - \frac{M_{d,L}}{\tau_{sp,L}}- \frac{M_{d,L}}{M_{gas}} \psi \\
    \frac{d { M_{d,S}}}{dt} & = \frac{M_{d,S}}{\tau_{acc}} + \frac{M_{d,L}}{\tau_{sh}} - \frac{M_{d,S}}{\tau_{co}} - \frac{M_{d,S}}{\tau_{SN,S}} - \frac{M_{d,S}}{ \tau_{sp,S}} - \frac{M_{d,S}}{M_{gas}}\psi
\end{split}
\end{equation}
where $\frac{d {M_{p*}}}{dt}$ is the dust production  rate by stars, and each ISM process is described by a corresponding timescales: $\tau_{sh}$ for shattering, $\tau_{co}$ for coagulation, $\tau_{SN,L}$ and $\tau_{SN,S}$ for SN shock destruction, $\tau_{sp,L}$ and $\tau_{sp,S}$ for sputtering, and $\tau_{acc}$ for accretion. The last term, wherein $\psi$ is the SFR, accounts for the dust mass loss due to star formation (astration). These equations are applied separately for each element entering into the dust grains. In the next subsections, we describe how we estimate in the simulations the various contributions.

\subsubsection{Dust production by stars} \label{dsynth}
A certain fraction of the elements concurring to dust composition is injected into the ISM by AGB winds, SNIa, and SNII as solid dust particles rather than as gas. Since we consider carbon and silicate dust, and we approximate the latter with olivine MgFeSiO$_4$, these elements are C, Si, O, Mg, and Fe.

\edi{For simplicity, we assume that for each of the three stellar dust formation channels, the fraction of each element {\it eligible} to condense to dust is independent of stellar mass and metallicity.
This approach, is partly similar to that introduced in one-zone models by \cite{dwek98}, and later adopted by many works  \citep[e.g.][]{calura08,hirashita15,aoyama17}. }
\rev{However, we note that the requirement of respecting the partition of MgFeSiO$_4$ for silicates, not considered in the papers mentioned above, introduces a dependence on stellar mass and metallicity of the actual dust condensation efficiencies of O, Mg, Fe and Si.  The reason is the dependence of stellar ejecta on the latter two parameters and is detailed in the rest of this Section.}

There are several studies on the dependency of dust condensation efficiencies on stellar mass, metallicity and ambient gas density \citep[see e.g.][and references therein]{nozawa07,schneider14}. In principle, it would be straightforward to incorporate these dependencies into our formalism. However, we have verified that our results are little affected by variations of the assumed efficiencies as large as 30\%, but for very high redshift $z \gtrsim 4$. This is because at late times most of the dust mass is produced by accretion of gas onto preexisting seed grains.

{\bf AGB}. \edi{Following \citet{dwek98}, we assume that the production of carbon or silicate dust is mutually exclusive in AGB winds, depending on the C/O number ratio in the ejecta. This view, supported by observations, relies on the assumption that AGB ejecta are mixed at microscopic scales. As a consequence, the maximum possible amount of CO forms. If C/O$>1$, all the oxygen is consumed to produce CO molecules and only the remaining carbon condenses to dust. When C/O$<1$, all the carbon ties to CO molecules. The leftover oxygen is then potentially available, but
usually not entirely consumed (see below), to condense into silicate grains, together with Mg, Si and Fe. }

\edi{We indicate with $M_{ej,O}^{AGB}$ and $M_{ej,C}^{AGB}$ the O and C masses ejected by the AGB winds of a star particle during a time-step.
If $M_{ej,C}^{AGB} > 0.75 M_{ej,O}^{AGB}$, where 0.75 is the ratio between O and C atomic weights, the particle produces carbon grains. Subtracting the C mass used to form CO molecules, we obtain that the corresponding mass of ejected carbon dust is
\begin{equation}
M_{dust,C}^{AGB}=
\max\left[\delta_{AGB,C}\left( M_{ej,C}^{AGB} - 0.75 \, M_{ej,O}^{AGB} \right)\, , \, 0 \right]
\end{equation}
In this work, we set
$\delta_{AGB,C}$, the {\it condensation efficiency} of carbon grains in AGB winds, to 1, following \cite{dwek98}  \citep[see also][]{calura08}.}

By converse, when $M_{ej,C}^{AGB} <  0.75 M_{ej,O}^{AGB}$, silicate grains condenses. \cite{dwek98} simply estimated the masses of the metals going into silicates by assuming that for each ejected atom of Si, Mg and Fe, an O atom will go into dust as well, as described by his equation 23.
As expected, by implementing this approach in our code we verified that it leads to the production of 'silicate grains' having a very variable mass ratios between its four elements, and in particular featuring a very large fraction of oxygen. These ratios are substantially different from those typically assumed by radiative transfer computations.
Therefore we adopted a different formulation, preserving the elemental mass partition of MgFeSiO\_4.
Our procedure is based on the idea that the availability of just one element sets the maximum number of "molecules" or groups of the assumed compound that is possible to form. This is the element for which, in the stellar ejecta, the number abundance divided by the number of its atoms entering into the compound is minimun.
This driving element has been dubbed {\it key element} by
\cite{zhukovska08}, and the concept has been used in later works to describe dust stellar production or accretion in the ISM, or both \citep[e.g.][see also Section \ref{sec:acc_cou} for the accretion process]{hirashita11,asano13,gjergo18,hou19,vijayan19}.

\edi{We indicate now with $M_{ej,X}^{AGB}$ the mass of the generic X element ejected by the AGB winds of a star particle during a time-step.
For olivine, X can be Mg, Fe, Si or O. Let
$N_{mol,sil}^{AGB}$ be the number of "molecules" of MgFeSiO$_4$ that can be formed during the time-step. This number is set by the element for which the number of atoms ejected divided by the number of its atoms in the compound $N_{ato}^X$ (1 for Mg, Fe, Si, and 4 for O) is minimum.} Then
\begin{equation}
\label{eq:nmol}
N_{mol,sil}^{AGB} = \delta_{AGB,sil} \min\limits_{X  \in \{Mg,Fe,Si,O\}} \left(\frac{M_{ej,X}^{AGB}}{\mu_X \, N_{ato}^X}\right)
\end{equation}
where $\mu_X$ is  the atomic weight of the X element,
and we have introduced an efficiency factor of condensation for silicate grains $\delta_{AGB,sil}$, set to 1 in our reference model.
Therefore, the mass of the X element locked into silicate grains is given by
\begin{equation}
\label{eq:mdust}
M_{dust,X}^{AGB}=
\begin{cases}
N_{mol,sil}^{AGB} \, \mu_X \, N_{ato}^X \; \; \mbox{for} \; \frac{M_{ej,C}^{AGB}}{M_{ej,O}^{AGB}} <  0.75  \\
0 \; \; \mbox{otherwise}
\end{cases}
\end{equation}
It is straightforward to adapt the above treatment to chemical compounds different from olivine.

{\bf SNae II and Ia}.
\edi{By converse, in the outflows produced by SNae explosions, carbon and silicate dust can condense at the same time, since such outflows are mixed only at macroscopically}
\citep[e.g.][and references therein]{dwek98}. Hence:
\begin{eqnarray}
M_{dust,C}^{SNx}&=&
\delta_{SNx} M_{ej,C}^{SNx}\\
M_{dust,X}^{SNx}&=&
N_{mol,sil}^{SNx} \, \mu_X \, N_{ato}^X\\
N_{mol,sil}^{SNx}&=& \delta_{SNx,sil} \min\limits_{X  \in \{Mg,Fe,Si,O\}} \left(\frac{M_{ej,X}^{SNx}}{\mu_X \, N_{ato}^X}\right)
\end{eqnarray}
where ${SNx}$ stands either for ${SNII}$ or ${SNIa}$,
and $M_{ej,X}^{SNx}$ is the mass of the generic $X$ element ejected by the ${SNx}$ explosions from a star particle during a time-step.
For SNae we assume a lower dust condensation efficiency of $\delta_{SNII,C} = \delta_{SNIa,C} = 0.5$ and $\delta_{SNII,sil}=\delta_{SNIa,sil}=0.8$ as in \cite{dwek98}. These values accounts for  incomplete condensation of available elements and for grain destruction by SN shock.
We point out that in the last years the role of individual SNIa in dust production has been substantially revised downward, based both on observational and theoretical studies \citep[see discussion in][and references therein]{gioannini17}. This could be easily accounted for by reducing the corresponding condensation efficiencies. However, since we found that even with the adopted values the global contribution of SNIa turns out to be minor in our simulations (see Section \ref{sec:results}), we do not perform this modification in the present work.

\subsubsection{Shattering}
\label{sec:shattering}
\edi{In the diffuse gas, large grains are decoupled from small-scale turbulent motions \citep{hirashita09}. Therefore they mutual collisions occurs at velocities high enough ($v \simeq 10\mbox{ km s}^{-1}$, \citealt{yan04}) to cause shattering into small grains.
Shattering originates small grains, without affecting the total dust mass. Its timescale is derived from the collision timescale:}
\begin{equation}
\tau_{\rm{coll}}=\frac{1}{v \sigma n}
\end{equation}
where $v$, $\sigma$ and $n$ are the typical collision velocity, the cross section and number density of colliding particles.
Based on the results by \cite{yan04}, we assume that $v = 10\mbox{ km s}^{-1}$ for gas density lower than $n_{gas}<1\mbox{ cm}^{-3}$, and that at higher densities $v$ decreases as $n_{gas}^{-2/3}$, so that  $v = 0.1\mbox{ km s}^{-1}$ for $n_{gas}=10^3\mbox{ cm}^{-3}$. Above the latter density, the shattering process is completely switched off because we assume instead that low velocity collisions result only in the coagulation of small grains to form large ones, as described in Section \ref{sec:acc_cou}.
Following Appendix B of \cite{aoyama17} this yields:
\begin{equation}
\label{eq:taush}
\tau_{sh} =
\begin{cases}
\tau_{sh,0}  \left(\dfrac{0.01}{D_L} \right) \left(\dfrac{1 \, \mbox{cm}^{-3}}{n_{gas}}\right) & \, \, \, \dfrac{n_{gas}}{1 \, \mbox{cm}^{-3}} < 1   \\
\tau_{sh,0}  \left(\dfrac{0.01}{D_L} \right) \left(\dfrac{1 \, \mbox{cm}^{-3}}{n_{gas}}\right)
\left(\dfrac{n_{gas}}{1 \, \mbox{cm}^{-3}}\right)^{2/3}
& \, \, \,   1 \le \dfrac{n_{gas}}{1 \, \mbox{cm}^{-3}} \le 10^3
\end{cases}
\end{equation}
where $\tau_{sh,0} = 5.41 \times 10^7 \mbox{yr}$ is obtained assuming a grain size of $0.1 \, \mu$m, and a material density of grains 3 g cm$^{-3}$. $D_L = M_{dust,L} / M_{gas}$ is the dust to gas ratio for large grains.

We explicitly note that at variance with previous implementations of the two-size approximation \citep{gjergo18,aoyama17}, we do not switch off abruptly shattering when $n_{gas} \ge 1 \mbox{ cm}^{-3}$. Besides being somewhat more realistic on physical grounds, we found that this smoother transition to the regime at which small velocities favour the opposite coagulation process (Section \ref{sec:acc_cou}) is required here to obtain a sufficient production of small grains\footnote{On the other hand, by repeating a couple of the galaxy cluster simulations presented in \cite{gjergo18}, we checked that the smooth transition introduced here has a negligible effect on them.}, as discussed in Section \ref{sec:sharp_sha}. In our simulations, densities above $1 \, \mbox{cm}^{-3}$ are not resolved, and most particles at density $>0.01 \, \mbox{cm}^{-3}$ are MP (Section \ref{sec:muppi}). In the latter case, we assume that shattering occurs only in the cold phase, so that we plug into Eq. \ref{eq:taush} the density of this component. The fact that we neglect shattering in the hot phase has two justifications. First of all it accounts typically for less than 1\% of the particle mass. Moreover, its temperature $\gtrsim 10^6$~K is high enough to quickly destroy the small grains (Section \ref{sec:spu}).

\rev{Finally we note that, in principle, a more gentle decline of the shattering efficiency above the chosen limit would be more aesthetic and physically accurate, but it is unnecessary. Indeed, we checked with a few test runs that the results are almost unaffected by the exact $n_{gas}$ shut-off, as long as it occurs in the range $10^2$ - $10^3$ cm$^{-3}$. Thus a smoother decline of the efficiency above some density in this range would produce indistinguishable results. The physical reason is that the cold phase is very seldom at densities above $10^2$ cm$^{-3}$, which approach the molecular cloud regime, where coagulation is at work}\footnote{In LR simulations and at $z \sim 0$ ($z \sim 2$) the cold phase density typically features a median $\sim 10$ ($\sim 15$) cm$^{-3}$ and  10\%-90\% percentiles of $\sim 2$ and $\sim 50$ ($\sim 2$ and $\sim 150$) cm$^{-3}$ respectively. In HR simulations these figures are about 1.5-2 times larger.}.

\subsubsection{Accretion and Coagulation in dense molecular gas}
\label{sec:acc_cou}

Accretion of gas metals onto grains as well as grain coagulation are relevant processes only in the densest regions of the cold ISM, $n_H \gtrsim 10^2 - 10^3$ cm$^{-3}$ \citep[e.g.][]{hirashita14}, where hydrogen is mostly in molecular form. These high densities are unresolved in most cosmological simulations, including our own. Therefore, we have to resort to a sub-resolution prescription to estimate the fraction of gas particles mass $F_{\rm dense}$ that is locally in this condition. To do this, we consistently rely on the MUPPI model used here to describe the unresolved processes of star formation and feedback (Sec. \ref{sec:muppi}). Therefore, we assume that accretion and coagulation occur only in MP (multi-phase) particles, specifically in the molecular fraction $f_{\rm mol}$ of the cold phase. Indeed, the typical density of the cold phase $n_{\rm cold}$ turns out to depend both on redshift and resolution, but in general it is much lower than the minimum density for the aforementioned processes to be relevant.
On the other hand MUPPI uses the  \cite{blitz06} scaling of molecular fraction with pressure to estimate $f_{\rm mol}$, that is used to compute the SFR (see Section~\ref{sec:muppi}).
Therefore, in the present application we simply set $F_{\rm dense}=f_{\rm cold} \times f_{\rm mol}$, where $f_{\rm cold}$ is the cold mass fraction of the SPH particle.
Gas metal accretion onto preexisting grains is a fundamental ISM process, strongly affecting the dust content of galaxies (see Section  \ref{sec:IIagbIa}).
Being a surface process, in the two-size approximations, it is directly taken into account only for small grains. Nevertheless it influences  also the amount of large grains, by enhancing the coagulation of small grains.

From equation 19 of  \cite{hirashita11}, the mass increase timescale due to accretion of the generic element $X$ on grains of radius $a$ is \begin{equation}
\tau_{acc, X}  = \frac{a \, f_X s \, \mu_X}{3 \, n\, Z_X \bar{\mu}\, S} \left(\frac{2 \pi}{m_X k T}\right)^{1/2}
F_{\rm dense}^{-1}
\label{eq:tauac}
\end{equation}
where $f_X$ is the mass fraction of element $X$ in the grain, $s$ is the density of the grain material, $\mu_X$ is the atomic weight of the element, $m_X$ its atom mass, $n$, $T$ and $\bar{\mu}$ are the gas number density, temperature and mean molecular weight respectively, $Z_X$ the mass fraction of the element $X$ in the gas phase, and $S$ is the sticking efficiency. A factor 3 in the denominator accounts for the fact that the timescale given by \cite{hirashita11} is for the grain radius growth, while here we are considering the mass growth.
For spherical grains the mass variations timescale is related to that for radius variations by $m / \dot m = a / 3 \dot a$.
The fraction $F_{dense}$ is included because we assume that accretion occurs only in the densest molecular regions of MP particles, as described above.

\begin{figure*}
    \includegraphics[width=18cm]{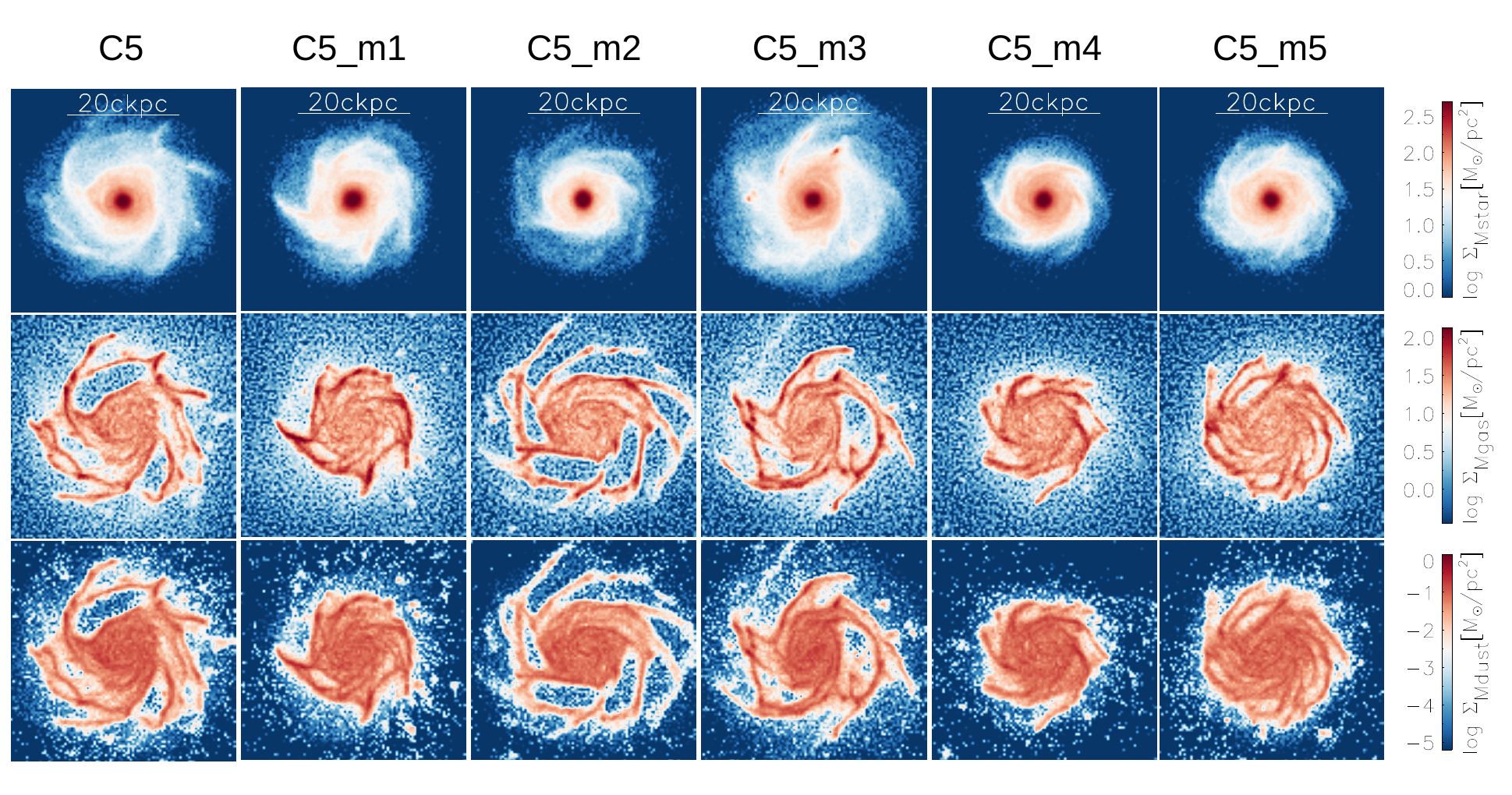}
    \vspace{-0.5truecm}
    \caption{Differences in the stellar (top), gas (middle) and dust (bottom) mass surface density among our C5 simulated galaxies. We show the fiducial model and its \rev{5 "butterfly effect" realizations m1 to m5. The latter are runs with minute variations in the initial conditions to demonstrate the variance due to chaotic processes (see text for details).} Images have been produced using a pixel size of $\sim$ 400 pc/h.}
    \label{fig:z0maps}
\end{figure*}

\begin{table}
\centering
\begin{tabular}{lccccc}
\hline
&\multicolumn{5}{c}{{ELEMENT}}\\
       & C   & O    & Mg   & Si   & Fe   \\
\hline
$f_X$  & 1.0 & 0.37 & 0.14 & 0.16 & 0.32 \\
A$_X${[}$10^3$ yr{]} & 2.5 & 1.56 & 0.73 & 0.91 & 2.52\\
\hline
\end{tabular}
\caption{Mass fractions in the grains and Normalization factors used to compute accretion timescale with Equation \ref{eq:tauacnum} for each element $X$ participating to grain composition. See text for assumptions.}
\label{tab:ax}
\end{table}

The previous expression can be rewritten in a convenient numerical form:
\begin{equation}
\tau_{acc, X} =
\frac{A_{X} \, a_{0.005} } {Z_{X} n_3 T^{1/2}_{50} S_{0.3} F_{\rm dense}  }
\label{eq:tauacnum}
\end{equation}
where the radius is in units of $0.005 \mu m$, the gas density in units of $10^3$ cm$^{-3}$, its temperature in units of 50 K and the sticking efficiency in units of 0.3, which are the fiducial values adopted in our computations. The normalization factor $A_{X}$ depends on the elements under consideration, via $f_X$ and $\mu_X$. The former fraction is 1 for C, which  forms pure carbonaceous grains, while for the silicate forming elements we compute $f_X$ by assuming  as usual the intermediate MgFeSiO$_4$ olivine composition. By further adopting a material density $s$ of 3.3 and 2.2 g cm$^{-3}$ for silicate and carbon grains respectively, we get the values of $A_X$ reported in Table \ref{tab:ax}.

As for silicate grains, we impose the condition that the accretion process maintains the same olivine-like mass fraction of the four elements adopted for stellar dust production (see Section \ref{dsynth}). We achieve this by adopting for all of them the accretion timescale of the element for which $\tau_{acc,X}$ is maximum.
This "leading" element has been commonly dubbed in previous work as {\it key element} \citep[e.g.][]{zhukovska08,hirashita11,asano13,hou19}. In our simulations, the key element is identified at each timestep and for each gas particle, based on the instantaneous metal abundance pattern. It usually, but not always, turns out to be Si for silicates, as {\it assumed} for instance by \cite{hirashita11} and \cite{hou19}.

To illustrate the importance of imposing the latter condition on the dust composition of accreting silicate grains, we also run cases where the accretion process is instead free, i.e. for each element we use its own value of $\tau_{acc, X}$ (Eq. \ref{eq:tauac}). In this case we give up to the very concept of key element. The corresponding model is dubbed
$\rm{C6\_AccFree}$  and its features will be discussed in Sections \ref{sec:fre}. 

In the densest ISM regions, low velocity small grain collisions result into coagulation, to form large grains. We adopt for this process the same timescale given by \cite{aoyama17}, but again taking into account that it occurs only in the fraction $F_{\rm dense}$ of the SPH particle mass:
\begin{equation}
\tau_{co} =
\tau_{co,0}
\, \left(\dfrac{0.01}{D_S} \right) \left(\dfrac{0.1 \, \mbox{km s}^{-1}}{v_{co}}\right)  F_{dense}^{-1}
\label{eq:tauco}
\end{equation}
Here $D_S = M_{dust,S} / M_{gas}$ is the dust to gas ratio for small grains and $v_{co}$ is the velocity dispersion of small grains. We set $v_{co}=0.2 \,\mbox{km s}^{-1}$ based on \cite{yan04}.  The normalization is $\tau_{co,0} = 2.71 \times 10^5 \, \mbox{yr}$, a value derived assuming a typical size of small grains of $0.005\, \mu$m, a material density of 3 g cm$^{-3}$. For simplicity, the latter figure is somewhat intermediate between that adopted above for silicate and carbon grains. We tested that the small related difference in the timescale for the two types of grains, has no practical consequences here.

\begin{figure*}
    \includegraphics[width=15cm]{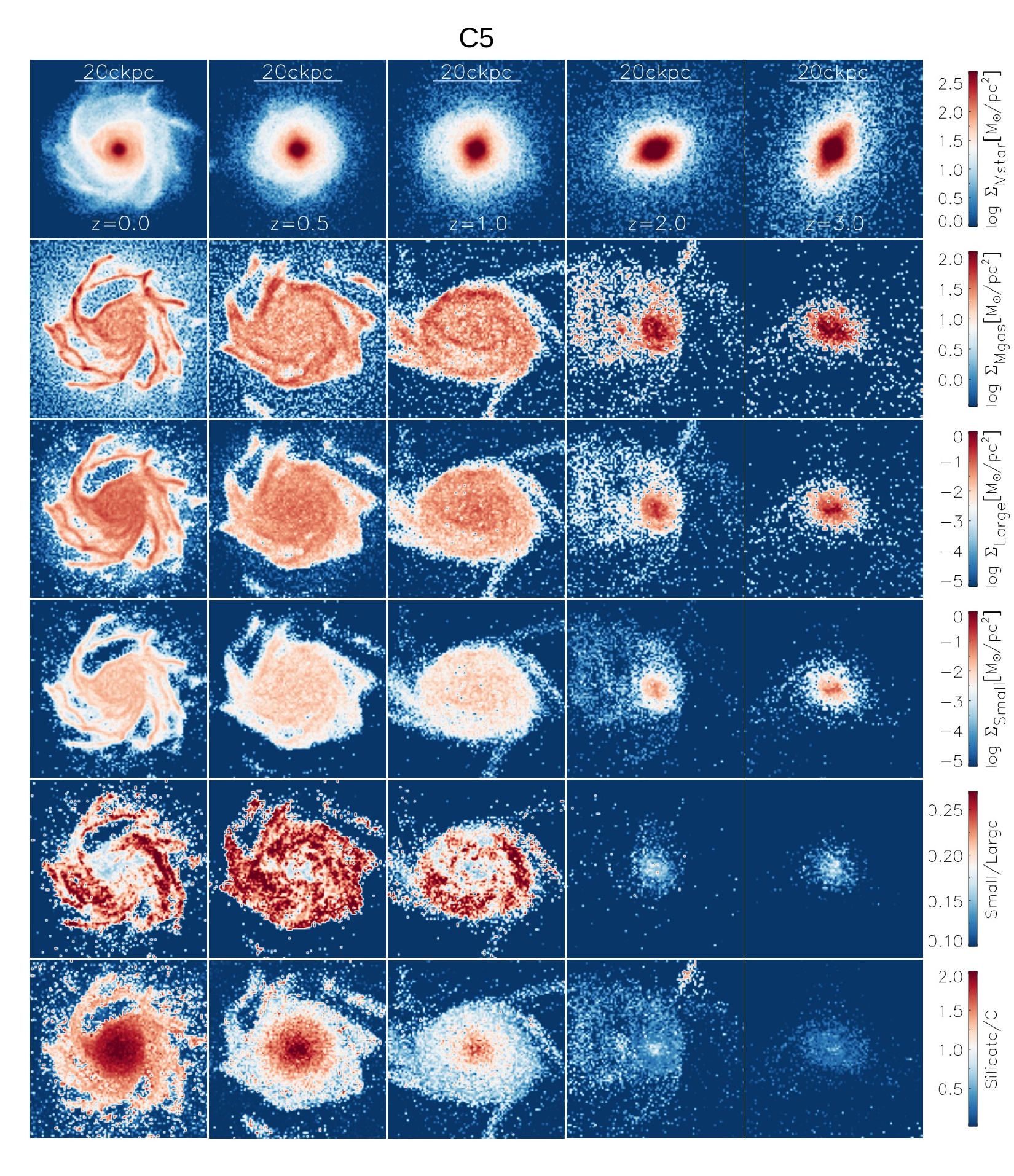}
    \vspace{-0.5truecm}
    \caption{Galaxy face-on maps of a few quantities at various redshifts for one of the C5 realizations. Columns from left to right correspond to z=0, 0.5, 1, 2 and 3. The horizontal rows from top to bottom show the surface mass density distribution of stars, gas, large dust grains, small dust grains,
    small-to-large grain \rev{mass} ratio and silicate-to-carbon \rev{mass} ratio. Images have been produced using a pixel size of $\sim$ 400 pc/h.}
    \label{fig:maps}
\end{figure*}

\subsubsection{SNae destruction}
\label{sec:SNdes}
Dust grains are eroded by thermal and non-thermal sputtering occurring in SNae shocks \citep[for a review see][]{mckee89}.
In this work we refer to destruction by SNae events as SNae destruction, while we reserve the term sputtering to the thermal sputtering occurring in the diffuse hot gas described in next Section. To treat the former process we follow \cite{aoyama17}.
We do not differentiate here the effects of SNII and SNIa, and we indicate $N_{SN}$ their total number over the timestep $\Delta t$. However the code is ready to take into account different values of the relevant parameters for the two SN types.
The process timescale for the process can be written as
\begin{eqnarray}
\tau_{SN} & = &  \frac{\Delta t}{1 - \left(1 - \eta\right)^{N_{SN}}} \\
\eta & = & \epsilon_{SN} \min \left( \frac{m_{\rm SW}}{m_{g}}, 1 \right) \nonumber
\end{eqnarray}
where $m_g$ is the mass of the SPH particle, $m_{\rm SW}$ is the gas mass swept by a SN event (in our simulations $m_{\rm SW}<<m_{g}$), and $\epsilon_{SN}$ is the grain destruction efficiency in the shock. We set $\epsilon_{SN}=0.1$ as in  \cite{aoyama17} and \cite{gjergo18}. The shocked gas mass has been estimated in \cite{mckee89} as:
\begin{equation}
m_{\rm SW} = 6800 M_{\odot} \left( \frac{E_{SN}}{10^{51} \mbox{erg}} \right) \left( \frac{v_s}{100\mbox{ km s}^{-1}}\right)^{-2}
\end{equation}
where $E_{SN}$ is the energy from a single SN explosion and $v_s$ the shock velocity. We adopt the fiducial value of $E_{SN}=10^{51}$ erg. \cite{mckee87} give for $v_s$ the formula
\begin{equation}
v_s = 200\mbox{ km s}^{-1} \left( n_{\rm gas} / 1 \mbox{cm}^{-3} \right)^{1/7} \left( E_{SN} / 10^{51} \mbox{erg}\right)^{1/14}
\end{equation}
However, since our simulations do not resolve the densities of SN blasts and the dependencies are very weak, we simply set $v_s = 200$ km s$^{-1}$, which implies $\eta = 170 \, M_{\odot}/M_g$.

\subsubsection{Thermal sputtering}
\label{sec:spu}
\edi{Grains surrounded by plasma at $T_g\gtrsim 5 \times 10^5$ K, are efficiently eroded by collisions with ions. To account for this {\it thermal sputtering} we use the {\it radius} variation timescale formula given for the  \cite{tsai95}, which approximates reasonably well the results of calculations for both C and silicate grains up to $T_g \simeq \mbox{a few}  \times 10^7$ K. The efficiency stalls above this temperature \citep{tielens94}.
From equations (14) and (15) of \cite{tsai95} we obtain, for the {\it mass} variation timescale:}
\begin{equation}
\tau_{sp} = \tau_{sp,0}  \, \left(\frac{a}{0.1 \, \mu\mbox{m}}\right) \left(\frac{0.01 \, \mbox{cm}^{-3}}{n_g}\right) \left[ \left(\frac{T_{sp,0}}{\min(T_g,3\times 10^7\mbox{K})}\right)^\omega+1 \right]
\label{eq:tausp}
\end{equation}
with $T_{sp,0} = 2\times 10^6 K$ and $\omega=2.5$.
In this equation the number density $n_g = \rho/\bar{\mu} m_p $ includes both ions and electrons. By using a mean molecular weight $\bar{\mu}=0.59$ of a fully ionized mixture of 75\% H and 25\% He, the normalization constant given in that paper would yield $\tau_{sp,0} = 5.5 \times 10^6 \mbox{yrs}$.
This value includes the relationship between the mass and the radius variations timescales $m / \dot m = a / 3 \dot a$.
However our previous simulations presented in \cite{gjergo18}, as well as the independent simulations of \cite{vogelsberger19}, both adopting the same sputtering parametrization,
suggest that to reproduce the relatively large amounts of dust observationally inferred in galaxy clusters \citep{43planck16}, a significantly longer sputtering timescale is required, by a factor 5 and 10 respectively.
Therefore in our fiducial model we adopt a five time larger value of the parameter, $\tau_{sp,0} = 2.7 \times 10^7 \mbox{yrs}$. However, we also run for comparison test simulations adopting the ''canonical'' normalization quoted above, as well as simulations adopting a ten time larger value (C6\_SpuTsai and C6\_Spu0.1Tsai respectively in Table \ref{tab:suite}). In the former case the results with respect to our fiducial value are almost identical, while in the latter case, which correspond to the preferred  normalization by \citep{vogelsberger19}, the increase of dust survival and its effect are quite noticeable.

We use $a=0.05 \, \mu$m  and $a=0.005 \, \mu$m for the effective radii of large and small grains, respectively. The former value is the average radius for a power law size distribution with index $-3.5$, extended from 0.03 $\mu$m (the adopted boundary between small and large grains) to $0.25$ $\mu$m \citep[e.g.][]{silva98}.

\begin{figure*}
    \includegraphics[width=\textwidth]{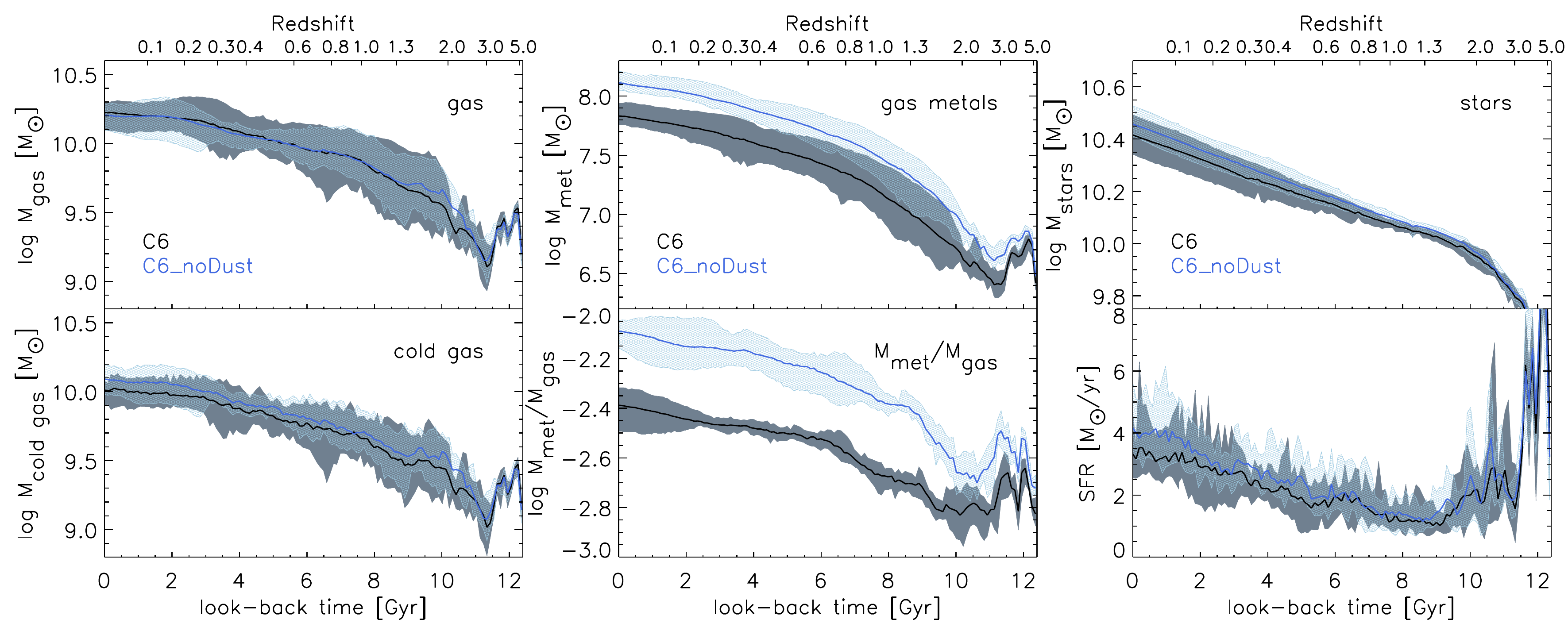}
    \vspace{-0.5truecm}
    \caption{Evolution of stars and gas related quantities for runs with dust processes on (C6, black) and off (C6\_nodust, blue).
    All quantities are computed considering particles in the galactic region, that is within $0.1 \rtwo$ from the halo centre and below a maximum height of $0.02 \rtwo$  from the disc plane.
    For each model, the line refers to the mean value of the six runs performed to estimate the butterfly effect, while the shaded area covers their whole dispersion. See text for details.
    Left panel: total and cold star forming gas masses.
    Middle panel: gas metal mass and the ratio between gas metals \rev{(i.e.\ excluding metals in the dust phase for the run C6)} and gas mass. Right panel: stellar mass and SFR.}
    \label{fig:nodust_evo}
\end{figure*}

\begin{figure*}
    \includegraphics[width=\columnwidth]{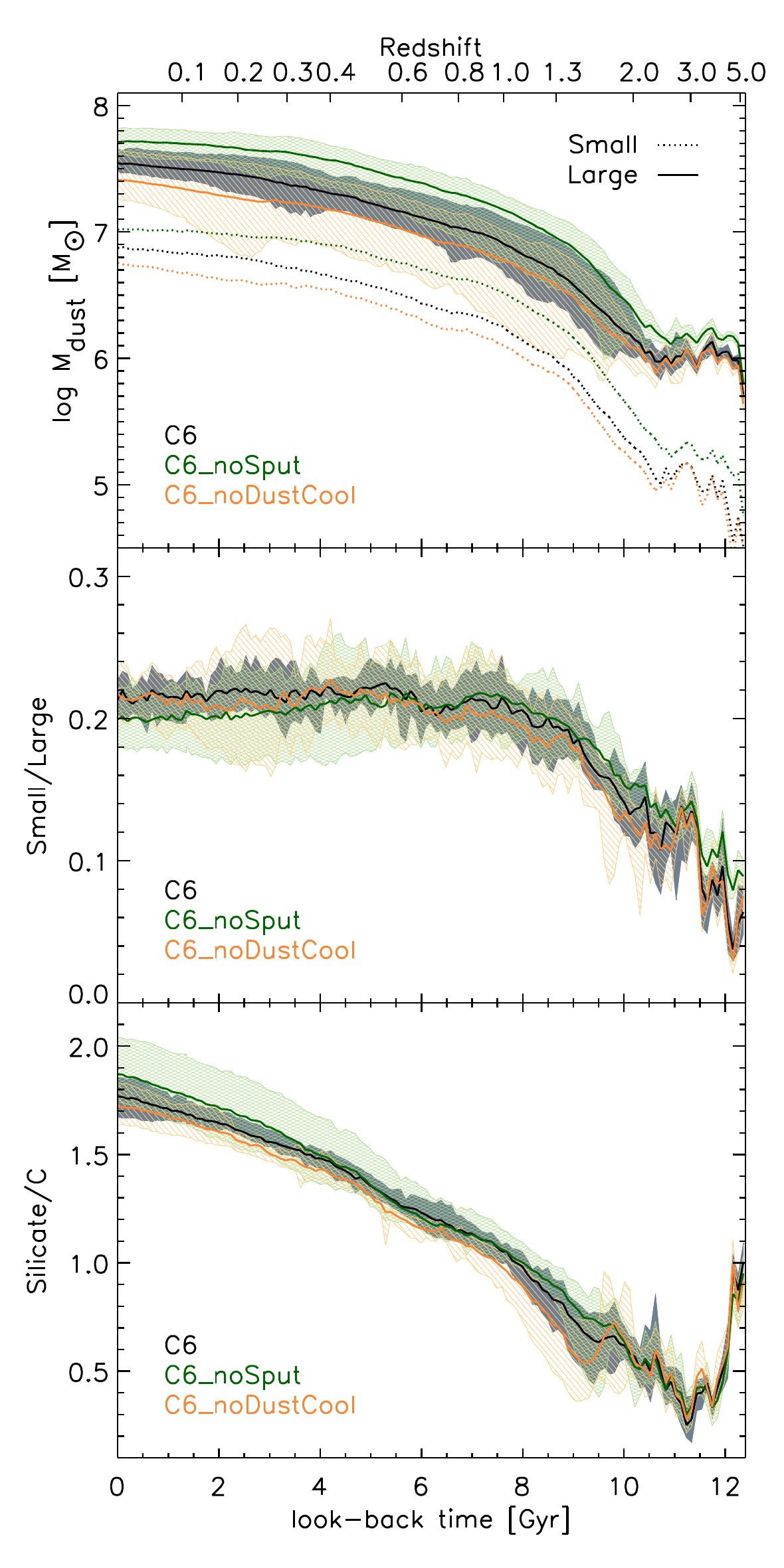}
    \includegraphics[width=\columnwidth]{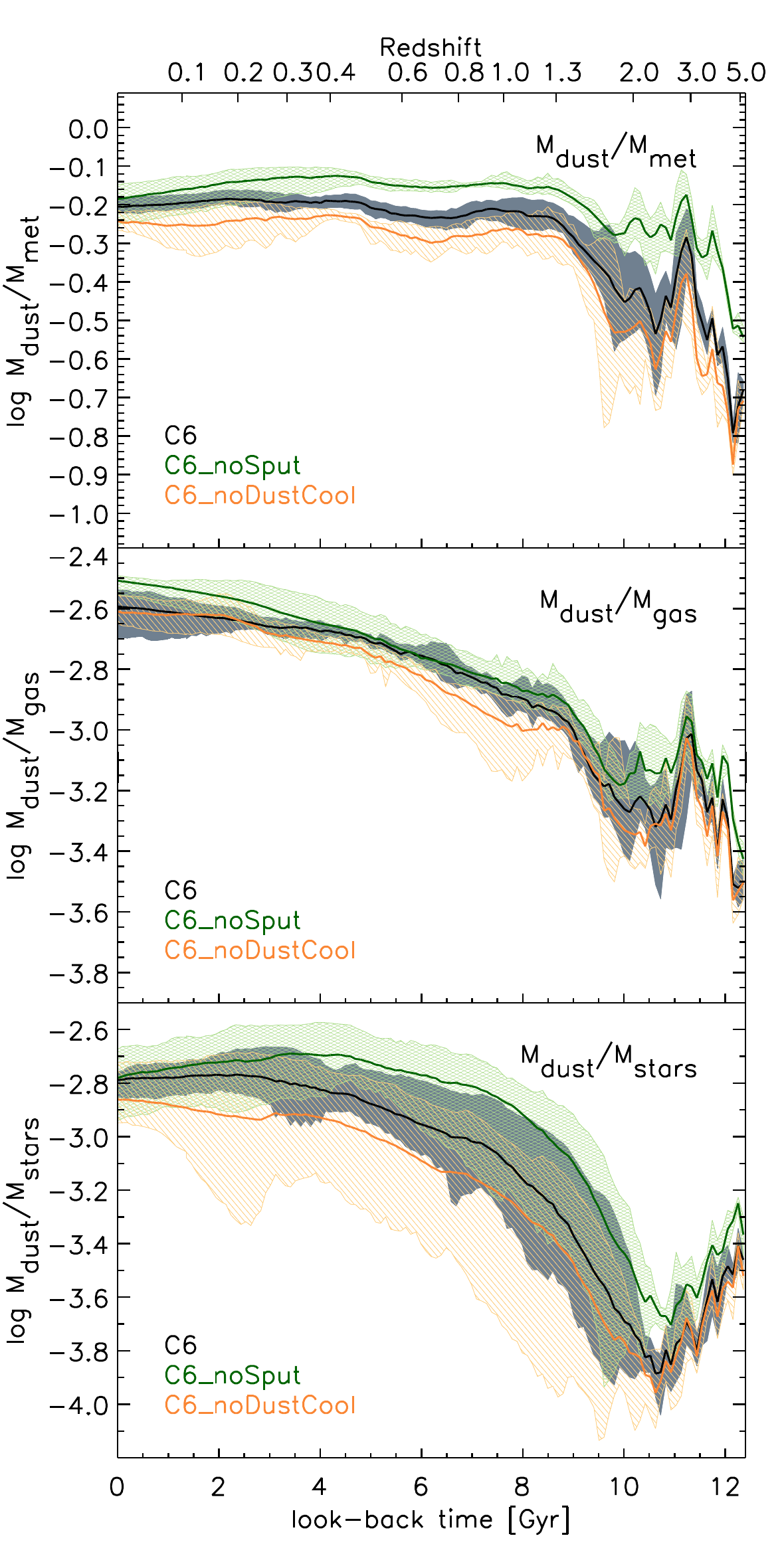}
    \vspace{-0.5truecm}
    \caption{The effect of the adopted sputtering efficiency and of dust cooling on the dust mass and on various mass ratios: small-to-large grains, Silicate-to-Carbon, dust-to-metals, dust-to-gas  and dust-to-stars.
    C6 is the fiducial model. In C6\_nospu sputtering is totally ignored. We also show a run, C6\_noDustCool, where cooling due to dust has been shut-off. The model with more efficient sputtering {C6\_SputTsai} produces results very similar to the fiducial one C6, and it is not shown to avoid cluttering.
    All quantities are computed considering particles in the galactic region, that is within $0.1 \rtwo$ from the halo centre and below a maximum height of $0.02 \rtwo$  from the disc plane.
    For each model, the solid line refers to the mean value of the six runs performed to estimate the butterfly effect and the shaded area covers their whole dispersion. See text for details and discussion.
}
    \label{fig:dust_evo_spu}
\end{figure*}

\begin{figure*}
    \includegraphics[width=\textwidth]{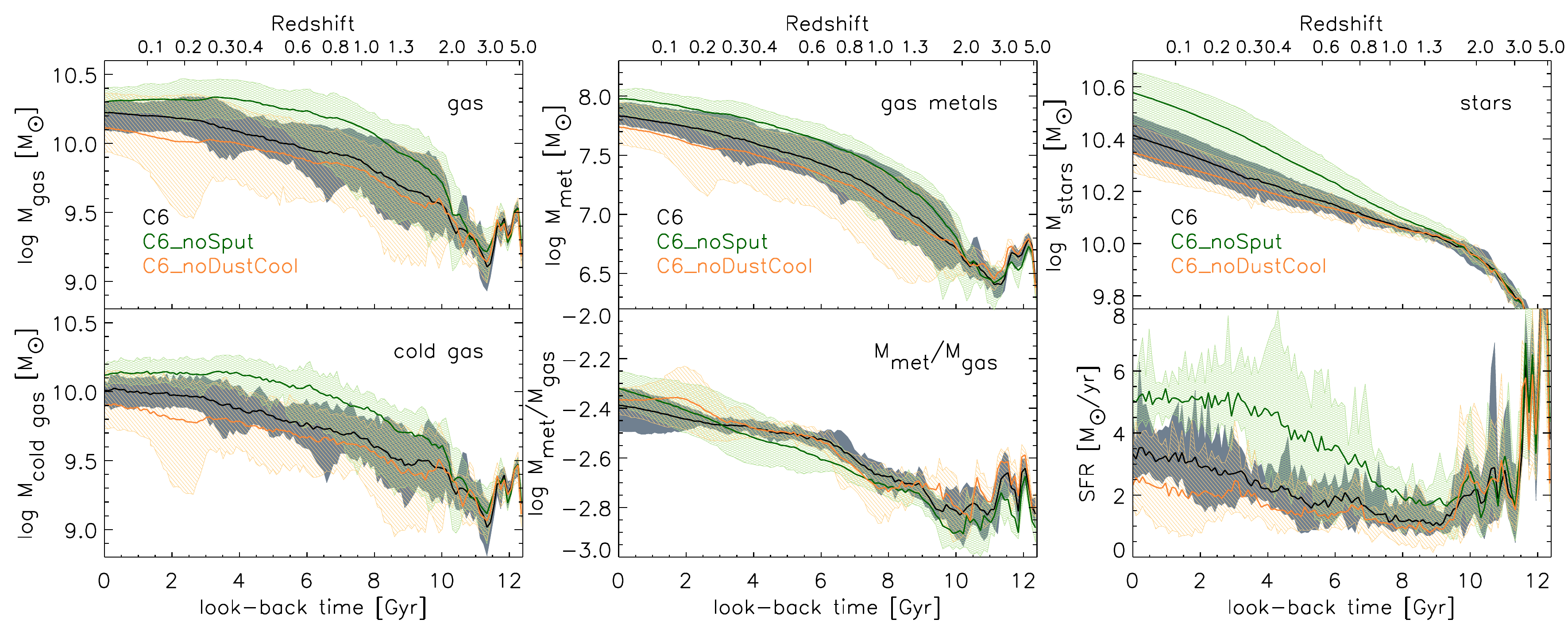}
    \vspace{-0.5truecm}
    \caption{Same as Figure \ref{fig:nodust_evo}, but comparing the fiducial model (C6) with that neglecting sputtering (C6\_noSput) and
    that with no dust cooling
    (C6\_noDustCool). The model with more efficient sputtering (C6\_SputTsai) produces result very similar to the fiducial one, and it is not shown to avoid cluttering.
    }
    \label{fig:baryons_evo_spu}
\end{figure*}

\subsubsection{Dust Cooling}
\label{sec:cool}
Besides being eroded by ion collisions, dust grains in a hot plasma are heated by collisions, mostly with electrons.
Ion collisions turn out to be comparatively negligible since they move much slower at the same energy, implying a much reduced collision rate \citep{montier04}. The energy absorbed by the grains is then efficiently radiated in the IR region, resulting in a net cooling of the gas.

Our treatment of this process is essentially the same as in \cite{vogelsberger19}, which is based on the computations by \cite{dwek81}. The only substantial difference is that we have two distinct populations of grains, as far as size is concerned. The latter paper provides the following expression for the heating rate in erg s$^{-1}$ of a single dust grain of radius $a$ in a thermal plasma with electron density $n_e$ :
\begin{equation}
H(a, T, n_e) =
    \begin{cases}
    5.38 \times 10^{-18} n_e a^2 T^{1.5} & \, \, x>4.5 \\
    3.37 \times 10^{-13} n_e a^{2.41} T^{0.88} & \, \, 1.5 <x \le 4.5 \\
    6.48 \times 10^{-6} n_e a^3  & \, \, x \le 1.5
    \end{cases}
\label{eq:duscooH}
\end{equation}
where $x=2.71 \times 10^8 a^{2/3}/T$. The contribution to the gas cooling function, arising from the population of grains with size $a$ and featuring a number density $n_d(a)$ can be calculated as
\begin{equation}
    \frac{\Lambda_d(a)}{n_H^2} =
    \frac{n_d(a)}{n_H^2} H(a, T, n_e)
\label{eq:duscoo}
\end{equation}
In our code we compute separately the contributions  from large and small grains, assuming for them the same radii used to calculate sputtering (Section \ref{sec:spu}), namely $a=0.05 \, \mu$m and $a=0.005 \, \mu$m respectively. For simplicity, when estimating the grain number density $n_d(a)$, we use for both silicate and carbon grains an intermediate material density of 3 g cm$^{-3}$.

\subsection{The set of simulations}
\label{sec:suite}
All the simulations of the present work adopt the parameters and assumptions of the model K3s–yA–IaB–kB of \cite{valentini19}, apart from those related to dust evolution, not included in that work. The adopted IMF is that by \cite{kroupa93}.
The latter simulation is that, among their models, which better agrees with several observational properties of the Milky Way (MW) (see their Table 2), including stellar chemical abundances.
Nevertheless, our results should not be considered as a model of the MW, as
no attempts to reproduce its accretion history have been made. We are rather interested in simulated galaxies with an extended disc component at $z=0$, and broadly speaking representative of spiral galaxies in the nearby Universe.
Moreover, we stress that since the inclusion of dust processes modifies to some extent the galaxy evolution (as shown in Section \ref{sec:results}), the calibration of  galaxy formation parameters should be reconsidered. This operation is clearly outside the scope of the present work.

Table \ref{tab:suite} lists the simulations discussed in the present work, with a brief remark on the difference of each of them with respect to the fiducial run.

\subsubsection{Simulations chaos and variance}
\label{sec:chaos}
A fairly underrated feature of numerical simulations is that, in most conditions, repeated runs of the same model produce results that can differ significantly from each other \citep[e.g.][]{keller19,genel19}. This form of {\it butterfly effect} is generated by the combination of purely numerical aspects with the physically chaotic nature of the system we aim to simulate. Tiny, apparently negligible, perturbations present in N-body systems at some point of the evolution are amplified to significant differences later on. In simulations, a typical, mostly unavoidable, source of the small perturbations is the reduction operations performed by parallel  protocols such as MPI or OpenMP. These operations are not fully deterministic. The chaotic nature of N-body systems becomes even more substantial due to the subgrid models of star formation and feedback, which usually relies on stochastic algorithms, for instance when spawning new stellar particles from gas particles.

Therefore, when evaluating the effect of different assumptions in simulations, in particular in zoom-in simulations of single objects such as those considered in this work, it would be necessary, albeit usually not performed \citep[but for some very recent work, see e.g.][]{kiat20,Davies2020}, to distinguish them from mere manifestations of the butterfly effect described above. \rev{To quantify this effect, we adopted the same approach as \cite{keller19}, namely} we repeated each run 6 times, introducing in 5 of them very small perturbations in the initial conditions, \rev{close to the machine precision level. These tiny perturbations produce differences in parallel simulations outcomes similar to those produced by changes in the hardware or software setup \citep{keller19}}.
Specifically, we randomly displace the initial position of each particle by an amount drawn from a uniform distribution ranging from $-dx$ to $+dx$, with $dx$ set to $10^{-10}$ times the gravitational softening adopted for gas particles. This translates to $dx \simeq 4.6 \times 10^{-7} h^{-1}$ cpc ($dx \simeq 2.3 \times 10^{-7} h^{-1}$ cpc) for C6 (C5) simulations. As a result, the last 3 decimal digits of the double precision particle position, which are stored in cMpc, are randomly modified in the initial conditions.
However, we checked that the results turn out to be rather independent of this choice, confirming the claim by \cite{genel19}. Indeed even runs adopting values of $dx$ greater by several orders of magnitude show similar differences when compared to the unperturbed run. Moreover, by running some models starting from six {\it additional} perturbed initial conditions, that is by doubling the sample, we checked that six repetitions of the run safely cover the spread of possible evolutions.
To give some visual feeling of the former variations, we show in Fig.\ \ref{fig:z0maps} $z=0$ maps of stars, gas and dust mass densities for all the six realizations of the same fiducial model.

Finally, we also verified that the run to run variations obtained by perturbing the initial conditions as described above, are comparable to those we get by running the same unperturbed initial conditions on different hardware.

\begin{figure}
    \includegraphics[width=\columnwidth]{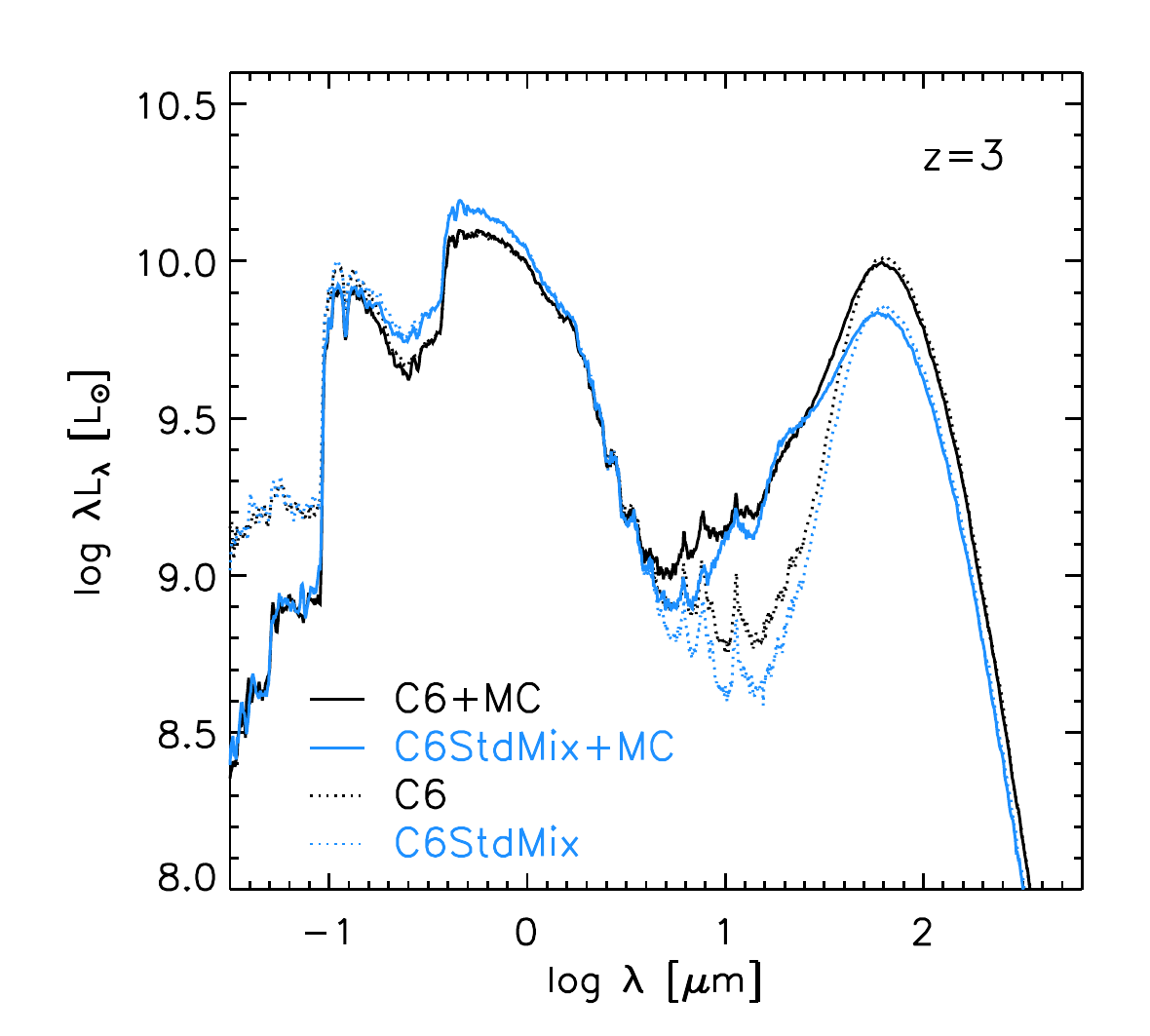}
    \vspace{-0.5truecm}
    \caption{Simulated galaxy C6 SEDs at $z = 3$ adopting a standard "Milky Way" dust  mixture (blue C6StdMix) or a mixture that takes into account the dust spatial distribution in the galaxy as predicted by our simulations (black C6) at that redshift. In solid lines we show the resulting SEDs when including a simple modelization for dust reprocessing due to molecular clouds (MCs), which are not resolved by the simulation. In this latter case it is assumed that stellar particles younger than 3 Myr are embedded in MCs optically thick to stellar optical and UV radiation. }
    \label{fig:seds}
\end{figure}

\begin{figure*}
    \includegraphics[width=\columnwidth]{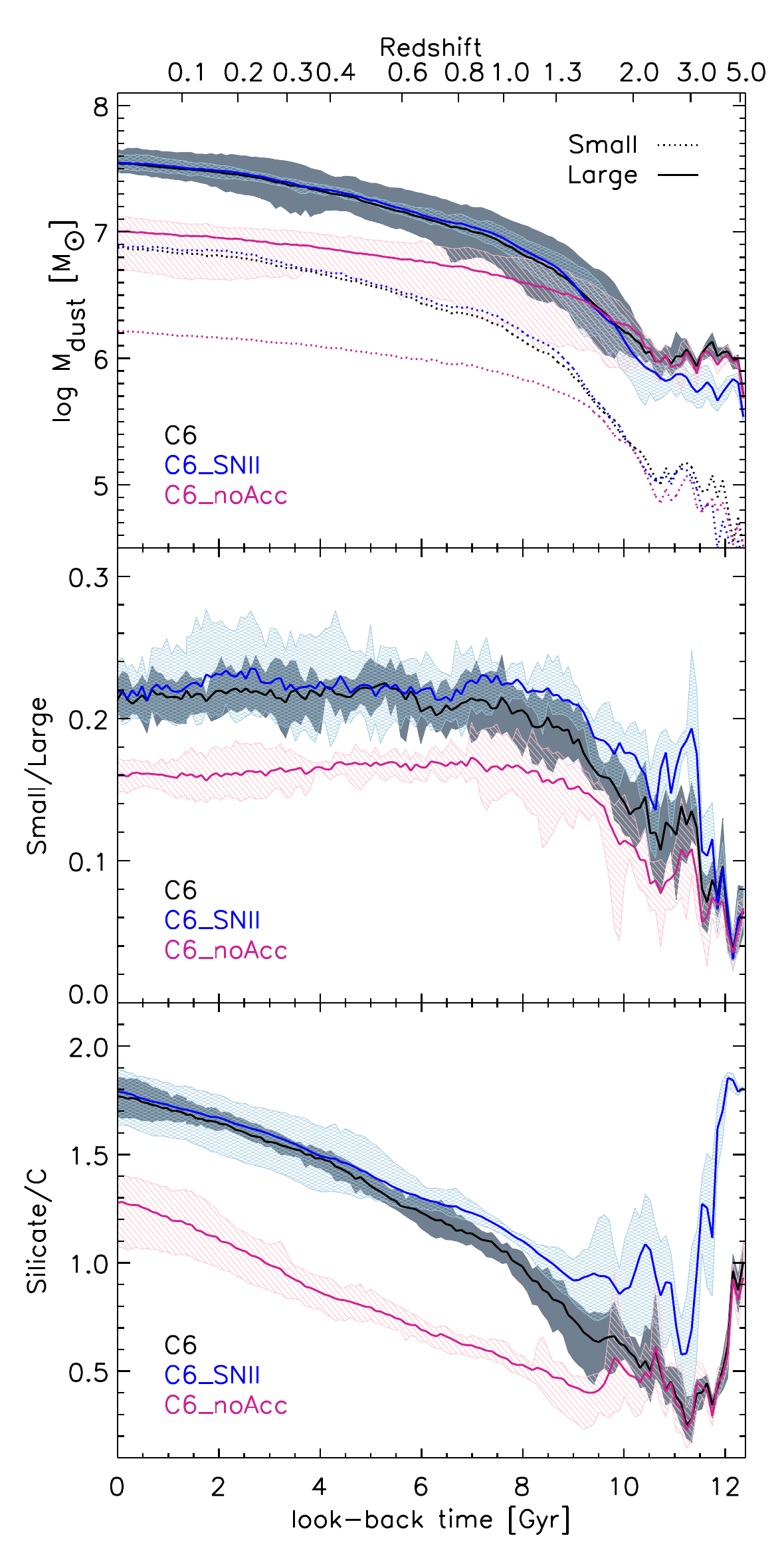}
    \includegraphics[width=\columnwidth]{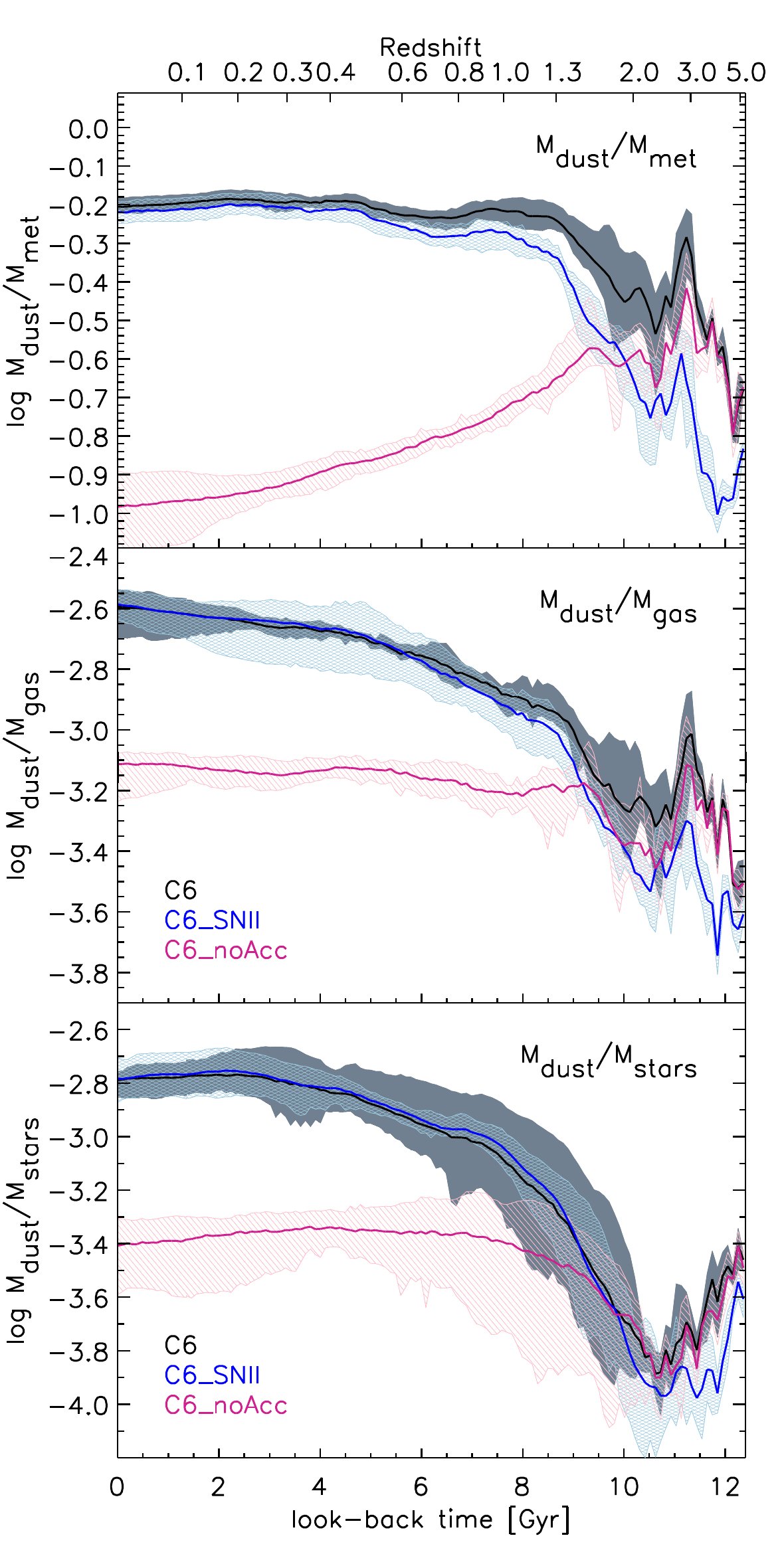}
    \vspace{-0.5truecm}
    \caption{Same as Fig \ref{fig:dust_evo_spu}, but comparing the fiducial model (C6) with that including dust production only from SNae II (C6\_SNII) and that
    in which the accretion of gas metals onto grains is not considered (C6\_noAcc).}
    \label{fig:dust_evo}
\end{figure*}

\begin{figure*}
    \includegraphics[width=\textwidth]{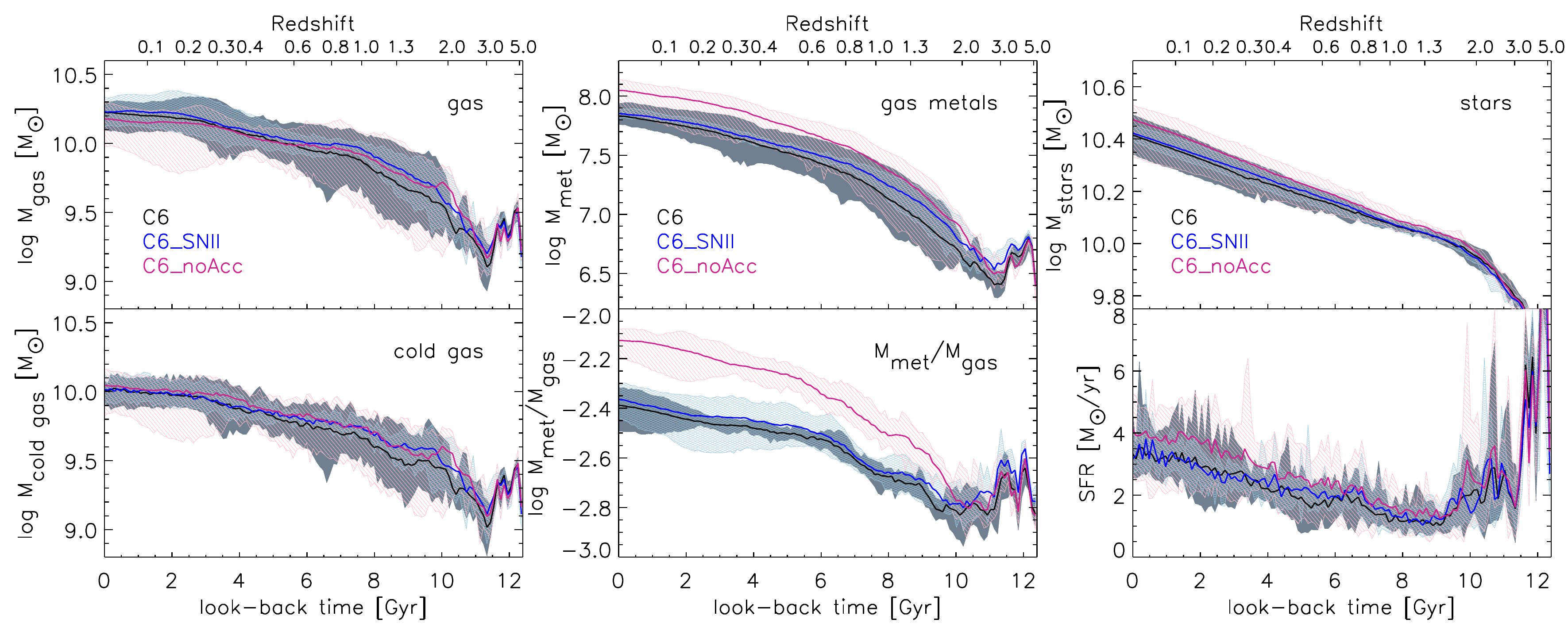}
    \vspace{-0.5truecm}
    \caption{Same as Figure \ref{fig:nodust_evo}, but comparing the fiducial model (C6)
    with the same set of models as in Fig. \ref{fig:dust_evo}.
    }
    \label{fig:baryons_evo}
\end{figure*}

\section{Results: Evolution of global quantities}
\label{sec:results}

\begin{table}
\centering
\caption{Relevant features of the simulation suite.
Column 1: simulation label.
Column 2: description.
Column 3: related figures.}
\renewcommand\tabcolsep{-0.8mm}
\begin{tabular}{@{}lcr@{}}
\hline
Label                & Description  & Fig    \\
\hline
C6                  & Fiducial LR       & \ref{fig:nodust_evo} - \ref{fig:prof_sigma}\\
C6\_noDust          & no dust processes & \ref{fig:nodust_evo} \\
C6\_SputTsai        & more sputtering   & not shown\\
C6\_noSput          & no sputtering     & \ref{fig:dust_evo_spu}, \ref{fig:baryons_evo_spu}\\
C6\_noDustCool      & no dust cooling   & \ref{fig:dust_evo_spu}, \ref{fig:baryons_evo_spu}\\
C6\_SNII            & only SNae II stars produce dust & \ref{fig:dust_evo}, \ref{fig:baryons_evo}\\
C6\_noAcc           & no accretion      & \ref{fig:dust_evo}, \ref{fig:baryons_evo} \\
C6\_AccFree         & unconstrained accretion of Silicate metals & \ref{fig:accfree} \\
C6\_SharpShat       & sharp shattering cut off  & \ref{fig:shatt}\\
C6\_noAst           & no astration      & \ref{fig:astsn}\\
C6\_noSNdes         & no SNae destruction & \ref{fig:astsn} \\
C5                  &  Fiducial HR & \ref{fig:z0maps}, \ref{fig:maps}, \ref{fig:prof_D}, \ref{fig:prof_sl}, \ref{fig:prof_sigma} \\
\hline
\end{tabular}
\label{tab:suite}
\end{table}

Unless otherwise specified, all the quantities discussed in this work refer to the central disc galaxy of the simulated box. They are computed considering the particles within $0.1 \rtwo$\footnote{$\rtwo$  is the radius enclosing a sphere whose mean density is 200 times the critical density at the considered redshift. The mass of this sphere is dubbed $\mtwo$} from the main halo centre and below a maximum height of $0.02 \rtwo$  from the disc plane. $\rtwo$ is about 250 and 180 kpc at $z=0$ and $z=1$ respectively.

For the fiducial model, we show in Fig.~\ref{fig:maps} face-on maps for one of the six C5 (HR) simulations at various redshifts, ranging from 0 to 3. From top to bottom, maps correspond to stellar mass density, gas mass density, large and small grain mass density, Small-to-Large dust mass ratio and Silicate-to-C dust mass ratio. The time sequence of the latter two maps shows, among other things, that the small grain population grows significantly later than that of big grains, and that silicates grow later than carbon grains. Along this Section we will discuss and justify in a more quantitative way this evolution of the dust mixture. We warn the reader that the appearance of the maps varies noticeably from one realization to another (see Fig.\ \ref{fig:z0maps}).

In Figs. \ref{fig:nodust_evo}, \ref{fig:dust_evo_spu}, \ref{fig:baryons_evo_spu}, \ref{fig:dust_evo}, \ref{fig:baryons_evo},\ref{fig:accfree}, \ref{fig:shatt} and \ref{fig:astsn}, we show the evolution of various quantities of the simulated galaxy, under different assumptions. In each case, the lines refer to the mean evolution of the six runs we perform for each flavour, while the shaded areas in the same colour cover the whole dispersion for the same set of runs. As discussed in Section \ref{sec:chaos}, multiple runs are required to cope with the butterfly effect, which could lead to misinterpretations of the impact of changing model assumptions. As a consequence, it would be extremely time consuming, and to some extent unnecessary, to conduct properly such a study at HR, that is evolving the C5 initial condition. In this case we run LR simulations, while the comparison of the fiducial setup with observations (Section \ref{sec:obs}) is performed also, and preferentially, at HR. To facilitate comparisons, our fiducial model $\rm{C6}$ is present in all the figures discussed in this section.

\subsection{Fiducial model}
In the present subsection, we concentrate on the main features of the fiducial model, while the following subsections are devoted to highlight the effects of changing some parameters or assumptions.

To begin with, Fig. \ref{fig:nodust_evo} compares the evolution of various baryonic components of the fiducial runs (C6) with those of identical runs,  but without dust (C6\_noDust). The fiducial runs show on average a somewhat lower star formation activity for a significant portion of their history and, as a consequence, terminate with a $\sim 10 \%$ less stellar mass.
Indeed, the gas metal content\footnote{\rev{We explicitly note that by "gas metal" we always refer to the metal content of SPH particles excluding that locked in dust grains.}}  is significantly lower in the run including dust, mostly \rev{(but not only, see below)} because an important fraction of metals are locked in grains. As a consequence, the associated metal gas cooling is less effective. This effect is only partly counterbalanced by the addition of cooling arising from grain collisions with hot gas particles. Although this latter contribution to cooling is non-negligible (Section \ref{subsec:cooling} and Fig. \ref{fig:baryons_evo_spu}), overall in the fiducial run less cold star forming gas is driven to the disk with respect to the run without dust C6\_noDust, as can be appreciated from the bottom left panel of Fig. \ref{fig:nodust_evo}. We remark that, given that dust cooling has a very different dependence on temperature than normal metal cooling, we expect that the relative balance between the two opposite contributions (decreased metal cooling by dust depletion and dust promoted gas cooling) depends on the evolutionary history of the galaxy.
We will return on the influence of dust cooling on the evolution of our simulated galaxies in Section \ref{subsec:cooling}. \rev{We also point out that the total metal content, including both gas and dust, is somewhat smaller in the fiducial model than in that without dust, ultimately due to the former's lower stellar content.}

The top left panel of Fig. \ref{fig:dust_evo_spu} shows the essentially steady increase of the dust mass in the galaxies. This increase parallels that of the other components of the galaxy (Fig. \ref{fig:baryons_evo_spu}) but is generally faster, mostly due to the presence of evolutionary processes in the ISM, which tend to amplify the direct dust production from stars. In the galactic environment, constructive dust processes dominate over destructive ones.
As a result, the dust over gas ratio (middle right panel in Fig. \ref{fig:dust_evo_spu}), after some oscillations at $z \gtrsim 2$ becomes a constantly increasing function of time, ending higher by a factor $\sim 4$ at $z=0$. The dust to metal ratio (top right panel), begins with a rapid increase from the initial values, which is mostly driven by the adopted dust condensation efficiencies.

The evolution of Small/Large  and that of the Silicate/Carbon  dust mass ratios can be appreciated in the middle left and bottom left panel of Fig. \ref{fig:dust_evo_spu} respectively. A quite reassuring result of our simulations is that the final values of these ratios are similar to those usually required by models that explain the observed properties of MW dust, which are commonly adopted in radiative transfer computations\footnote{For instance, in model 4 by \cite{weingartner01} the mass ratio of small to large is about 0.21,  while the mass ratio of Silicate to Carbon grains is 2.5.
These values are similar in the model proposed by \cite{silva98} for the MW dust, adopted by default in their GRASIL code. \label{foo:ratios}}.
However, the ratios evolve substantially over cosmic time. At high redshift $z\gtrsim 1$ the relative abundance of large grains is predicted to be higher because a more substantial fraction of the dust is produced by stars. Stars are believed (and assumed in our model) to pollute the ISM with large grains. ISM evolution requires time to modify the dust distribution. At lower redshift, the ratio reaches and maintains a value very similar to the final one.
By converse, the relative abundance of Silicate over C grains evolves steadily from values four times smaller than the present at $z\sim 2$. Silicate grains begin to dominate over carbon grains below $z \sim 1$. This \rev{later assembly} of silicates in the galaxy is related to the fact that both for stellar dust production as well as for accretion process, they are limited by the availability of a key element (it could be Mg, Si or Fe but never O), which is significantly less abundant than C.
\rev{Consequently, during the first few Gyr of the galaxy evolution, the formation of silicates is slower than that of C dust.}
Indeed, in the run C6$\_$AccFree, discussed in Section \ref{sec:fre} and where this condition is artificially released for the accretion process, "silicates" mass is larger than that of carbon dust from the beginning, thanks to the high and early oxygen production from stars. We remind that oxygen, being produced mostly by SNII, is released into the ISM by stellar generations already a few Myr after their formation, at variance with respect to C and Fe, which are mostly contributed by AGBs and SNIa respectively.
Our results differ from those by \cite{dwek98} and \cite{calura08}, who found with their one zone models that silicates dominate from the beginning of the galaxy evolution. This happens mainly because their treatment of accretion, as well as that of stellar production, does not force the grain composition to a specific compound, as our C6\_AccFree run does for the former process (see Section \ref{sec:fre} and Fig. \ref{fig:accfree}).

\subsection{On the effects of mixture variations on SEDs at early time}
\label{sec:SED}
The redshift variations of dust optical properties should be taken into account when interpreting observations by means of dust reprocessing models. Some possible consequences of neglecting them can be appreciated from Fig. \ref{fig:seds}, where we show synthetic SEDs of the C6 model galaxy at $z=3$ computed with the public radiative transport code SKIRT\footnote{http://www.skirt.ugent.be} \citep{camps20}, under different assumptions.

In one case (C6StdMix in the figure) we simply adopted a {\it standard} dust mixture, calibrated to match the average extinction and emission properties of the MW\footnote{More specifically, the standard mixture is that used so far in most GRASIL \citep{silva98} and all GRASIL3D \citep{dominguez14} applications, to compute synthetic SED of galaxies predicted by semi-analytic models and galaxy formation simulations respectively. They both allow global variations of the mixture, albeit this feature has been seldom used due to lack of information. However, a spatial dependence on these properties, which is required here, is yet not implemented}, that is (see previous Section) characterised by Sil/C and Small/Large ratios close to those predicted by our reference model at $z=0$.
In this case, the only information on dust derived from the simulation is the total dust content of each SPH particle.
In another computation (C6 in the figure) we have instead exploited all the information on the local dust abundances for the four categories of grains we follow (graphite and silicate, small and large), by adjusting the mixture at the position of each SPH particle\footnote{
\edi{In this case, we passed to SKIRT a superposition of four spatial distribution of dust densities, one for each of our model's grain types. The adopted size distributions are those by \cite{silva98}, but having 0.03 $\mu$m as the limit between large and small grains. The local normalizations of the four size distributions are calculated according to the corresponding grain type density.}}. Let us first of all consider the SEDs obtained using only the resolved density fields information provided by the simulations, i.e.\ without any further attempt to model the sub-resolution geometry (see below).
These are the blue and black dotted lines for the two above assumptions on the dust mixture respectively.
The differences between the two cases are rather important, particularly around the far IR peak and in the mid IR regime, where the use of the standard mixture under-predicts the specific luminosity by up to $\sim 50\%$. Adopting the mixture predicted by the simulation, the optical and UV stellar flux turns out to be about $10 \%$ more absorbed, and the IR power increases correspondingly. Also, the optical and UV spectral slopes are affected.

However, none of these two SEDs can be a fair representation of the real situation, because, as it has been discussed many times in literature, stars in galaxies are affected by {\it age dependent} dust reprocessing \citep[e.g.][]{silva98,charlot00,granato00,panuzzo07,jonsson10,dominguez14,camps16,goz17}. Indeed, stars younger than a few Myr are still embedded or very close to their parent molecular clouds, and therefore the radiation they emit suffers much more dust reprocessing than that coming from older stars. In general, cosmological simulations of galaxy formation cannot resolve the molecular clouds (MCs) structure, and as a consequence they cannot provide to dust reprocessing codes the full spatial information required to predict the model SED realistically. To show the possible implications of the age-dependent dust reprocessing we have also computed two more cases, one for each of the two dust compositions (blue and black solid lines respectively), emulating to some extent the same method used in GRASIL and GRASIL3D to treat MCs and the associated age dependent reprocessing, with typical values for the relevant parameters adopted in their applications. Thus we have assumed that stellar particles younger than 3 Myr are in the centre of idealized spherical MCs.  The additional assumption is that 5\%  of the galactic gas is organized in molecular clouds of mass $10^6 M_\odot$ and radius 15 pc.
The  two latter parameters are degenerate in the sense that different values yield the same result provided that $M/r^2$ in unchanged \citep[see][]{silva98}.
\rev{
For simplicity, in this exploratory test we assumed
for all the MCs in the galaxy the same averaged dust to gas ratio and dust composition, rather than adopting local values.
The MCs turn out to be optically thick to most of the  radiation emitted by young stars embedded inside them, both when we use the dust composition reproducing the average MW properties, as well as when we adopt the average dust mixture predicted by the simulated galaxy.} The SED of the MC system and that produced by stars older than 3 Myr, embedded in the general resolved ISM, have been computed with SKIRT.
Again the result (solid lines in the figure) highlights the substantial differences of the predicted IR SEDs with the two assumptions on the dust properties. Nevertheless, the differences introduced by the sub-resolution modelling of dust reprocessing of MCs are by far more important.

We plan to explore in detail the observational consequences of the dust properties evolution in the near future.

\subsection{Sputtering}
\label{subsec:sputtering}
The effect of the inclusion of sputtering on the various global properties can be assessed by means of Figs. \ref{fig:dust_evo_spu} and \ref{fig:baryons_evo_spu}.

From the top left panel of Fig. \ref{fig:dust_evo_spu} we notice that on average in our runs ignoring sputtering (C6\_nospu, green) the galaxy always contains about $\sim$50\% more dust than in the fiducial runs (C6, black). The same is true when comparing to the run adopting the canonical sputtering efficiency by \cite{tsai95} (C6\_SputTsai), which we do not show in the figure, for a better readability, because it features only very minor differences from C6.
On the other hand, the middle and bottom left panels of Fig. \ref{fig:dust_evo_spu} show that the ratios of small over large grains, as well as that of silicate over carbon grains, are instead little affected by sputtering.

As it will be discussed also in the next Section \ref{subsec:cooling}, we point out that the higher dust mass predicted by models with lower sputtering is not just the trivial and direct consequence of its reduced efficiency under the same gas conditions (density and temperature).
If that were the case, models with less sputtering would have less gas metals, whilst the opposite is true. This is due to the closely related process of hot dust cooling, which becomes more important when dust survives longer in a hot gas, increasing the gas flow toward the galaxy and in particular the availability of cool star forming gas. As a consequence also the metal and dust production from stars is enhanced. Moreover, in a more metal enriched ISM, dust accretes more.
In conclusion, favoured dust survival produces further back-reaction effects that indirectly promote the presence of even more dust.

\begin{figure*}
\includegraphics[width=\columnwidth]{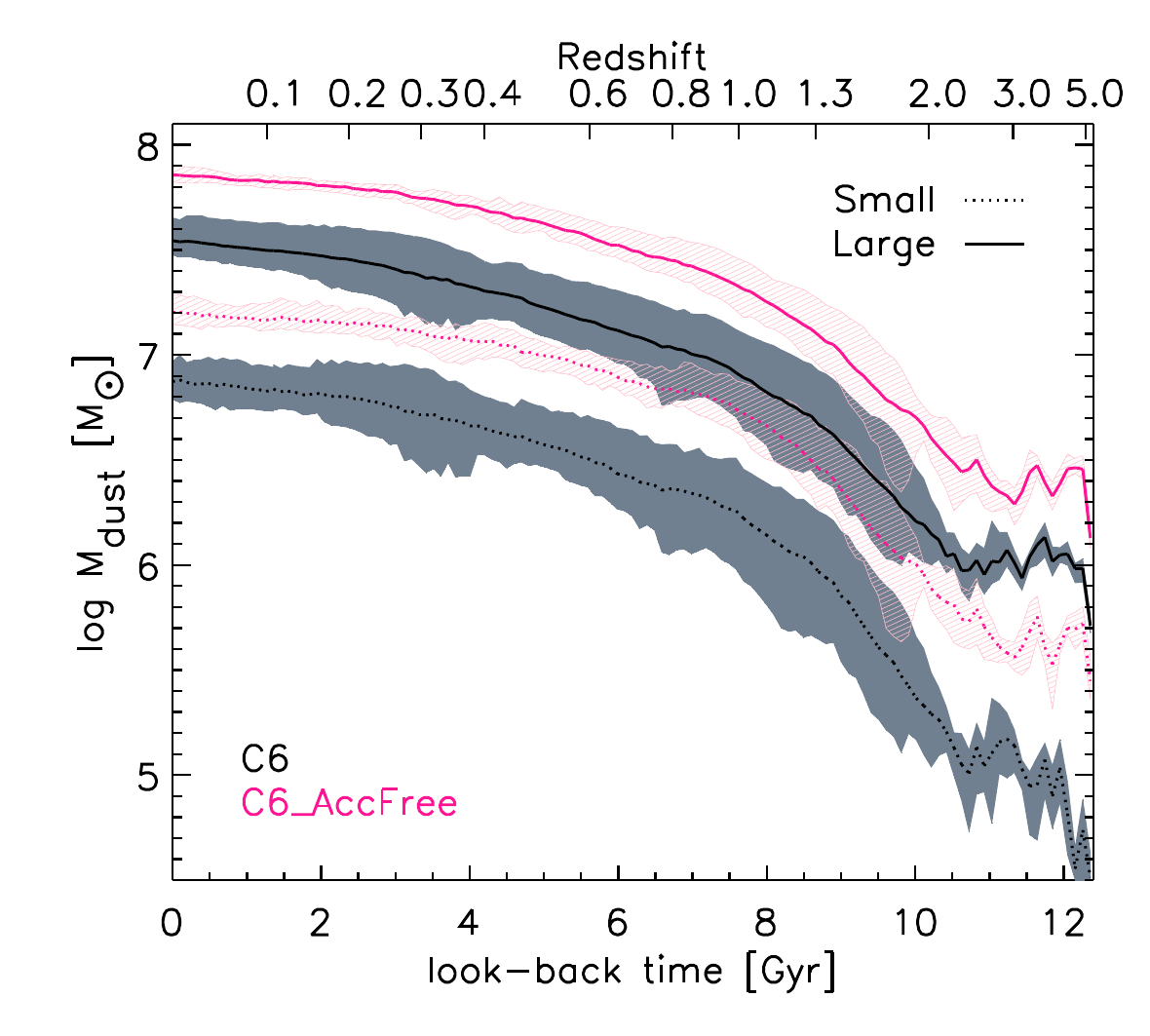}
\includegraphics[width=\columnwidth]{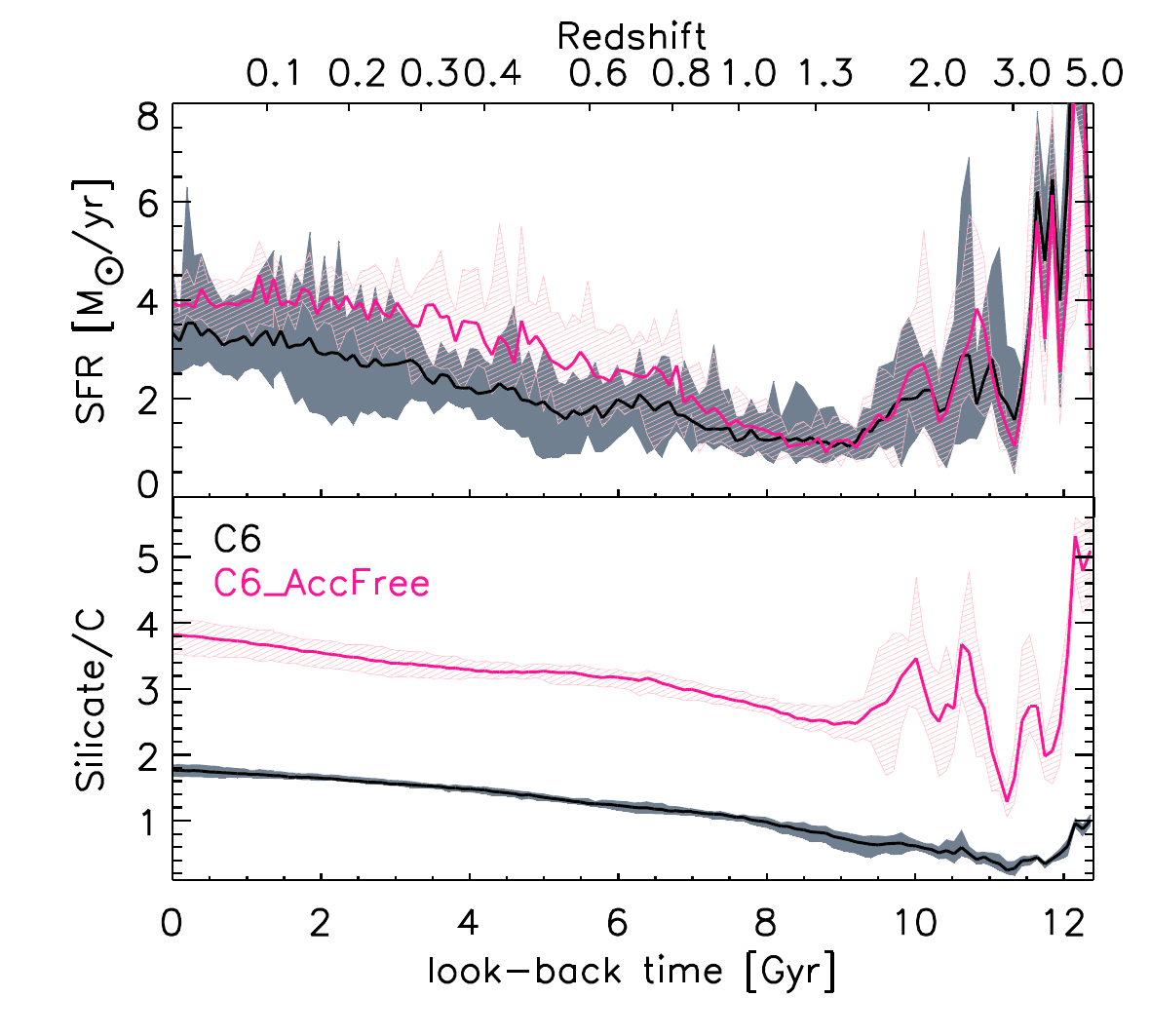}
\vspace{-0.5truecm}
\caption{Evolution of the dust content, the silicate  over carbon dust mass ratio and the SFR for the run  C6\_AccFree, in which the accretion process of metals over silicate grains is not constrained to maintain the proportion of olivine MgFeSiO$_4$.}
\label{fig:accfree}
\end{figure*}

\begin{figure*}
  \includegraphics[width=\columnwidth]{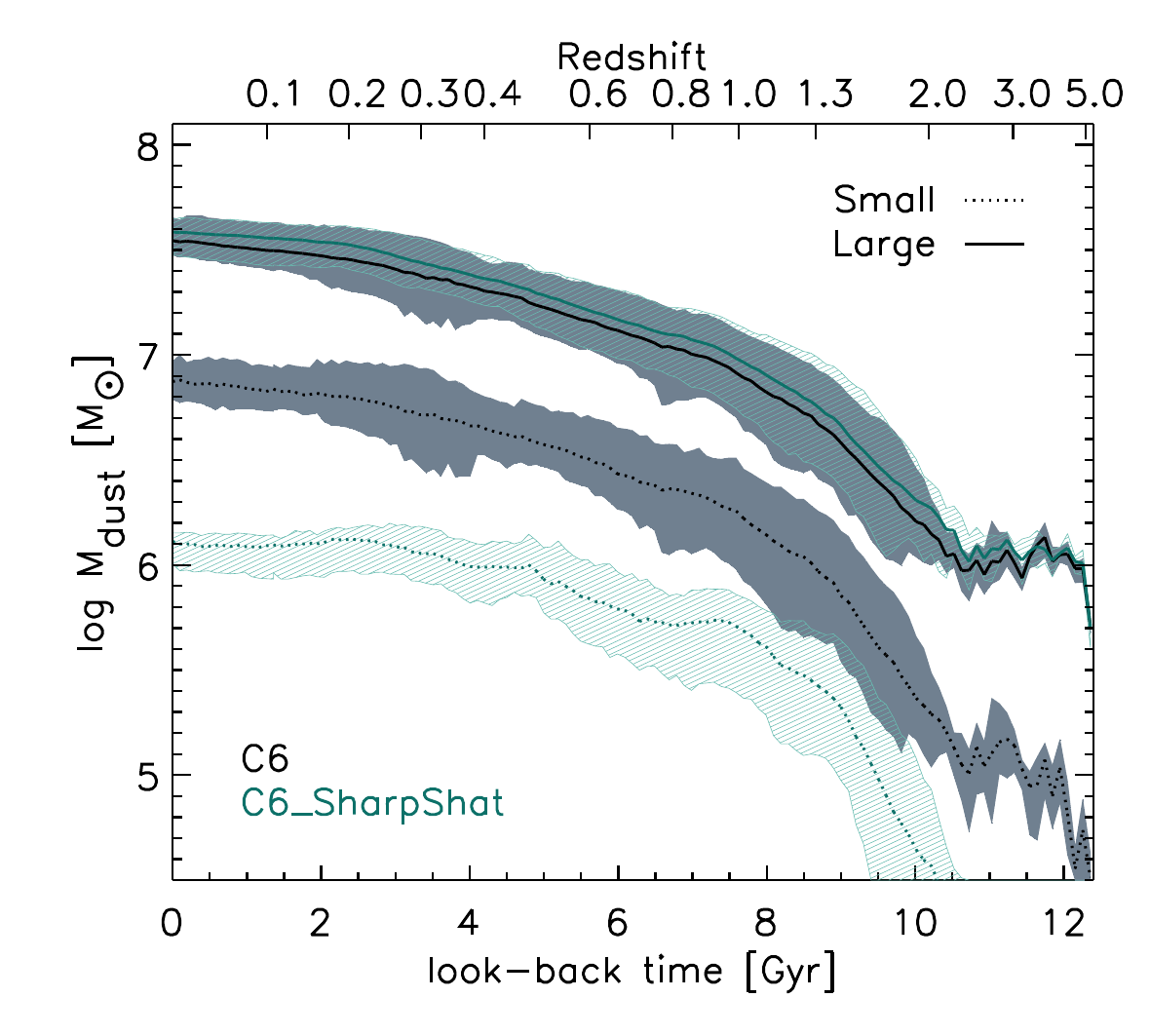}
  \includegraphics[width=\columnwidth]{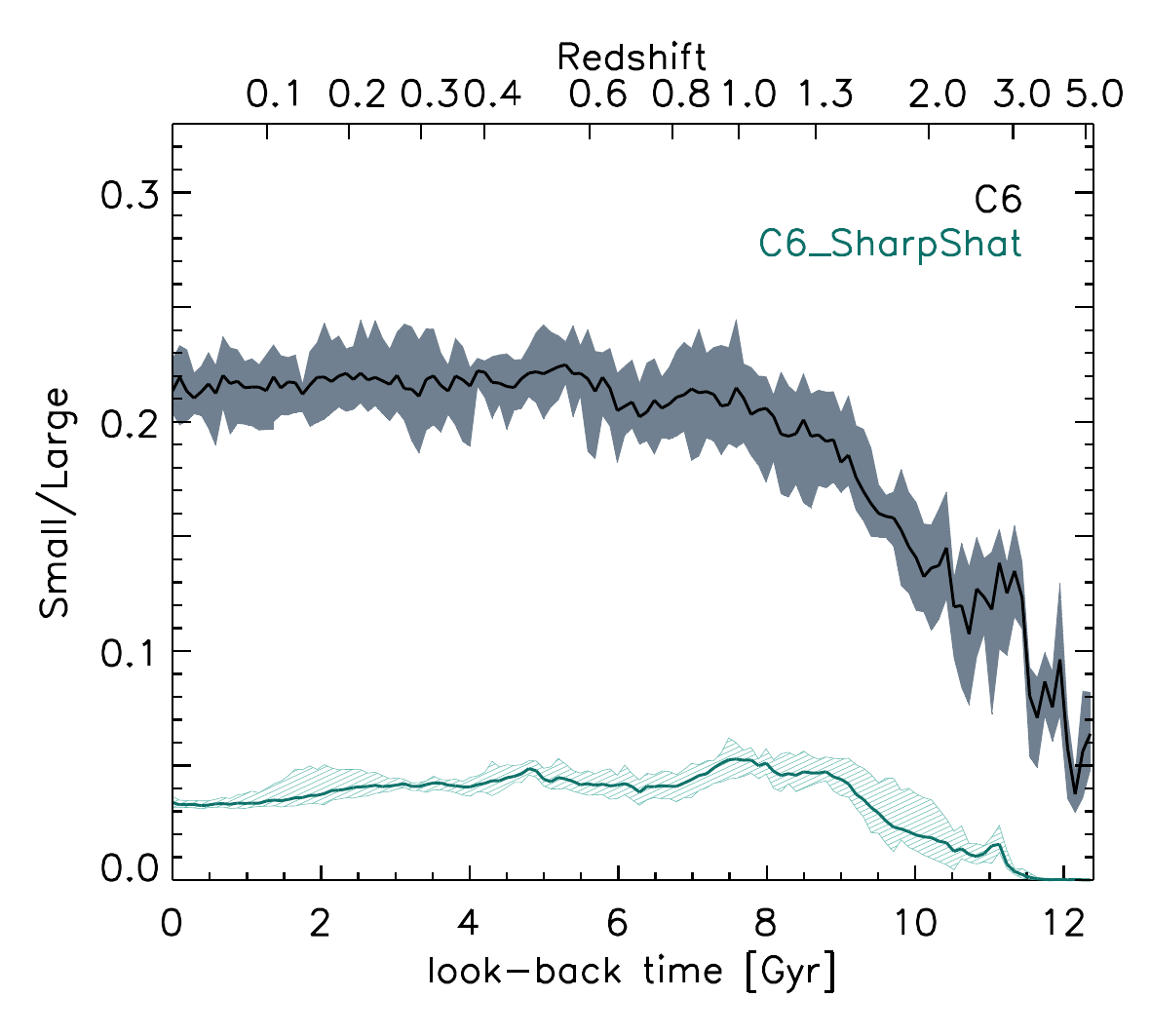}
  \vspace{-0.5truecm}
  \caption{The run C6\_SharpShat in which the shattering process is switch-off abruptly at $\rm{n_{gas}}>1 \, \rm{cm}^{-3}$ results in a too low abundance of small grains.}
  \label{fig:shatt}
\end{figure*}

\begin{figure}
  \includegraphics[width=\columnwidth]{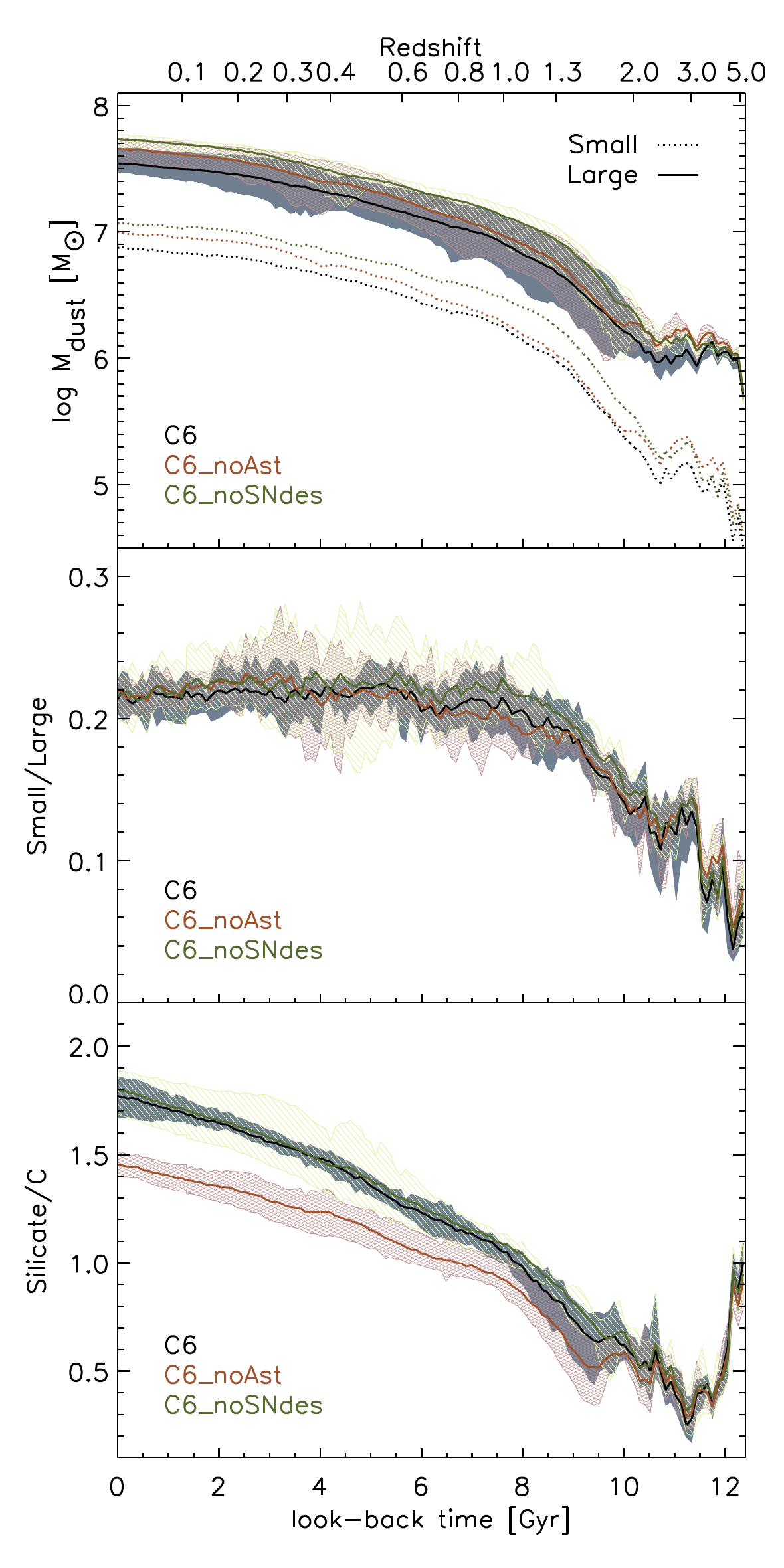}
  \vspace{-0.5truecm}
  \caption{The effects of ignoring the dust destruction processes of astraction (C6\_noAst) or SNae destruction (C6\_noSNdes). In both cases about 50 \% more dust mass than the fiducial model (C6) is predicted at $z=0$. Moreover, by ignoring astraction Silicate over Carbon dust mass ratio decreases by $\sim 20 \%$.}
  \label{fig:astsn}
\end{figure}

\begin{figure*}
    \includegraphics[width=0.95\textwidth]{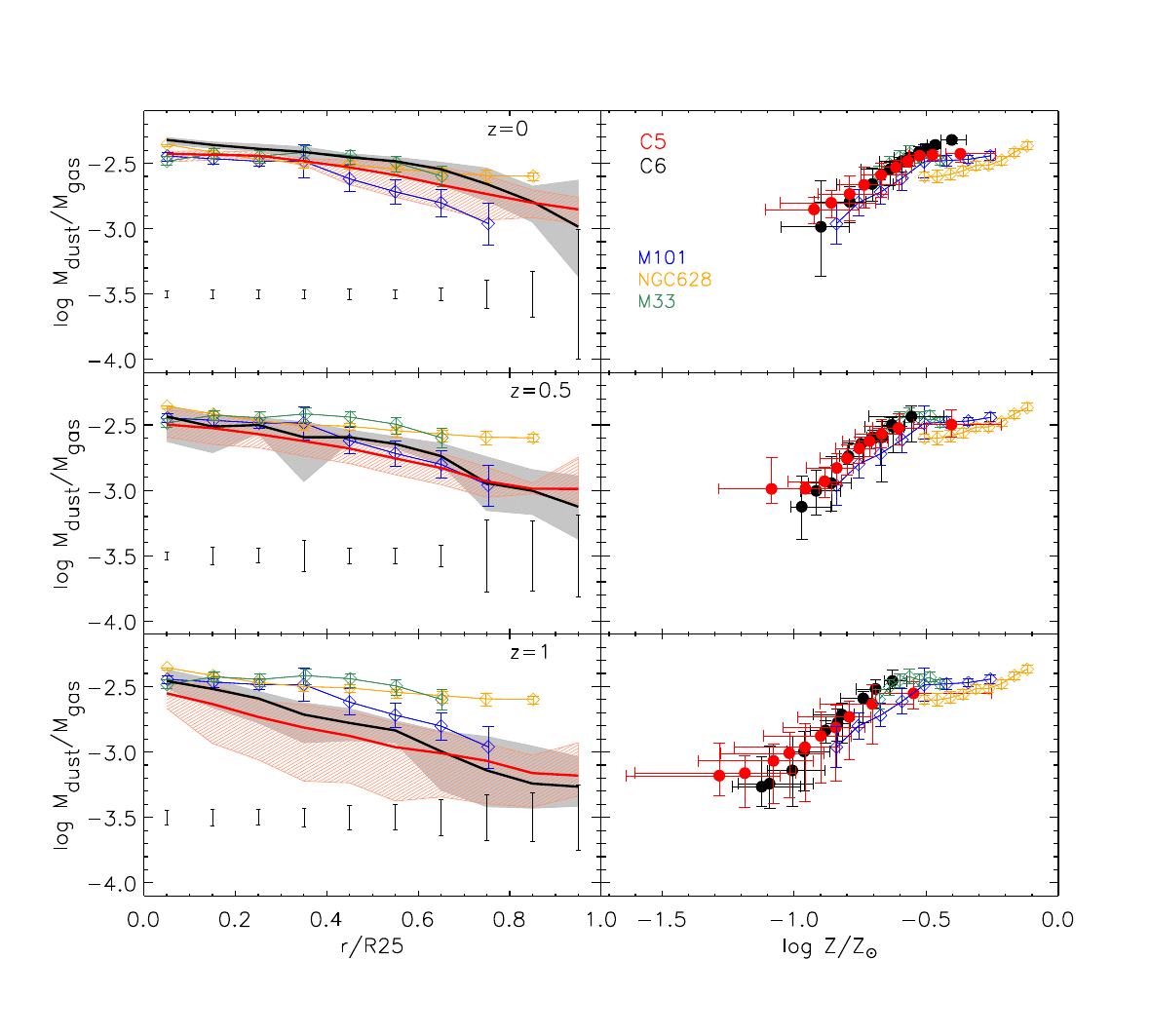}
    \vspace{-1truecm}
    \caption{
    Left panels: Mean $\dtog$ ratio per bin of galacto-centric distance. We project the simulated galaxies in a 2D grid on the disc plane, and by obtaining the amount of gas and dust in each pixel we compute the $\dtog$ profiles for each one of the 6 realizations of our C6 and C5 models. The means of the 6 profiles are shown with black and red solid lines, respectively; while the associated shaded areas cover the maximum dispersion in each bin of distance. Vertical bars at the bottom of the panels depict instead the typical dispersion of the pixels $\dtog$.
    Top, middle and bottom panels show the simulated $\dtog$ profiles at redshift z=0, 0.5 and 1, respectively.
    Empty diamonds show the profiles of three spiral galaxies (M101, NGC628 and M31) reported in \citet{relano20}
    in the local Universe, but for sake of comparison we display them also in the middle and bottom panels (z=0.5 and 1 respectively).
    Right panels: Black and red dots show the C6 and C5 mean $\dtog$ ratio but as a function of the mean gas metallicity in the same bins of distance. The horizontal and vertical error bars show the 6 realizations maximum dispersion in each of those bins. See text for more details.
    }
    \label{fig:prof_D}
\end{figure*}

\begin{figure*}
    \includegraphics[width=0.95\textwidth]{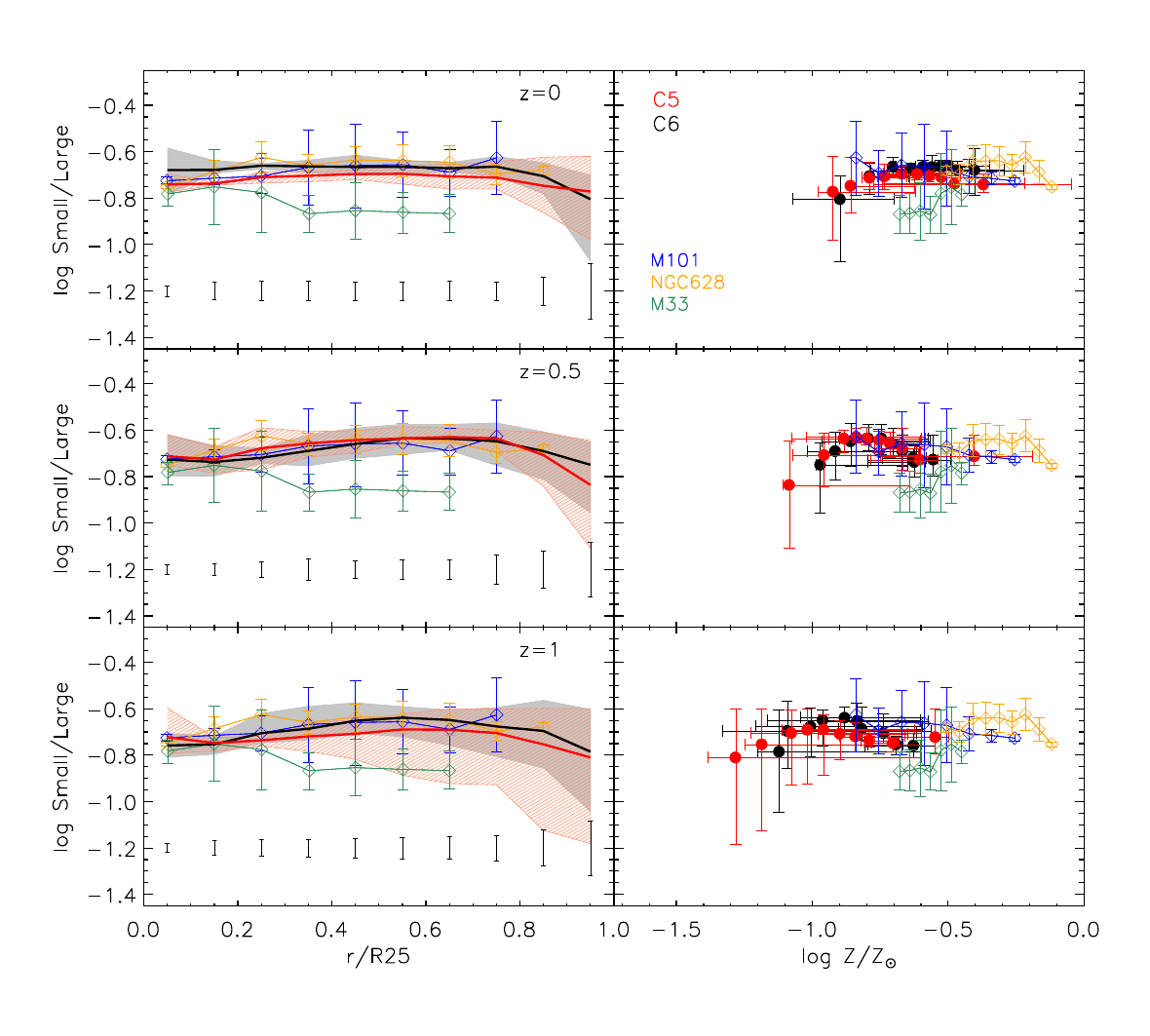}
    \vspace{-1truecm}
    \caption{Same as Fig. \ref{fig:prof_D} but for the Small-to-Large grain mass ratio.
    }
    \label{fig:prof_sl}
\end{figure*}

\begin{figure}
     \vspace{-0.7truecm}
    \includegraphics[width=\columnwidth]{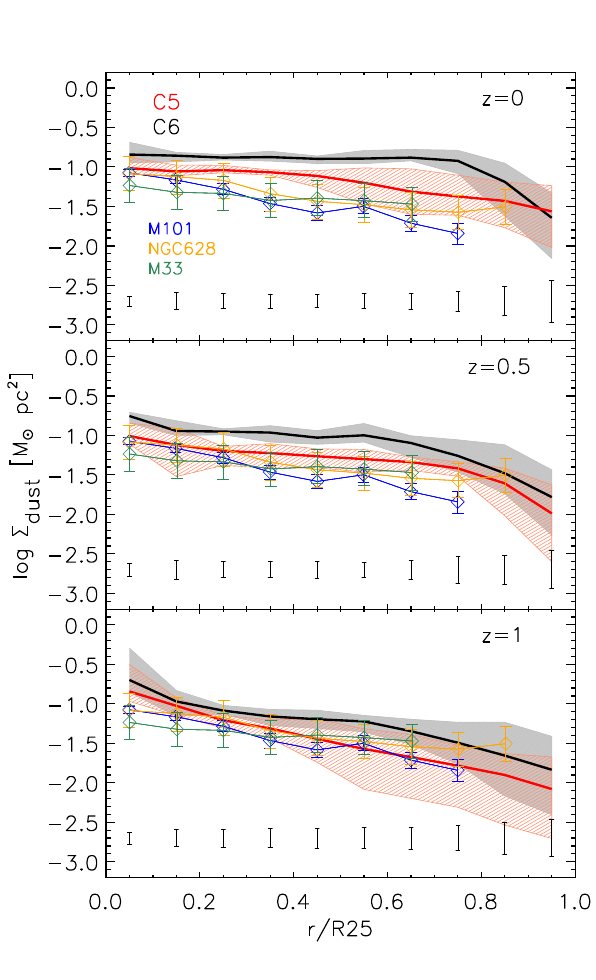}
    \vspace{-1.0truecm}
    \caption{Same as left panels of Fig. \ref{fig:prof_D} but for the dust mass surface density profiles.
    }
    \label{fig:prof_sigma}
\end{figure}

\subsection{Cooling due to Dust}
\label{subsec:cooling}
In this subsection we compare our fiducial model ($\rm{C6}$) with a model where dust cooling has been shut off ($\rm{C6\_noDustCool}$). The related figures are again Figs. \ref{fig:dust_evo_spu} and \ref{fig:baryons_evo_spu}.
The latter run is characterized by a lower dust and metal content, SFR and final stellar mass. Dust cooling promotes the presence of more dust via the combination of three effects: (i)
it decreases the temperature of hot gas and thus dust destruction by sputtering; the resulting larger amount of dust in hot gas tend to increase even more cooling; however this "loop" is progressively braked and stopped by the steep temperature decrease of both dust cooling and sputtering efficiency (Eqs. \ref{eq:duscoo} and \ref{eq:tausp});   (ii) it favours star formation (right top panel of the same figure) and therefore dust production by stars as well as (iii) metal pollution of the ISM, which increases dust growth by accretion. The inclusion of dust cooling is significant enough to enhance by $\sim 30 \%$ the SFR in the last $\sim 8$ Gyr. The increased  SFR translates into proportionally higher pollution of the ISM with metals and dust produced by the stars. As always, the SFR related increase of dust mass in the ISM wins against the slighter dust reduction due to the corresponding enhanced SNae rate and dust astration.

It is also worth to point out in this context the side effect of another, quite unrealistic, way of increasing the dust content, namely allowing an accretion of gas metals onto silicate dust grains not constrained by any specific proportions of Mg, Fe, Si and O. In other words, without the use of a {\it key element}, as described in Section \ref{sec:acc_cou}. As it can be appreciated by inspecting Figure \ref{fig:accfree}, the runs C6\_AccFree feature on average a dust content higher by a factor $\gtrsim 2$ than the fiducial one C6, again partly due to positive back-reaction, resulting into a significantly more active SF history.
These $\rm{C6\_AccFree}$ runs will be further discussed in Section \ref{sec:fre}.

These findings along with those presented in Subsection \ref{subsec:sputtering}, show that the two closely related processes of thermal sputtering and cooling due to dust, although often neglected in different kinds of galaxy formation computations, have a significant effect.

\subsection{Dust Production by SNII only}
\label{sec:IIagbIa}
Some previous simulations incorporating dust production, included for simplicity only the contribution of SNII \citep[e.g.][]{aoyama17,hou17}. According to our result (compare model $\rm{C6\_SNII}$ with the fiducial run $\rm{C6}$ in Figs.
\ref{fig:dust_evo} and \ref{fig:baryons_evo}), this approximation is good enough at  $z \lesssim 1.5$ to predict the total dust mass. At higher redshift, it leads to an underestimate of the dust mass by up to a factor two, while the Silicate/C dust mass ratio turns out to be substantially different down to a somewhat lower redshift $z\gtrsim 1$.
In particular, the importance of AGBs in affecting the dust content at early times implies that at $z \gtrsim 1.5$ the fiducial simulation features a silicate over carbon dust ratio smaller by a factor $\ge 2$, with respect to the simulation incorporating only SNII dust production. Indeed, AGB stars are essential producers of C in chemical evolution models.

The approximation of neglecting other dust production channels than SNII becomes more and more accurate at lower and lower redshift because the contribution of accretion to the total dust content increases with time, becoming dominant at $z \lesssim 0.8$. This is shown in the same Fig, \ref{fig:dust_evo}, where we included the results of the runs in  which accretion has been switched off (C6\_noAcc).

\subsection{Free Accretion of Metals}
\label{sec:fre}
The run dubbed $\rm{C6\_Accfree}$ is characterized by unconstrained accretion of metals over dust particles (Section \ref{sec:acc_cou}). In this case the accretion efficiency of each of the four elements entering into the assumed silicate composition is not regulated to maintain the same proportion as in the compound MgFeSiO$_4$, but just by its own gas abundance.
\rev{Although in the $\rm{C6\_AccFree}$ runs grains are still originally produced by stars respecting the latter composition, these runs
soon feature} an unrealistic and almost complete depletion ($\gtrsim 99 \%$) of the four elements, a too large Silicate/C dust mass ratio (Fig. \ref{fig:accfree}, see footnote \ref{foo:ratios}), and, as expected, a nonphysical "silicate" grain composition.
Indeed, we found that during most of the evolution the number of Oxygen atoms locked up into "silicate" grains is three to four time larger than that of olivine family (Mg$_X$Fe$_{(2-X)}$SiO$_4$).
The effect of this approximation on the dust content and star formation have been discussed in Section \ref{subsec:cooling}.

\subsection{Sharp Shattering}
\label{sec:sharp_sha}
Depending on the relative velocity, grain-grain collisions can cause shattering (\ref{sec:shattering}), producing smaller grains, or coagulation (\ref{sec:acc_cou}), which yields larger ones. The typical grain velocity is in turn a function of the gas density, being significantly higher in less dense environments. Previous implementations of the two-size approximation in simulations \citep[e.g.][]{aoyama17}, including that of our group for galaxy clusters \citep{gjergo18}, modelled for simplicity the transition between the two regimes by switching off shattering for $n_{gas} > 1 \, \rm{cm}^{-3}$, and turning on coagulation when $n_{gas} > 10^3 \, \rm{cm}^{-3}$.
In this work, as explained in Section \ref{sec:shattering}, we use instead a more gentle, and possibly somewhat more realistic, transition.  As we anticipated in that section, in the context of the MUPPI sub-resolution star formation and feedback model adopted here, it turned out unfeasible to get a reasonably high abundance of small grains maintaining the sharp shattering threshold previously adopted.

Fig. \ref{fig:shatt} illustrates the importance of a smooth transition. While the dust mass in large grains (solid lines in the left panel) is insensitive to the sharp density cut-off of shattering, the mass in small grains (dotted lines) results much lower for the $\rm{C6\_SharpShat}$ model than for the fiducial one, being the difference more evident at lower redshifts. As already remarked, with the present implementation  we obtain low redshift values of $S/L$ similar to those inferred from dust reprocessing models $\simeq 0.2 - 0.3$ \citep[e.g.][]{silva98,weingartner01,relano20}, while $\rm{C6\_SharpShat}$ runs predict in the last few Gyrs unsatisfactory values $S/L \lesssim 0.05$.

\rev{Before adopting the smooth transition, we performed several test runs changing the efficiency of the three processes that, in principle, may affect the relative abundance of small and large grains, namely accretion, shattering and coagulation. However, the mass ratio turns out to be little affected, remaining well below 0.1 even by increasing (decreasing) the shattering (coagulation) efficiency by a factor three, or by increasing the accretion efficiency by a factor 10. Globally speaking, coagulation dominates over shattering in the galaxy, thus most of the mass acquired by small grains by accretion is transferred to large ones. However shattering never occurs at the same time as accretion and coagulation in a particle.
Only by allowing large grain to shatter into small ones under less restrictive conditions, as a result of the smooth, and physically motivated, shattering cut-off, we obtained a satisfactory relative amount of small grains.}

\subsection{Astration versus SNae destruction}
\label{sec:no_ast}
\cite{mckinnon18}  suggested that the ISM dust loss due to the formation of stars, that is the so-called astration, can be neglected in a full dust evolution model because other dust destruction mechanisms, in particular SNae destruction (Section \ref{sec:SNdes}),  would largely dominate it. This simplification could be useful for numerical schemes like that adopted by these authors, where gas is represented by cells while specific simulation particles represent dust. By converse, with our SPH code, the implementation of the process is straightforward. To assess the absolute and relative importance of these two dust destruction mechanisms, we have run test cases switching off dust astration, meaning that SPH particles retain their full dust content when spawning new stars\footnote{\rev{In our simulations, a gas particle can generate up to four star particles before total consumption \citep[for details see][]{valentini19}.  In the C6\_noAst runs, the dust mass remains in the former particle for the first three generations. If this SPH particle needs to generate the fourth gas particle, it becomes a stellar particle, and its dust content is distributed to neighbouring gas particles.}}, as well as switching off dust SNae destruction. As can be seen in Fig. \ref{fig:astsn}, this test indicates that astration and SNae destruction are actually both important in our simulations, and at a comparable level. Neglecting astration or SNae destruction yields $\sim 50$\% more dust mass at $z=0$.
In the former case the overproduction is somewhat lower, but also the final silicate over carbon dust ratio is reduced by about 30\%.
\rev{We do not show the star formation histories of these two runs, because they turn out to be very similar to that of the fiducial runs. Thus the difference in dust content are dominated in both cases by the exclusion of the two dust destruction processes, with a minor role of the different stellar production of metals and dust.}
The difference in Silicate/C between C6\_noSNdes and C6\_noAst, is related to the fact that when dust grains are destroyed by SNae explosion, their metals are immediately available to be reincorporated in grains by accretion, while the same is not true when grains are astrated.

The different conclusion by \cite{mckinnon18} can be ascribed to their more top-heavy IMF, to their smaller lower mass limit for SNII explosion ($6 \msun{}$ instead of $8 \msun{}$), and to other minor differences in the parameter values adopted on their analytic estimate. Indeed, by using our values in the analytical estimate of their appendix C, the SNae destruction rate decreases by a factor $\simeq 6$, becoming comparable to, and somewhat smaller than, the astration destruction rate. In conclusion, none of the two processes can be neglected \rev{as a general rule, and their relative importance depends on assumptions concerning the stellar populations.}

\section{Results: Comparison with Observations}
\label{sec:obs}
In the previous sections, we have seen that our fiducial low resolution run C6 produces properties of dust content in broad keeping with what is known for local disc galaxies. We devote this section to a preliminary comparison with recent information derived from resolved observations of three nearby disc galaxies by \cite{relano20}, to get a feeling on the generic reasonableness of our results. Interestingly these observations have been already compared, in that paper, to the simulation of an isolated disk galaxy by \cite{aoyama17}, which includes a somewhat simpler treatment of dust processes based on the two-size approximation.

A few points should be taken into account in evaluating our data comparison. First, we evolve just one initial condition, selected mainly because it has been used often in literature, typically under the belief that it should produce as a final result an object not too dissimilar from the MW. Moreover, we made the choice of not adjusting the star formation and feedback parameters (the MUPPI model) with respect to the "best" ones selected in a previous work of our group \citep{valentini19}. In principle, some tuning would be required to compensate the effects of the dust related processes included now, such as the depletion of gas metals which affects the metal cooling, and the dust promoted cooling of hot gas. Finally, for the parameters related to dust modelling, we generally accepted the values suggested by previous studies of the various processes.

A related issues is numerical convergence.
\cite{scannapieco12}, compared the outcome of evolving C5 and C6 initial conditions with 13 different simulations codes, finding that numerical convergence was not good for any of them, with differences in galactic global quantities sometimes exceeding 100\%. This holds true partly for our code, which implies that it is conceivable that at even higher resolution than C5 (HR) the results would change by a non-negligible amount. On the other hand, it is now widely recognized that it is in general unrealistic to pretend strict numerical convergence of simulations involving complex sub-resolution modelling, for observables that depend strongly on it \cite[see discussion in ][]{schaye15}. This leads to the concept of {\it weak convergence}, consisting in  "convergence" provided some re-calibration of the sub-resolution model parameters is performed when resolution is changed. This task is clearly outside the scope of the present work.
In this section, our reference model is C5 of Table \ref{tab:suite}, which is that adopting fiducial parameters, but applied to the higher resolution initial conditions C5. Nevertheless, to give an idea of the level of stability of the results against resolution change, without parameter re-calibration, we plot also the result for the C6-LR fiducial simulations.

In a next paper, we plan to perform a more complete comparison of our modelling with available observational constraints, evolving also different initial conditions, and to adjust the parameters to get the best possible agreement.

\cite{relano20} fitted, on a pixel-by-pixel basis, near to far-IR maps of three spiral galaxies of different masses, M101, NGC628, and M33, to derive dust mass maps over the discs of these galaxies. The fit assumed optically thin dust emission, which could be inappropriate in the mid-IR regime where the emission can be substantially contributed by dense molecular clouds \cite[e.g.][]{silva98}.
They used the classical dust model by \cite{desert90}, which
consists of three grain populations: polycyclic aromatic
hydrocarbons (PAHs), very small grains (VSGs) and big silicate grains (BGs).
By identifying the first two components with the small grains and the third one with large grains of the two-size approximation, they also produced mass maps separately for small and large dust. We warn the reader that this identification should be regarded with some caution, since in the \cite{desert90} model the transition between VSGs and BGs occurs at 0.15 $\mu$m,
while that between small and large grains in the two-size approximation is at 0.03 $\mu$m. Moreover, the former model adopts optical properties of graphite for VSG, while small grains in our work include also a contribution from silicates. Nevertheless, \cite{relano20} compared
their results with the non-cosmological SPH simulation by \cite{aoyama17}, which implements the two-size approximation for dust evolution. Despite the aforementioned caveats, we think it is interesting to repeat the same exercise here for our model.

Figs. \ref{fig:prof_D} and \ref{fig:prof_sl} compare the radial profile and the metallicity dependence of the dust over gas mass ratio $\dtog$ (also named D/G), and small over large dust mass ratio Small/Large predicted by our  simulations at redshift 0, 0.5 and 1, with that derived from observation of the three spiral galaxies by \cite{relano20}. Of course in principle the data should be compared with the model at $z=0$, but we report them in the three panels mostly as a reference.

We mimicked as close as possible the procedure used by \cite{relano20} to compute these profiles. We projected the simulated galaxies in a 2D grid on the disc plane, and compute the amount of gas, small grains and large grains in each pixel. This allows us to compute the G/D and Small/Large \rev{mass} ratios on a pixel-by-pixel basis\footnote{The size of our pixels are 890 pc and 500 pc for C6 and C5 respectively. The latter value is used for C5 (HR) because it is very close to the data resolution for M101 and NGC 628 (509 and 490 pc respectively), and still somewhat larger than the gravitational softening ($\sim$ 445 pc). The former value used in C6 (LR) is its gravitational softening, not dramatically larger than the data resolution for these two galaxies. By converse, M33, because of its proximity, has a much better physical resolution in the data (56 pc), which is far beyond our reach.}.
We then build the {\it radial profile} for each simulated galaxy by computing 1/mean(G/D) and mean(Small/Large) among all pixels that fall in rings of 0.1 R25\footnote{R25 is the radius at which the B band surface brightness falls to 25 mag arcsec$^{-2}$. We estimate R25 by means of the stellar disc scale length, $\rm{R_d}$, using R25$\simeq$4$\rm{R_d}$ \citep{elmegreen98}.
In turn, $\rm{R_d}$ is obtained by fitting an exponential function $\Sigma(r)=\Sigma_0 \rm{exp}(-r/\rm{R_d})$ (where $\Sigma_0$ is the central surface density) to the stellar surface density profile of each simulated galaxy.} width.
The solid black and red lines in the left panels of Figs. \ref{fig:prof_D} and \ref{fig:prof_sl} correspond to the mean profile among the 6 different realizations of C6 and C5 respectively, while the shaded areas, as always, cover the full dispersion of the corresponding 6 runs. Note however that the bars presented at the bottom of this and the various logarithmic plots represent instead the dispersion found by \cite{relano20} for the pixels in each ring of distance, which they estimated as the root mean square of the quantities multiplied by the derivative of the logarithm (Rela\~no, private communication). We computed in the same way these dispersions for the simulations, which  turn out to be quite comparable to those obtained for the data. For sake of clarity we show in the plot the mean of these dispersions computed among the six realizations of C6. Those for C5 are very similar.

The \rev{gas} metallicity dependence of D/G and Small/Large \rev{mass} ratios shown in the right panels of Figs. \ref{fig:prof_D} and \ref{fig:prof_sl} are instead constructed as follows, again reproducing as much as possible the procedure used for the data. We first compute the \rev{gas (excluding dust)} metallicity (Z) in each 2D pixel, namely the ratio between total gas metal per pixel and total gas mass per pixel. Then with these values we build the Z gradient for each galaxy, associating to each radial ring the mean Z of all pixels that fall within the ring\footnote{\rev{The model gas metallicity we show considers all the gas particles in the pixels without any weight. However, since observed metallicities are obtained by means of spectroscopic observations if HII regions, we also checked that the differences introduced by considering only the multi-phase star-forming particles or weighting the ring mean by the SFR in each pixel are negligible.}}. At this point we have, for each radial bin, a mean value of D/G, Small/Large (\rev{in mass}) and Z, which are the quantities plotted in the right hand panels of Figs. \ref{fig:prof_D} and \ref{fig:prof_sl}. Horizontal and vertical error bars in these figures cover the full dispersion of the 6 different D/G, Small/Large and Z values we obtain per radial bin.

The general agreement of the model relationships at $z=0$ with those inferred from observations, albeit with substantial modelling, is remarkable, both for the gradients as well as for the normalization. The agreement is better when considering the higher resolution runs C5, but the differences between C6 and C5 are moderate. As for the surface density profiles, shown in Fig. \ref{fig:prof_sigma}, they seem to be slightly overproduced particularly in the external regions by up to $\sim 0.3$ dex, due to a shallower gradient than the "observed" one. Our possible overproduction of dust can be likely related to the shallow increase of the SFR at $z\lesssim 1$ in the simulated galaxy. Indeed the agreement with the surface density data is very good for the model at $z=1$, where the disk structure of the model galaxy is already well developed, but before the later rise of the SF activity.
This latter feature of our simulation is to some extent undesirable, with respect to the common wisdom on the star formation history of disk galaxies. It was already present in most recent MUPPI runs starting from the same initial condition but not including dust evolution \citep[see Figure 3 in][]{valentini19}.


Overall our simulation compare better than that by \cite{aoyama17} with the observational determinations by \cite{relano20}. Indeed, that simulation show (a) a dust surface density profile sharply rising toward the centre \citep[see Fig.\ 2 in ][]{relano20};   (b) to a lesser extent, also their profile of dust to gas result steeper than the observed ones \citep[Fig.\ 3 in ][]{relano20}; (c) the dust to gas ratio as a function of metallicity is under-predicted, particularly at low $Z$ \citep[Fig.\ 3 in ][]{relano20}; (d) finally,  the Small/Large grain ratio is over-predicted and shows a clear radial increase, while that inferred from observations is flat \citep[Fig.\ 3 in ][]{relano20}. Our simulated galaxy is not affected by any of these serious problems.
This could be in some sense surprising because the dust modelling by \cite{aoyama17} is relatively similar. However there are a few general differences that could contribute to their less satisfactory results: (i) they simulate an isolated galaxy rather than a cosmological initial condition; in particular, as remarked by \cite{relano20}, their initial condition produce a galaxy with a more prominent bulge than the observed ones; (ii) they do not include dust cooling nor sputtering; (iii) their treatment of chemical evolution is much less detailed, in that it includes only the SNII channel both for metal as well for dust production, adopting instant recycling approximation. Moreover it follows only the total metallicity, and not the contribution of several elements. Thus, they cannot differentiate carbon from silicate dust and they are forced to a simplified treatment of metal accretion.

\section{Summary and prospects}
\label{sec:summary}

In the context of galaxy formation simulations, we coupled a treatment of ISM dust evolution with the sub-resolution star formation and feedback model MUPPI \citep{murante10,murante15,Valentini2017,valentini19}. The dust model includes predictions for the grain size distribution, obtained by adopting the two-size approximation \citep{hirashita15}, and their chemical composition. Thus it allows, in principle, to perform more reliable post processing computations of dust reprocessing, as briefly discussed in the Section \ref{sec:SED}.
We have also added to the computation of gas cooling the contribution arising from collisions of plasma particles with grains, following  \cite{dwek81}.
The present paper was mostly devoted to present and discuss a suite of simulations conceived to study the role of various dust processes. \revrev{Within the present framework, the galaxy dust content is dominated by direct stellar production only at relatively early times, $z \gtrsim 1$. In contrast, at late time ISM dust processes are more important.
Accretion of gas metals onto preexisting seed grains, promoted by the collisional shattering of large grains into smaller ones, becomes the primary source of dust mass.
This sequence produces a significant depletion of the gas metals entering the dust composition (C, O, Mg, Fe and Si in our computations).
The relatively stable equilibrium between shattering and coagulation, which occurs under different physical conditions, sets the ratio between small and large grains to values in good keeping with observational constraints.}

It is remarkable that although our simulations include only a few of the possible consequences of the presence of dust in the ISM, namely dust promoted hot gas cooling and gas metal depletion,  they already demonstrate the relevance of dust processes in affecting galaxy evolution. In particular, sputtering and dust cooling are all but negligible in determining the dust content of the disc galaxies, and also their star formation history. The higher dust mass predicted by models with lower sputtering efficiency is not only directly driven by the slower destruction of grains in hot plasma, but also, if not mainly, by the higher star formation resulting from a faster dust promoted cooling, occurring when more grains are present in the hot ISM.

To asses the general reasonableness of our fiducial model we compared it with spatially resolved observations of three disc galaxies.
Despite the fact that we have not adjusted in any way our fiducial model, simply adopting previously selected parameter values,
we found it to be in nice agreement with resolved estimates of the dust content and of the small over large grains ratios of three nearby disc galaxies \citep{relano20}

As a natural continuation of the present work, we plan to apply the modelling described here to a cosmological volume. A preliminary step will be a careful calibration of the various parameters,  by means of an extensive comparison with available data. Clearly, even the parameters and assumptions pertinent to the star formation and feedback scheme MUPPI, which here we inherited from previous studies \cite[moslty][]{valentini19}, need some reconsideration, since the introduction of dust processes affects the system evolution.

Simulating a cosmological box will allow to investigate in statistical sense the evolution of scaling relations involving dust in galaxies, to study how the evolution of dust properties may affect the spectral energy distribution of galaxies, as well as the properties of dust in the intergalactic medium.  While these issues has been already considered by other groups, at least to some extent, \citep[][in the first case without dust size distribution]{mckinnon18,aoyama18,hou19}, the results in this field depend substantially on the adopted prescriptions for the unresolved physical processes, and it is therefore necessary to investigate their robustness against different treatments.

We also plan to take advantage of the information on the dust content, as provided by our modelling, to improve the self-consistency of other aspects of the simulations. Examples include the abundance of molecular hydrogen which enters into the MUPPI star formation and feedback model, or the role of dust in influencing the gas dynamics of galactic winds by radiation pressure.

\section*{Acknowledgements}
We are grateful to the referee James Trayford for carefully reading our manuscript and for providing constructive comments which substantially helped to improve the paper.
We warmly thank Mónica Relaño for helping us in comparing our model with the results of her group, and Peter Camps for advice in using the radiative transfer code SKIRT.
This project has received funding from the Consejo Nacional de Investigaciones Cient\'ificas y T\'ecnicas de la Rep\'ublica Argentina (CONICET) and from
the European Union's Horizon 2020 Research and Innovation Programme under the Marie Sklodowska-Curie grant agreement No 734374.
Simulations have been carried out at the computing centre of INAF (Italia). We acknowledge the computing centre of INAF-Osservatorio Astronomico di Trieste, under the coordination of the CHIPP project \citep{bertocco2019,taffoni20}, for the availability of computing resources and support.
MV is supported by the Excellence Cluster ORIGINS, which is funded by the Deutsche Forschungsgemeinschaft
(DFG, German Research Foundation) under Germany's Excellence Strategy - EXC-2094 - 390783311.
SB acknowledges financial support from Progetti di Rilevante Interesse Nazionale  (PRIN) funded by the Minisitero della Istruzione, della Università e della Ricerca (MIUR) 2015W7KAWC, the Istituto Nazionale di Fisica Nucleare (INFN) InDark grant. LS acknowledges the support from grant PRINMIUR 2017 - 20173ML3WW\_001.

\section*{Data Availability}
The data used for this article will be shared on reasonable request to the corresponding author.



\bibliographystyle{mnras}

\begin{thebibliography}{}
\makeatletter
\relax
\def\mn@urlcharsother{\let\do\@makeother \do\$\do\&\do\#\do\^\do\_\do\%\do\~}
\def\mn@doi{\begingroup\mn@urlcharsother \@ifnextchar [ {\mn@doi@}
  {\mn@doi@[]}}
\def\mn@doi@[#1]#2{\def\@tempa{#1}\ifx\@tempa\@empty \href
  {http://dx.doi.org/#2} {doi:#2}\else \href {http://dx.doi.org/#2} {#1}\fi
  \endgroup}
\def\mn@eprint#1#2{\mn@eprint@#1:#2::\@nil}
\def\mn@eprint@arXiv#1{\href {http://arxiv.org/abs/#1} {{\tt arXiv:#1}}}
\def\mn@eprint@dblp#1{\href {http://dblp.uni-trier.de/rec/bibtex/#1.xml}
  {dblp:#1}}
\def\mn@eprint@#1:#2:#3:#4\@nil{\def\@tempa {#1}\def\@tempb {#2}\def\@tempc
  {#3}\ifx \@tempc \@empty \let \@tempc \@tempb \let \@tempb \@tempa \fi \ifx
  \@tempb \@empty \def\@tempb {arXiv}\fi \@ifundefined
  {mn@eprint@\@tempb}{\@tempb:\@tempc}{\expandafter \expandafter \csname
  mn@eprint@\@tempb\endcsname \expandafter{\@tempc}}}

\bibitem[\protect\citeauthoryear{{Aoyama}, {Hou}, {Shimizu}, {Hirashita},
  {Todoroki}, {Choi}  \& {Nagamine}}{{Aoyama} et~al.}{2017}]{aoyama17}
{Aoyama} S.,  {Hou} K.-C.,  {Shimizu} I.,  {Hirashita} H.,  {Todoroki} K.,
  {Choi} J.-H.,   {Nagamine} K.,  2017, \mn@doi [\mnras]
  {10.1093/mnras/stw3061}, \href
  {http://adsabs.harvard.edu/abs/2017MNRAS.466..105A} {466, 105}

\bibitem[\protect\citeauthoryear{{Aoyama}, {Hou}, {Hirashita}, {Nagamine}  \&
  {Shimizu}}{{Aoyama} et~al.}{2018}]{aoyama18}
{Aoyama} S.,  {Hou} K.-C.,  {Hirashita} H.,  {Nagamine} K.,   {Shimizu} I.,
  2018, \mn@doi [\mnras] {10.1093/mnras/sty1431}, \href
  {https://ui.adsabs.harvard.edu/abs/2018MNRAS.478.4905A} {478, 4905}

\bibitem[\protect\citeauthoryear{{Aoyama} et~al.,}{{Aoyama}
  et~al.}{2019}]{aoyama19}
{Aoyama} S.,  et~al., 2019, \mn@doi [\mnras] {10.1093/mnras/stz021}, \href
  {https://ui.adsabs.harvard.edu/abs/2019MNRAS.484.1852A} {484, 1852}

\bibitem[\protect\citeauthoryear{{Aoyama}, {Hirashita}  \& {Nagamine}}{{Aoyama}
  et~al.}{2020}]{aoyama20}
{Aoyama} S.,  {Hirashita} H.,   {Nagamine} K.,  2020, \mn@doi [\mnras]
  {10.1093/mnras/stz3253}, \href
  {https://ui.adsabs.harvard.edu/abs/2020MNRAS.491.3844A} {491, 3844}

\bibitem[\protect\citeauthoryear{{Asano}, {Takeuchi}, {Hirashita}  \&
  {Nozawa}}{{Asano} et~al.}{2013}]{asano13}
{Asano} R.~S.,  {Takeuchi} T.~T.,  {Hirashita} H.,   {Nozawa} T.,  2013,
  \mn@doi [\mnras] {10.1093/mnras/stt506}, \href
  {http://adsabs.harvard.edu/abs/2013MNRAS.432..637A} {432, 637}

\bibitem[\protect\citeauthoryear{{Beck} et~al.,}{{Beck} et~al.}{2016}]{beck16}
{Beck} A.~M.,  et~al., 2016, \mn@doi [\mnras] {10.1093/mnras/stv2443}, \href
  {http://adsabs.harvard.edu/abs/2016MNRAS.455.2110B} {455, 2110}

\bibitem[\protect\citeauthoryear{{Bekki}}{{Bekki}}{2013}]{bekki13}
{Bekki} K.,  2013, \mn@doi [\mnras] {10.1093/mnras/stt589}, \href
  {http://adsabs.harvard.edu/abs/2013MNRAS.432.2298B} {432, 2298}

\bibitem[\protect\citeauthoryear{{Bertocco} et~al.,}{{Bertocco}
  et~al.}{2019}]{bertocco2019}
{Bertocco} S.,  et~al., 2019, arXiv e-prints, \href
  {https://ui.adsabs.harvard.edu/abs/2019arXiv191205340B} {p. arXiv:1912.05340}

\bibitem[\protect\citeauthoryear{{Bianchi} \& {Schneider}}{{Bianchi} \&
  {Schneider}}{2007}]{bianchi07}
{Bianchi} S.,  {Schneider} R.,  2007, \mn@doi [\mnras]
  {10.1111/j.1365-2966.2007.11829.x}, \href
  {http://adsabs.harvard.edu/abs/2007MNRAS.378..973B} {378, 973}

\bibitem[\protect\citeauthoryear{{Blitz} \& {Rosolowsky}}{{Blitz} \&
  {Rosolowsky}}{2006}]{blitz06}
{Blitz} L.,  {Rosolowsky} E.,  2006, \mn@doi [\apj] {10.1086/505417}, \href
  {http://adsabs.harvard.edu/abs/2006ApJ...650..933B} {650, 933}

\bibitem[\protect\citeauthoryear{{Burke} \& {Silk}}{{Burke} \&
  {Silk}}{1974}]{burke74}
{Burke} J.~R.,  {Silk} J.,  1974, \mn@doi [\apj] {10.1086/152840}, \href
  {https://ui.adsabs.harvard.edu/abs/1974ApJ...190....1B} {190, 1}

\bibitem[\protect\citeauthoryear{{Calura}, {Pipino}  \& {Matteucci}}{{Calura}
  et~al.}{2008}]{calura08}
{Calura} F.,  {Pipino} A.,   {Matteucci} F.,  2008, \mn@doi [\aap]
  {10.1051/0004-6361:20078090}, \href
  {http://adsabs.harvard.edu/abs/2008A%26A...479..669C} {479, 669}

\bibitem[\protect\citeauthoryear{{Camps} \& {Baes}}{{Camps} \&
  {Baes}}{2020}]{camps20}
{Camps} P.,  {Baes} M.,  2020, \mn@doi [Astronomy and Computing]
  {10.1016/j.ascom.2020.100381}, \href
  {https://ui.adsabs.harvard.edu/abs/2020A&C....3100381C} {31, 100381}

\bibitem[\protect\citeauthoryear{{Camps}, {Trayford}, {Baes}, {Theuns},
  {Schaller}  \& {Schaye}}{{Camps} et~al.}{2016}]{camps16}
{Camps} P.,  {Trayford} J.~W.,  {Baes} M.,  {Theuns} T.,  {Schaller} M.,
  {Schaye} J.,  2016, \mn@doi [\mnras] {10.1093/mnras/stw1735}, \href
  {https://ui.adsabs.harvard.edu/abs/2016MNRAS.462.1057C} {462, 1057}

\bibitem[\protect\citeauthoryear{{Charlot} \& {Fall}}{{Charlot} \&
  {Fall}}{2000}]{charlot00}
{Charlot} S.,  {Fall} S.~M.,  2000, \mn@doi [\apj] {10.1086/309250}, \href
  {https://ui.adsabs.harvard.edu/abs/2000ApJ...539..718C} {539, 718}

\bibitem[\protect\citeauthoryear{{Davies}, {Crain}  \& {Pontzen}}{{Davies}
  et~al.}{2020}]{Davies2020}
{Davies} J.~J.,  {Crain} R.~A.,   {Pontzen} A.,  2020, arXiv e-prints, \href
  {https://ui.adsabs.harvard.edu/abs/2020arXiv200613221D} {p. arXiv:2006.13221}

\bibitem[\protect\citeauthoryear{{Desert}, {Boulanger}  \& {Puget}}{{Desert}
  et~al.}{1990}]{desert90}
{Desert} F.~X.,  {Boulanger} F.,   {Puget} J.~L.,  1990, \aap, \href
  {https://ui.adsabs.harvard.edu/abs/1990A&A...237..215D} {500, 313}

\bibitem[\protect\citeauthoryear{{Dom{\'{\i}}nguez-Tenreiro}, {Obreja},
  {Granato}, {Schurer}, {Alpresa}, {Silva}, {Brook}  \&
  {Serna}}{{Dom{\'{\i}}nguez-Tenreiro} et~al.}{2014}]{dominguez14}
{Dom{\'{\i}}nguez-Tenreiro} R.,  {Obreja} A.,  {Granato} G.~L.,  {Schurer} A.,
  {Alpresa} P.,  {Silva} L.,  {Brook} C.~B.,   {Serna} A.,  2014, \mn@doi
  [\mnras] {10.1093/mnras/stu240}, \href
  {http://esoads.eso.org/abs/2014MNRAS.439.3868D} {439, 3868}

\bibitem[\protect\citeauthoryear{{Draine}}{{Draine}}{2003}]{draine03}
{Draine} B.~T.,  2003, \mn@doi [\araa]
  {10.1146/annurev.astro.41.011802.094840}, \href
  {http://esoads.eso.org/abs/2003ARA%26A..41..241D} {41, 241}

\bibitem[\protect\citeauthoryear{{Draine} \& {Salpeter}}{{Draine} \&
  {Salpeter}}{1979}]{draine79}
{Draine} B.~T.,  {Salpeter} E.~E.,  1979, \mn@doi [\apj] {10.1086/157206},
  \href {http://adsabs.harvard.edu/abs/1979ApJ...231..438D} {231, 438}

\bibitem[\protect\citeauthoryear{{Dwek}}{{Dwek}}{1998}]{dwek98}
{Dwek} E.,  1998, \mn@doi [\apj] {10.1086/305829}, \href
  {http://esoads.eso.org/abs/1998ApJ...501..643D} {501, 643}

\bibitem[\protect\citeauthoryear{{Dwek} \& {Werner}}{{Dwek} \&
  {Werner}}{1981}]{dwek81}
{Dwek} E.,  {Werner} M.~W.,  1981, \mn@doi [\apj] {10.1086/159138}, \href
  {https://ui.adsabs.harvard.edu/abs/1981ApJ...248..138D} {248, 138}

\bibitem[\protect\citeauthoryear{{Elmegreen}}{{Elmegreen}}{1998}]{elmegreen98}
{Elmegreen} D.~M.,  1998, {Galaxies and galactic structure}

\bibitem[\protect\citeauthoryear{{Feldmann}}{{Feldmann}}{2015}]{Feldmann2015}
{Feldmann} R.,  2015, \mn@doi [\mnras] {10.1093/mnras/stv552}, \href
  {https://ui.adsabs.harvard.edu/abs/2015MNRAS.449.3274F} {449, 3274}

\bibitem[\protect\citeauthoryear{{Genel} et~al.,}{{Genel}
  et~al.}{2019}]{genel19}
{Genel} S.,  et~al., 2019, \mn@doi [\apj] {10.3847/1538-4357/aaf4bb}, \href
  {https://ui.adsabs.harvard.edu/abs/2019ApJ...871...21G} {871, 21}

\bibitem[\protect\citeauthoryear{{Gioannini}, {Matteucci}, {Vladilo}  \&
  {Calura}}{{Gioannini} et~al.}{2017}]{gioannini17}
{Gioannini} L.,  {Matteucci} F.,  {Vladilo} G.,   {Calura} F.,  2017, \mn@doi
  [\mnras] {10.1093/mnras/stw2343}, \href
  {https://ui.adsabs.harvard.edu/abs/2017MNRAS.464..985G} {464, 985}

\bibitem[\protect\citeauthoryear{{Gjergo}, {Granato}, {Murante},
  {Ragone-Figueroa}, {Tornatore}  \& {Borgani}}{{Gjergo}
  et~al.}{2018}]{gjergo18}
{Gjergo} E.,  {Granato} G.~L.,  {Murante} G.,  {Ragone-Figueroa} C.,
  {Tornatore} L.,   {Borgani} S.,  2018, \mn@doi [\mnras]
  {10.1093/mnras/sty1564}, \href
  {https://ui.adsabs.harvard.edu/abs/2018MNRAS.479.2588G} {479, 2588}

\bibitem[\protect\citeauthoryear{{Goz}, {Monaco}, {Granato}, {Murante},
  {Dom{\'\i}nguez-Tenreiro}, {Obreja}, {Annunziatella}  \& {Tescari}}{{Goz}
  et~al.}{2017}]{goz17}
{Goz} D.,  {Monaco} P.,  {Granato} G.~L.,  {Murante} G.,
  {Dom{\'\i}nguez-Tenreiro} R.,  {Obreja} A.,  {Annunziatella} M.,   {Tescari}
  E.,  2017, \mn@doi [\mnras] {10.1093/mnras/stx869}, \href
  {https://ui.adsabs.harvard.edu/abs/2017MNRAS.469.3775G} {469, 3775}

\bibitem[\protect\citeauthoryear{{Granato}, {Lacey}, {Silva}, {Bressan},
  {Baugh}, {Cole}  \& {Frenk}}{{Granato} et~al.}{2000}]{granato00}
{Granato} G.~L.,  {Lacey} C.~G.,  {Silva} L.,  {Bressan} A.,  {Baugh} C.~M.,
  {Cole} S.,   {Frenk} C.~S.,  2000, \mn@doi [\apj] {10.1086/317032}, \href
  {http://esoads.eso.org/abs/2000ApJ...542..710G} {542, 710}

\bibitem[\protect\citeauthoryear{{Granato}, {De Zotti}, {Silva}, {Bressan}  \&
  {Danese}}{{Granato} et~al.}{2004}]{granato04}
{Granato} G.~L.,  {De Zotti} G.,  {Silva} L.,  {Bressan} A.,   {Danese} L.,
  2004, \mn@doi [\apj] {10.1086/379875}, \href
  {https://ui.adsabs.harvard.edu/abs/2004ApJ...600..580G} {600, 580}

\bibitem[\protect\citeauthoryear{{Graziani}, {Schneider}, {Ginolfi}, {Hunt},
  {Maio}, {Glatzle}  \& {Ciardi}}{{Graziani} et~al.}{2020}]{graziani20}
{Graziani} L.,  {Schneider} R.,  {Ginolfi} M.,  {Hunt} L.~K.,  {Maio} U.,
  {Glatzle} M.,   {Ciardi} B.,  2020, \mn@doi [\mnras] {10.1093/mnras/staa796},
  \href {https://ui.adsabs.harvard.edu/abs/2020MNRAS.494.1071G} {494, 1071}

\bibitem[\protect\citeauthoryear{{Hirashita}}{{Hirashita}}{1999}]{hirashita99}
{Hirashita} H.,  1999, \mn@doi [\apjl] {10.1086/311806}, \href
  {http://adsabs.harvard.edu/abs/1999ApJ...510L..99H} {510, L99}

\bibitem[\protect\citeauthoryear{{Hirashita}}{{Hirashita}}{2015}]{hirashita15}
{Hirashita} H.,  2015, \mn@doi [\mnras] {10.1093/mnras/stu2617}, \href
  {http://esoads.eso.org/abs/2015MNRAS.447.2937H} {447, 2937}

\bibitem[\protect\citeauthoryear{{Hirashita} \& {Aoyama}}{{Hirashita} \&
  {Aoyama}}{2019}]{hirashita19}
{Hirashita} H.,  {Aoyama} S.,  2019, \mn@doi [\mnras] {10.1093/mnras/sty2838},
  \href {https://ui.adsabs.harvard.edu/abs/2019MNRAS.482.2555H} {482, 2555}

\bibitem[\protect\citeauthoryear{{Hirashita} \& {Kuo}}{{Hirashita} \&
  {Kuo}}{2011}]{hirashita11}
{Hirashita} H.,  {Kuo} T.-M.,  2011, \mn@doi [\mnras]
  {10.1111/j.1365-2966.2011.19131.x}, \href
  {http://esoads.eso.org/abs/2011MNRAS.416.1340H} {416, 1340}

\bibitem[\protect\citeauthoryear{{Hirashita} \& {Voshchinnikov}}{{Hirashita} \&
  {Voshchinnikov}}{2014}]{hirashita14}
{Hirashita} H.,  {Voshchinnikov} N.~V.,  2014, \mn@doi [\mnras]
  {10.1093/mnras/stt1997}, \href
  {http://esoads.eso.org/abs/2014MNRAS.437.1636H} {437, 1636}

\bibitem[\protect\citeauthoryear{{Hirashita} \& {Yan}}{{Hirashita} \&
  {Yan}}{2009}]{hirashita09}
{Hirashita} H.,  {Yan} H.,  2009, \mn@doi [\mnras]
  {10.1111/j.1365-2966.2009.14405.x}, \href
  {http://adsabs.harvard.edu/abs/2009MNRAS.394.1061H} {394, 1061}

\bibitem[\protect\citeauthoryear{{Hou}, {Hirashita}, {Nagamine}, {Aoyama}  \&
  {Shimizu}}{{Hou} et~al.}{2017}]{hou17}
{Hou} K.-C.,  {Hirashita} H.,  {Nagamine} K.,  {Aoyama} S.,   {Shimizu} I.,
  2017, \mn@doi [\mnras] {10.1093/mnras/stx877}, \href
  {http://adsabs.harvard.edu/abs/2017MNRAS.469..870H} {469, 870}

\bibitem[\protect\citeauthoryear{{Hou}, {Aoyama}, {Hirashita}, {Nagamine}  \&
  {Shimizu}}{{Hou} et~al.}{2019}]{hou19}
{Hou} K.-C.,  {Aoyama} S.,  {Hirashita} H.,  {Nagamine} K.,   {Shimizu} I.,
  2019, \mn@doi [\mnras] {10.1093/mnras/stz121}, \href
  {https://ui.adsabs.harvard.edu/abs/2019MNRAS.485.1727H} {485, 1727}

\bibitem[\protect\citeauthoryear{{Jenkins}}{{Jenkins}}{2009}]{jenkins09}
{Jenkins} E.~B.,  2009, \mn@doi [\apj] {10.1088/0004-637X/700/2/1299}, \href
  {http://esoads.eso.org/abs/2009ApJ...700.1299J} {700, 1299}

\bibitem[\protect\citeauthoryear{{Jonsson}, {Groves}  \& {Cox}}{{Jonsson}
  et~al.}{2010}]{jonsson10}
{Jonsson} P.,  {Groves} B.~A.,   {Cox} T.~J.,  2010, \mn@doi [\mnras]
  {10.1111/j.1365-2966.2009.16087.x}, \href
  {https://ui.adsabs.harvard.edu/abs/2010MNRAS.403...17J} {403, 17}

\bibitem[\protect\citeauthoryear{{Keller}, {Wadsley}, {Wang}  \&
  {Kruijssen}}{{Keller} et~al.}{2019}]{keller19}
{Keller} B.~W.,  {Wadsley} J.~W.,  {Wang} L.,   {Kruijssen} J.~M.~D.,  2019,
  \mn@doi [\mnras] {10.1093/mnras/sty2859}, \href
  {https://ui.adsabs.harvard.edu/abs/2019MNRAS.482.2244K} {482, 2244}

\bibitem[\protect\citeauthoryear{{Kroupa}, {Tout}  \& {Gilmore}}{{Kroupa}
  et~al.}{1993}]{kroupa93}
{Kroupa} P.,  {Tout} C.~A.,   {Gilmore} G.,  1993, \mn@doi [\mnras]
  {10.1093/mnras/262.3.545}, \href
  {https://ui.adsabs.harvard.edu/abs/1993MNRAS.262..545K} {262, 545}

\bibitem[\protect\citeauthoryear{{Li}, {Narayanan}  \& {Dav{\'e}}}{{Li}
  et~al.}{2019}]{li19}
{Li} Q.,  {Narayanan} D.,   {Dav{\'e}} R.,  2019, \mn@doi [\mnras]
  {10.1093/mnras/stz2684}, \href
  {https://ui.adsabs.harvard.edu/abs/2019MNRAS.490.1425L} {490, 1425}

\bibitem[\protect\citeauthoryear{{Mattsson}}{{Mattsson}}{2016}]{mattsson16}
{Mattsson} L.,  2016, \mn@doi [\planss] {10.1016/j.pss.2016.05.002}, \href
  {https://ui.adsabs.harvard.edu/abs/2016P&SS..133..107M} {133, 107}

\bibitem[\protect\citeauthoryear{{McKee}}{{McKee}}{1989}]{mckee89}
{McKee} C.,  1989, in {Allamandola} L.~J.,  {Tielens} A.~G.~G.~M.,  eds,  IAU
  Symposium Vol. 135, Interstellar Dust. p.~431

\bibitem[\protect\citeauthoryear{{McKee}, {Hollenbach}, {Seab}  \&
  {Tielens}}{{McKee} et~al.}{1987}]{mckee87}
{McKee} C.~F.,  {Hollenbach} D.~J.,  {Seab} G.~C.,   {Tielens} A.~G.~G.~M.,
  1987, \mn@doi [\apj] {10.1086/165403}, \href
  {http://adsabs.harvard.edu/abs/1987ApJ...318..674M} {318, 674}

\bibitem[\protect\citeauthoryear{{McKinnon}, {Torrey}  \&
  {Vogelsberger}}{{McKinnon} et~al.}{2016}]{mckinnon16}
{McKinnon} R.,  {Torrey} P.,   {Vogelsberger} M.,  2016, \mn@doi [\mnras]
  {10.1093/mnras/stw253}, \href
  {http://adsabs.harvard.edu/abs/2016MNRAS.457.3775M} {457, 3775}

\bibitem[\protect\citeauthoryear{{McKinnon}, {Torrey}, {Vogelsberger},
  {Hayward}  \& {Marinacci}}{{McKinnon} et~al.}{2017}]{mckinnon17}
{McKinnon} R.,  {Torrey} P.,  {Vogelsberger} M.,  {Hayward} C.~C.,
  {Marinacci} F.,  2017, \mn@doi [\mnras] {10.1093/mnras/stx467}, \href
  {http://adsabs.harvard.edu/abs/2017MNRAS.468.1505M} {468, 1505}

\bibitem[\protect\citeauthoryear{{McKinnon}, {Vogelsberger}, {Torrey},
  {Marinacci}  \& {Kannan}}{{McKinnon} et~al.}{2018}]{mckinnon18}
{McKinnon} R.,  {Vogelsberger} M.,  {Torrey} P.,  {Marinacci} F.,   {Kannan}
  R.,  2018, \mn@doi [\mnras] {10.1093/mnras/sty1248}, \href
  {https://ui.adsabs.harvard.edu/abs/2018MNRAS.478.2851M} {478, 2851}

\bibitem[\protect\citeauthoryear{{Montier} \& {Giard}}{{Montier} \&
  {Giard}}{2004}]{montier04}
{Montier} L.~A.,  {Giard} M.,  2004, \mn@doi [\aap]
  {10.1051/0004-6361:20034365}, \href
  {http://adsabs.harvard.edu/abs/2004A%26A...417..401M} {417, 401}

\bibitem[\protect\citeauthoryear{{Murante}, {Monaco}, {Giovalli}, {Borgani}  \&
  {Diaferio}}{{Murante} et~al.}{2010}]{murante10}
{Murante} G.,  {Monaco} P.,  {Giovalli} M.,  {Borgani} S.,   {Diaferio} A.,
  2010, \mn@doi [\mnras] {10.1111/j.1365-2966.2010.16567.x}, \href
  {https://ui.adsabs.harvard.edu/abs/2010MNRAS.405.1491M} {405, 1491}

\bibitem[\protect\citeauthoryear{{Murante}, {Monaco}, {Borgani}, {Tornatore},
  {Dolag}  \& {Goz}}{{Murante} et~al.}{2015}]{murante15}
{Murante} G.,  {Monaco} P.,  {Borgani} S.,  {Tornatore} L.,  {Dolag} K.,
  {Goz} D.,  2015, \mn@doi [\mnras] {10.1093/mnras/stu2400}, \href
  {http://adsabs.harvard.edu/abs/2015MNRAS.447..178M} {447, 178}

\bibitem[\protect\citeauthoryear{{Murray}, {Quataert}  \& {Thompson}}{{Murray}
  et~al.}{2005}]{Murray2005}
{Murray} N.,  {Quataert} E.,   {Thompson} T.~A.,  2005, \mn@doi [\apj]
  {10.1086/426067}, \href
  {https://ui.adsabs.harvard.edu/abs/2005ApJ...618..569M} {618, 569}

\bibitem[\protect\citeauthoryear{{Nozawa}, {Kozasa}, {Habe}, {Dwek}, {Umeda},
  {Tominaga}, {Maeda}  \& {Nomoto}}{{Nozawa} et~al.}{2007}]{nozawa07}
{Nozawa} T.,  {Kozasa} T.,  {Habe} A.,  {Dwek} E.,  {Umeda} H.,  {Tominaga} N.,
   {Maeda} K.,   {Nomoto} K.,  2007, \mn@doi [\apj] {10.1086/520621}, \href
  {http://adsabs.harvard.edu/abs/2007ApJ...666..955N} {666, 955}

\bibitem[\protect\citeauthoryear{{Oh}, {Smith}, {Peacock}  \& {Khochfar}}{{Oh}
  et~al.}{2020}]{kiat20}
{Oh} B.~K.,  {Smith} B.~D.,  {Peacock} J.~A.,   {Khochfar} S.,  2020, \mn@doi
  [\mnras] {10.1093/mnras/staa2318}, \href
  {https://ui.adsabs.harvard.edu/abs/2020MNRAS.497.5203O} {497, 5203}

\bibitem[\protect\citeauthoryear{{Osman}, {Bekki}  \& {Cortese}}{{Osman}
  et~al.}{2020}]{osman2020}
{Osman} O.,  {Bekki} K.,   {Cortese} L.,  2020, arXiv e-prints, \href
  {https://ui.adsabs.harvard.edu/abs/2020arXiv200908051O} {p. arXiv:2009.08051}

\bibitem[\protect\citeauthoryear{{Panuzzo}, {Granato}, {Buat}, {Inoue},
  {Silva}, {Iglesias-P{\'a}ramo}  \& {Bressan}}{{Panuzzo}
  et~al.}{2007}]{panuzzo07}
{Panuzzo} P.,  {Granato} G.~L.,  {Buat} V.,  {Inoue} A.~K.,  {Silva} L.,
  {Iglesias-P{\'a}ramo} J.,   {Bressan} A.,  2007, \mn@doi [\mnras]
  {10.1111/j.1365-2966.2006.11337.x}, \href
  {https://ui.adsabs.harvard.edu/abs/2007MNRAS.375..640P} {375, 640}

\bibitem[\protect\citeauthoryear{{Planck Collaboration (XLIII)}
  et~al.,}{{Planck Collaboration (XLIII)} et~al.}{2016}]{43planck16}
{Planck Collaboration (XLIII)} et~al., 2016, \mn@doi [\aap]
  {10.1051/0004-6361/201628522}, \href
  {http://adsabs.harvard.edu/abs/2016A%26A...596A.104P} {596, A104}

\bibitem[\protect\citeauthoryear{{Popping}, {Somerville}  \&
  {Galametz}}{{Popping} et~al.}{2017}]{popping16}
{Popping} G.,  {Somerville} R.~S.,   {Galametz} M.,  2017, \mn@doi [\mnras]
  {10.1093/mnras/stx1545}, \href
  {http://adsabs.harvard.edu/abs/2017MNRAS.471.3152P} {471, 3152}

\bibitem[\protect\citeauthoryear{{Rela{\~n}o}, {Lisenfeld}, {Hou}, {De Looze},
  {V{\'\i}lchez}  \& {Kennicutt}}{{Rela{\~n}o} et~al.}{2020}]{relano20}
{Rela{\~n}o} M.,  {Lisenfeld} U.,  {Hou} K.~C.,  {De Looze} I.,  {V{\'\i}lchez}
  J.~M.,   {Kennicutt} R.~C.,  2020, \mn@doi [\aap]
  {10.1051/0004-6361/201937087}, \href
  {https://ui.adsabs.harvard.edu/abs/2020A&A...636A..18R} {636, A18}

\bibitem[\protect\citeauthoryear{{Scannapieco} et~al.,}{{Scannapieco}
  et~al.}{2012}]{scannapieco12}
{Scannapieco} C.,  et~al., 2012, \mn@doi [\mnras]
  {10.1111/j.1365-2966.2012.20993.x}, \href
  {https://ui.adsabs.harvard.edu/abs/2012MNRAS.423.1726S} {423, 1726}

\bibitem[\protect\citeauthoryear{{Schaye} et~al.,}{{Schaye}
  et~al.}{2015}]{schaye15}
{Schaye} J.,  et~al., 2015, \mn@doi [\mnras] {10.1093/mnras/stu2058}, \href
  {http://adsabs.harvard.edu/abs/2015MNRAS.446..521S} {446, 521}

\bibitem[\protect\citeauthoryear{{Schneider}, {Valiante}, {Ventura},
  {dell'Agli}, {Di Criscienzo}, {Hirashita}  \& {Kemper}}{{Schneider}
  et~al.}{2014}]{schneider14}
{Schneider} R.,  {Valiante} R.,  {Ventura} P.,  {dell'Agli} F.,  {Di
  Criscienzo} M.,  {Hirashita} H.,   {Kemper} F.,  2014, \mn@doi [\mnras]
  {10.1093/mnras/stu861}, \href
  {http://adsabs.harvard.edu/abs/2014MNRAS.442.1440S} {442, 1440}

\bibitem[\protect\citeauthoryear{{Silva}, {Granato}, {Bressan}  \&
  {Danese}}{{Silva} et~al.}{1998}]{silva98}
{Silva} L.,  {Granato} G.~L.,  {Bressan} A.,   {Danese} L.,  1998, \mn@doi
  [\apj] {10.1086/306476}, \href
  {http://esoads.eso.org/abs/1998ApJ...509..103S} {509, 103}

\bibitem[\protect\citeauthoryear{{Springel}}{{Springel}}{2005}]{springel05}
{Springel} V.,  2005, \mn@doi [\mnras] {10.1111/j.1365-2966.2005.09655.x},
  \href {http://esoads.eso.org/abs/2005MNRAS.364.1105S} {364, 1105}

\bibitem[\protect\citeauthoryear{{Springel} \& {Hernquist}}{{Springel} \&
  {Hernquist}}{2003}]{springel03}
{Springel} V.,  {Hernquist} L.,  2003, \mn@doi [\mnras]
  {10.1046/j.1365-8711.2003.06206.x}, \href
  {http://esoads.eso.org/abs/2003MNRAS.339..289S} {339, 289}

\bibitem[\protect\citeauthoryear{{Springel} et~al.,}{{Springel}
  et~al.}{2008}]{Springel2008}
{Springel} V.,  et~al., 2008, \mn@doi [\mnras]
  {10.1111/j.1365-2966.2008.14066.x}, \href
  {https://ui.adsabs.harvard.edu/abs/2008MNRAS.391.1685S} {391, 1685}

\bibitem[\protect\citeauthoryear{{Sumpter} \& {Van Loo}}{{Sumpter} \& {Van
  Loo}}{2020}]{Sumpter2020}
{Sumpter} R.,  {Van Loo} S.,  2020, \mn@doi [\mnras] {10.1093/mnras/staa846},
  \href {https://ui.adsabs.harvard.edu/abs/2020MNRAS.494.2147S} {494, 2147}

\bibitem[\protect\citeauthoryear{{Taffoni}, {Becciani}, {Garilli}, {Maggio},
  {Pasian}, {Umana}, {Smareglia}  \& {Vitello}}{{Taffoni}
  et~al.}{2020}]{taffoni20}
{Taffoni} G.,  {Becciani} U.,  {Garilli} B.,  {Maggio} G.,  {Pasian} F.,
  {Umana} G.,  {Smareglia} R.,   {Vitello} F.,  2020, arXiv e-prints, \href
  {https://ui.adsabs.harvard.edu/abs/2020arXiv200201283T} {p. arXiv:2002.01283}

\bibitem[\protect\citeauthoryear{{Tielens}, {McKee}, {Seab}  \&
  {Hollenbach}}{{Tielens} et~al.}{1994}]{tielens94}
{Tielens} A.~G.~G.~M.,  {McKee} C.~F.,  {Seab} C.~G.,   {Hollenbach} D.~J.,
  1994, \mn@doi [\apj] {10.1086/174488}, \href
  {http://adsabs.harvard.edu/abs/1994ApJ...431..321T} {431, 321}

\bibitem[\protect\citeauthoryear{{Tornatore}, {Borgani}, {Dolag}  \&
  {Matteucci}}{{Tornatore} et~al.}{2007}]{tornatore07}
{Tornatore} L.,  {Borgani} S.,  {Dolag} K.,   {Matteucci} F.,  2007, \mn@doi
  [\mnras] {10.1111/j.1365-2966.2007.12070.x}, \href
  {http://esoads.eso.org/abs/2007MNRAS.382.1050T} {382, 1050}

\bibitem[\protect\citeauthoryear{{Triani}, {Sinha}, {Croton}, {Pacifici}  \&
  {Dwek}}{{Triani} et~al.}{2020}]{triani20}
{Triani} D.~P.,  {Sinha} M.,  {Croton} D.~J.,  {Pacifici} C.,   {Dwek} E.,
  2020, \mn@doi [\mnras] {10.1093/mnras/staa446}, \href
  {https://ui.adsabs.harvard.edu/abs/2020MNRAS.493.2490T} {493, 2490}

\bibitem[\protect\citeauthoryear{{Tsai} \& {Mathews}}{{Tsai} \&
  {Mathews}}{1995}]{tsai95}
{Tsai} J.~C.,  {Mathews} W.~G.,  1995, \mn@doi [\apj] {10.1086/175943}, \href
  {http://esoads.eso.org/abs/1995ApJ...448...84T} {448, 84}

\bibitem[\protect\citeauthoryear{{Valentini} \& {Brighenti}}{{Valentini} \&
  {Brighenti}}{2015}]{Valentini2015}
{Valentini} M.,  {Brighenti} F.,  2015, \mn@doi [\mnras]
  {10.1093/mnras/stv090}, \href
  {https://ui.adsabs.harvard.edu/abs/2015MNRAS.448.1979V} {448, 1979}

\bibitem[\protect\citeauthoryear{{Valentini}, {Murante}, {Borgani}, {Monaco},
  {Bressan}  \& {Beck}}{{Valentini} et~al.}{2017}]{Valentini2017}
{Valentini} M.,  {Murante} G.,  {Borgani} S.,  {Monaco} P.,  {Bressan} A.,
  {Beck} A.~M.,  2017, \mn@doi [\mnras] {10.1093/mnras/stx1352}, \href
  {https://ui.adsabs.harvard.edu/abs/2017MNRAS.470.3167V} {470, 3167}

\bibitem[\protect\citeauthoryear{{Valentini}, {Borgani}, {Bressan}, {Murante},
  {Tornatore}  \& {Monaco}}{{Valentini} et~al.}{2019}]{valentini19}
{Valentini} M.,  {Borgani} S.,  {Bressan} A.,  {Murante} G.,  {Tornatore} L.,
  {Monaco} P.,  2019, \mn@doi [\mnras] {10.1093/mnras/stz492}, \href
  {https://ui.adsabs.harvard.edu/abs/2019MNRAS.485.1384V} {485, 1384}

\bibitem[\protect\citeauthoryear{{Valiante}, {Schneider}, {Salvadori}  \&
  {Bianchi}}{{Valiante} et~al.}{2011}]{valiante11}
{Valiante} R.,  {Schneider} R.,  {Salvadori} S.,   {Bianchi} S.,  2011, \mn@doi
  [\mnras] {10.1111/j.1365-2966.2011.19168.x}, \href
  {http://adsabs.harvard.edu/abs/2011MNRAS.416.1916V} {416, 1916}

\bibitem[\protect\citeauthoryear{{Vijayan}, {Clay}, {Thomas}, {Yates},
  {Wilkins}  \& {Henriques}}{{Vijayan} et~al.}{2019}]{vijayan19}
{Vijayan} A.~P.,  {Clay} S.~J.,  {Thomas} P.~A.,  {Yates} R.~M.,  {Wilkins}
  S.~M.,   {Henriques} B.~M.,  2019, \mn@doi [\mnras] {10.1093/mnras/stz1948},
  \href {https://ui.adsabs.harvard.edu/abs/2019MNRAS.489.4072V} {489, 4072}

\bibitem[\protect\citeauthoryear{{Vogelsberger}, {McKinnon}, {O'Neil},
  {Marinacci}, {Torrey}  \& {Kannan}}{{Vogelsberger}
  et~al.}{2019}]{vogelsberger19}
{Vogelsberger} M.,  {McKinnon} R.,  {O'Neil} S.,  {Marinacci} F.,  {Torrey} P.,
    {Kannan} R.,  2019, \mn@doi [\mnras] {10.1093/mnras/stz1644}, \href
  {https://ui.adsabs.harvard.edu/abs/2019MNRAS.487.4870V} {487, 4870}

\bibitem[\protect\citeauthoryear{{Wakelam} et~al.,}{{Wakelam}
  et~al.}{2017}]{wakelam17}
{Wakelam} V.,  et~al., 2017, \mn@doi [Molecular Astrophysics]
  {10.1016/j.molap.2017.11.001}, \href
  {https://ui.adsabs.harvard.edu/abs/2017MolAs...9....1W} {9, 1}

\bibitem[\protect\citeauthoryear{{Weingartner} \& {Draine}}{{Weingartner} \&
  {Draine}}{2001}]{weingartner01}
{Weingartner} J.~C.,  {Draine} B.~T.,  2001, \mn@doi [\apj] {10.1086/318651},
  \href {http://adsabs.harvard.edu/abs/2001ApJ...548..296W} {548, 296}

\bibitem[\protect\citeauthoryear{{Yan}, {Lazarian}  \& {Draine}}{{Yan}
  et~al.}{2004}]{yan04}
{Yan} H.,  {Lazarian} A.,   {Draine} B.~T.,  2004, \mn@doi [\apj]
  {10.1086/425111}, \href {http://adsabs.harvard.edu/abs/2004ApJ...616..895Y}
  {616, 895}

\bibitem[\protect\citeauthoryear{{Zhukovska}, {Gail}  \&
  {Trieloff}}{{Zhukovska} et~al.}{2008}]{zhukovska08}
{Zhukovska} S.,  {Gail} H.-P.,   {Trieloff} M.,  2008, \mn@doi [\aap]
  {10.1051/0004-6361:20077789}, \href
  {http://adsabs.harvard.edu/abs/2008A%26A...479..453Z} {479, 453}

\bibitem[\protect\citeauthoryear{{Zhukovska}, {Dobbs}, {Jenkins}  \&
  {Klessen}}{{Zhukovska} et~al.}{2016}]{zhukovska16}
{Zhukovska} S.,  {Dobbs} C.,  {Jenkins} E.~B.,   {Klessen} R.~S.,  2016,
  \mn@doi [\apj] {10.3847/0004-637X/831/2/147}, \href
  {http://adsabs.harvard.edu/abs/2016ApJ...831..147Z} {831, 147}

\makeatother
\end{thebibliography}





\bsp	
\label{lastpage}
\end{document}